\documentclass[amsmath,amssymb,10pt,aps,notitlepage,prx,twocolumn,longbibliography]{revtex4-2} 
\usepackage[colorlinks=true,linkcolor=violet,citecolor=violet,urlcolor=violet]{hyperref}
 \usepackage{lipsum}
 \usepackage{graphicx}
\usepackage{latexsym}
\usepackage{amsmath}
\usepackage[table]{xcolor}
\usepackage{amsthm}
\usepackage{amssymb}
\usepackage{epstopdf} 
\usepackage{enumitem}
\usepackage{setspace}
\usepackage{dcolumn}
\usepackage{bm}
\usepackage{setspace} 
\usepackage{slashed}
\usepackage{color}
\usepackage{youngtab}
\usepackage{tikz}
\usepackage{tikz-3dplot}
\usepackage{tikz-3dplot}
\usepackage{braket}
\usepackage{subfigure}
\usepackage{xcolor}
\usepackage{makecell}
\usepackage{mathtools}

\usepackage[thicklines]{cancel}
\newcommand*{\mynot}[1]{\renewcommand{\CancelColor}{\color{#1}}\xcancel}

\DeclareMathAlphabet{\mathpzc}{OT1}{pzc}{m}{it}

\usepackage{IEEEtrantools}
 \usepackage{moreverb}
\allowdisplaybreaks

\begin{document}

\title{Anyon condensation and confinement transition in a Kitaev spin liquid bilayer}

\author{Kyusung Hwang}
\email{khwang@kias.re.kr}
\affiliation{School of Physics, Korea Institute for Advanced Study, Seoul 02455, Korea}

\date{\today}
\begin{abstract}
Transitions between quantum spin liquids (QSLs) are fundamental problems lying beyond the Landau paradigm and requiring a deep understanding of the entanglement structures of QSLs called topological orders. The novel concept of anyon condensation has been proposed as a theoretical mechanism, predicting various possible transitions between topological orders, but it has long been elusive to confirm the mechanism in quantum spin systems. Here, we introduce a concrete spin model that incarnates the mechanism of anyon condensation transition. 
Our model harbors two topological QSLs in different parameter regions, a non-abelian Kitaev spin liquid (KSL) bilayer state and a resonating valence bond (RVB) state.
The bilayer-KSL-to-RVB transition indeed occurs by the mechanism of anyon condensation, which we identify by using parton theories and exact diagonalization studies.
Moreover, we observe ``anyon confinement'' phenomena in our numerical results, akin to the quark confinement in high energy physics.
Namely, non-abelian Ising anyons of the bilayer KSL are confined in the transition to the RVB state.
Implications and extensions of this study are discussed in various aspects such as (i) anyon-condensed multilayer construction of the Kitaev's sixteenfold way of anyon theories, (ii) additional vison condensation transition from the RVB to a valence bond solid (VBS) in the Kitaev bilayer system, (iii) dynamical anyon condensation in a non-Hermitian Kitaev bilayer, (iv) generalizations of our model to other lattice geometries, and (v) experimental realizations.
This work puts together the two fascinating QSLs that are extensively studied in modern condensed matter and quantum physics into a concrete spin model, offering a comprehensive picture that unifies the anyon physics of the Kitaev spin liquids and the resonating valence bonds.
\end{abstract}

\maketitle

\tableofcontents

\section{Introduction}

Anyons are exotic particles that exhibit fractional statistics beyond conventional bosons and fermions~\cite{Wilczek1982}.
Such nontrivial particles can emerge in topological phases with long-range entanglement such as fractional quantum Hall liquids~\cite{WenBook,FradkinBook} and quantum spin liquids (QSLs)~\cite{WenBook,FradkinBook,SachdevBook,Kitaev2003,Kitaev2006,LevinWen2005,Savary2016,Norman2016,Ng2017,Knolle2019,Broholm2020,Motome2020,Trebst2022,Takagi2019}.
Recently, various QSLs and anyons have been active subjects of research due to potential applications in quantum technologies such as quantum memories and quantum computations~\cite{Kasahara2018,Satzinger2021,Semeghini2021,Google2023,Iqbal2023,Deng2023,GoogleQEC2023,Lukin2024,Preskill2002,Freedman2003,Nayak2008,Bonesteel2005,Trebst2008,Bombin2010,Pachos2017,Pachos2012,Simon2023book,Tchernyshyov2013,Tchernyshyov2014,Jiang2015,Pachos2020,Tantivasadakarn2023}.
Kitaev's spin liquid states of the toric code and the honeycomb spin model~\cite{Kitaev2003,Kitaev2006,Kee2019,Hickey2019,Trivedi2019,Batista2019} and resonating valence bond states of quantum dimer models~\cite{Rokhsar1988,Moessner2001,Misguich2002,Samajdar2020,Verresen2021,Verresen2022} are promising examples. Both have not only exact solutions but also available experimental platforms such as the quantum magnet $\alpha$-RuCl$_3$~\cite{Motome2020,Trebst2022,Kasahara2018,Takagi2019} and various quantum processors based on superconducting qubits, trapped-ions, and reconfigurable Rydberg atom arrays~\cite{Satzinger2021,Semeghini2021,Deng2023,GoogleQEC2023,Google2023,Iqbal2023,Lukin2024}.

Transitions between QSLs with anyons are less studied, yet outstanding problems in modern condensed matter and quantum physics that promise a profound understanding of quantum entanglement and anyons.
A theoretical mechanism for such topological transitions has been formulated using the concept of anyon condensation in their seminal work by Bais and Slingerland~\cite{Bais2009}.
In this mechanism, anyon condensation reconstructs many-body quantum entanglement and anyons leading to a new topological phase and new anyons, where the braiding statistics with the condensed anyon plays a key role for the determination of the fate of each anyon, i.e., ``confined''~vs.~``deconfined''.
The anyon condensation mechanism predicts a variety of transitions among topological phases with anyons, providing global insights on a wide class of topological quantum matter~\cite{Bais2009,Bais2009,Bombin2008,Barkeshli2010,Burnell2011,Burnell2012,Bais2014,Neupert2016,Burnell2018,Kong2014,Teo2015,Barkeshli2019,Schmidt2020,Schmidt2022,Meng2021,Schuch2021,Schuch2022,Huxford2023,Doherty2024,Burnell2023}.
In the field of quantum information, anyon condensation has been an important concept that underlies in designing and manipulating topological quantum codes for fault-tolerant quantum computations~\cite{Brown2017,Brown2018,Browne2022,Ellison2022,Brown2022}.
Unfortunately, it has been elusive to confirm the mechanism in quantum spin systems, i.e., QSL-to-QSL transitions by anyon condensation, because of the scarcity of appropriate microscopic models and the difficulty with defining an order parameter for anyon condensation in terms of local spin operators.

In quantum magnetism, studies of anyon physics have been focused on symmetry breaking transitions from QSLs to long-range orders in frustrated magnets including kagome and triangular lattice antiferromagnets.
$\mathbb{Z}_2$ spin liquids such as short-ranged resonating valence bonds and the toric code may be the simplest cases, described by $\mathbb{Z}_2$ gauge theories and four types of anyon excitations: trivial boson with no fractionalization ($1$), bosonic spinon ($e$), bosonic vison ($m$), and fermionic spinon ($\epsilon=e\times m$)~\cite{Kitaev2003,Wen1991,Sachdev1992,Senthil2000,Fradkin2001,Balents2002,Wang2006,Poilblanc2010,White2011,Depenbrock2012,Lu2011,Iqbal2011,Read1989,Huh2011,Zhang2012,Jiang2012,Messio2012,Hermele2013,Wan2013,Hwang2015,Zaletel2015,Lu2017,Meng2018BFG,Wessel2018BFG,Jiang2012,White2015,Sheng2015,Mila2007,Misguich2008,Slagle2014,Meng2021QDM,Meng2022QDM}. Mutual statistics exist between different anyon species.
The bosonic spinon carries a spin-1/2 quantum number, thus condensing the bosonic spinon leads to a long-range magnetic order~\cite{Sachdev1992,Wang2006,Messio2012,Wessel2018BFG}.
Unlike the spinon, the vison is a spin-0 excitation, preserving time-reversal symmetry. Vison condensation breaks lattice symmetries yielding a crystalline order of spin-singlet dimers, i.e., valence bond solid (VBS)~\cite{Moessner2001,Mila2007,Misguich2008,Huh2011,Slagle2014,Hwang2015,Meng2021QDM,Meng2018BFG}. 
In such transitions to VBS orders (or magnetic orders), the system becomes trivial in the sense that every nontrivial anyon gets either condensed or confined.
Essential features of the transitions can be summarized as follows.
\begin{equation}
\color{violet}
\left\{
\begin{array}{c}
{\bf {\mathbb{\bf Z}}_2~{\rm \bf spin~liquid}}
\\
\\
1,e,m,\epsilon
\end{array}
\right\}
\xRightarrow[]{\langle m \rangle \ne 0} 
\left\{
\begin{array}{c}
{{\rm \bf VBS}}
\\
\\
1
\end{array}
\right\}
:~~\makecell{\rm symmetry\\ \rm breaking}
\nonumber
\end{equation}

\begin{figure}[tb]
\includegraphics[width=\linewidth]{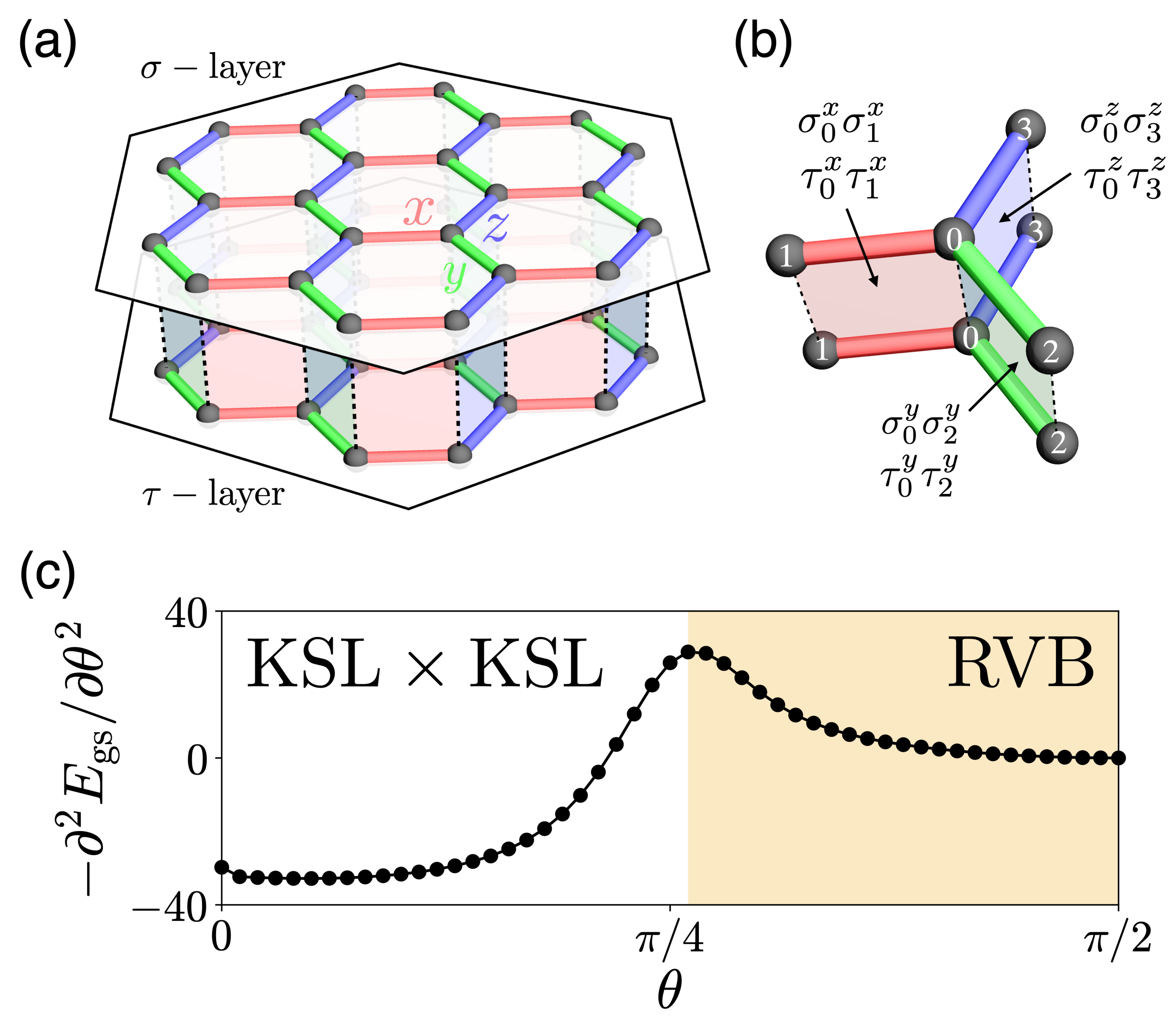}
\caption{Bilayer spin model.
(a) Honeycomb lattice bilayer with the AA stacking.
Pauli spins, $\sigma$ and $\tau$, reside on the upper and lower layers coupled by the intra-layer Kitaev interactions ($K_\sigma$ and $K_\tau$) and the four-spin inter-layer interactions ($G$). The $x,y,z$-bonds are denoted by different colors of red, green, and blue.
The figure depicts the $(24+24)$-site bilayer cluster used in the numerical exact diagonalization.
(b) Illustration of the four-spin inter-layer interactions.
(c) Phase diagram of the model.
The KSL$\times$KSL and RVB states are connected by a continuous transition at $\theta_{\rm c}\simeq0.26\pi$; indicated by a small peak in the derivative of the ground state energy, $-\partial^2 E_{\rm gs}/\partial \theta^2$.
}
\label{fig:1}
\end{figure}

In this paper, we extend the scope of anyon condensation physics to quantum phase transitions between QSLs.
Especially, we construct a spin model that establishes a transition from a QSL (having non-abelian anyons) to another QSL (having only abelian anyons) via anyon condensation.
We find such a topological transition in a bilayer system of Kitaev spin liquids.
To be specific, our spin model harbors a non-abelian Kitaev spin liquid (KSL) bilayer state and a resonating valence bond (RVB) state in different parameter regions.
The RVB state is induced by entangling the KSL bilayer with bond-dependent inter-layer interactions similar to the Kitaev interactions~(Fig.~\ref{fig:1}).

The nature of the transition between the KSL bilayer state and the RVB state is identified by computing the condensed anyon and order parameters.
In the KSL bilayer state, each layer supports three different anyons: trivial boson ($1$), non-abelian Ising anyon ($\sigma$) and fermion ($\psi$). There is nontrivial braiding between $\sigma$ and $\psi$; a $\psi$-particle sees a $\sigma$-particle as a $\mathbb{Z}_2$ flux. In the bilayer-KSL-to-RVB transition, the fermion-pair ($\psi_{\rm I}\boxtimes\psi_{\rm II}$) between layer-I and layer-II is condensed by inter-layer interactions.
The anyon physics realized by our spin model is outlined as follows.
\begin{eqnarray}
&&
\color{violet}
\left\{
\begin{array}{c}
{\rm \bf KSL}_{\bf I}
\\
\\
1_{\rm I},\sigma_{\rm I},\psi_{\rm I}
\end{array}
\right\}
\boxtimes
\left\{
\begin{array}{c}
{\rm \bf KSL}_{\bf II}
\\
\\
1_{\rm II},\sigma_{\rm II},\psi_{\rm II}
\end{array}
\right\}
\nonumber\\
\color{violet}
\xRightarrow[]{\langle \psi_{\rm I}\boxtimes\psi_{\rm II} \rangle \ne 0} 
&&
\color{violet}
\left\{
\begin{array}{c}
{\rm \bf RVB}
\\
\\
1,e,m,\epsilon
\end{array}
\right\}
: ~~\makecell{\rm symmetry \\ \rm preserved}
\nonumber
\end{eqnarray}
The two points, (i) $\psi_{\rm I}\boxtimes\psi_{\rm II}$ is the condensed anyon and (ii) nontrivial braiding exists between $\psi$ and $\sigma$ in each layer, are the key elements to understand the anyon physics of the bilayer-KSL-to-RVB transition as we shall see in Sec.~\ref{sec:anyon}.

\begin{figure*}[tb]
\includegraphics[width=\linewidth]{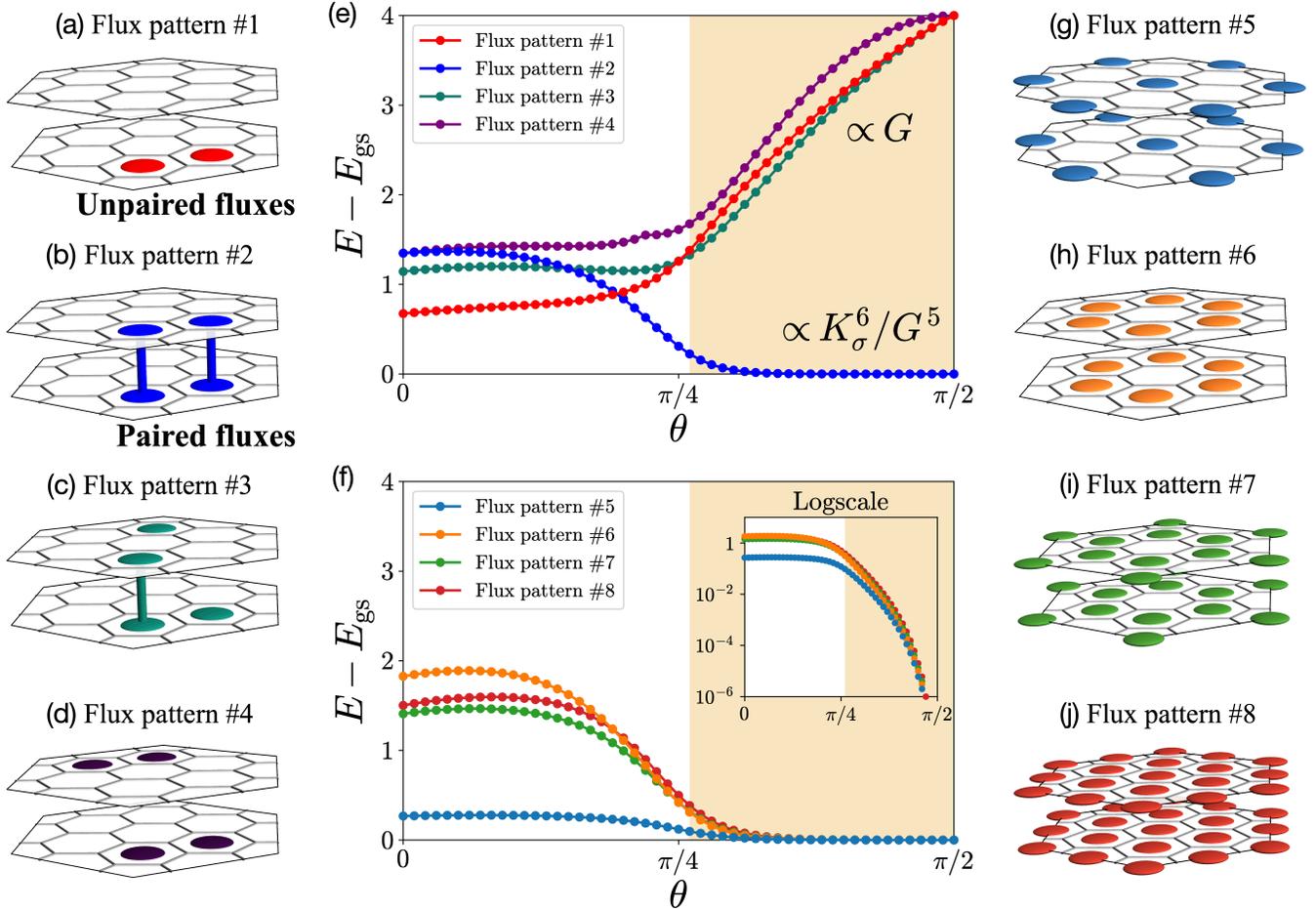}
\caption{Confinement of the Ising anyons.
(a)-(d),(g)-(j) Visualizations of eight different flux patterns (\#1,2,3,4,5,6,7,8). 
Colored disks denote the fluxes with $W=-1$ or $Z=-1$. The dumbbell-shaped disks highlight the paired fluxes between the two layers located at same positions ($W_p=Z_p=-1$).
(e),(f) The lowest excitation energy profiles in the eight flux patterns.
The inset of panel (f) shows the excitation energies in a log scale.
In the RVB phase, the paired fluxes ($\sigma_{\rm I}\boxtimes\sigma_{\rm II}$) have vanishingly small excitation energies ($\propto K_\sigma^6/G^5$) whereas the unpaired fluxes ($\sigma_{\rm I}$ \& $\sigma_{\rm II}$) have extremely large energy costs ($\propto G$).
}
\label{fig:2}
\end{figure*}

Remarkably, our model allows to study the phenomena of ``anyon confinement'' akin to the quark confinement in high energy physics. Namely, non-abelian Ising anyons ($\sigma_{\rm I}~\&~\sigma_{\rm II}$) of the KSL bilayer get confined, i.e., they cannot be directly observed in the low energy physics of the RVB state. We confirm the confinement physics by our numerical exact diagonalization calculations~(Fig.~\ref{fig:2}).

Unlike transitions from $\mathbb{Z}_2$ spin liquids to VBS orders, the bilayer-KSL-to-RVB transition does not break any symmetries of the system.
The anyon condensation only reconstructs the underlying entanglement structure and supported anyons (i.e., topological order).

Our main results are displayed in Figs.~\ref{fig:1},~\ref{fig:2},~and~\ref{fig:9}.
Figure~\ref{fig:1} shows the phase diagram of the model obtained by numerical exact diagonalization. 
In particular, Fig.~\ref{fig:2} highlights the confinement of Ising anyons in the RVB state by demonstrating extremely large energy costs for the excitations.
In Fig.~\ref{fig:9}, we provide a numerical evidence on the existence of the anyon condensation transition.

This work builds a bridge between the two intriguing QSLs extensively studied over the fields of topological quantum matter, quantum magnetism, and quantum information, and offers a comprehensive picture unifying the anyon physics of the Kitaev spin liquids and the resonating valence bonds.

\subsection{Outline of the paper}

The paper is organized as follows.
In Sec.~\ref{sec:2}, we introduce the bilayer spin model with an emphasis on the conserved quantities. The RVB state and the KSL$\times$KSL state arise in the strong coupling limit and the weak coupling limit of inter-layer interactions, respectively. In Sec.~\ref{sec:RVB}, an effective quantum dimer model is developed for the RVB state by conducting a sixth order degenerate perturbation theory. We introduce a dimer representation for the original spin model and the hardcore dimer constraint defining the dimer Hilbert space, which plays a crucial role in understanding the anyon condensation and anyon confinement phenomena from our numerical results later. In Sec.~\ref{sec:KSL2}, we construct a Majorana mean-field theory for the KSL$\times$KSL state and show that time-reversal can be spontaneously broken in each layer due to inter-layer interactions. It is shown that both layers have finite energy gaps and nonzero Chern numbers with opposite signs, keeping the whole system achiral.
In Sec.~\ref{sec:ED}, we investigate the full phase diagram of the model by numerical exact diagonalization methods on a (24+24)-site bilayer cluster.
The RVB state and the KSL$\times$KSL state are characterized and distinguished by computing various quantities including entanglement entropy, hardcore dimer constraint, chirality structure factor, and topological degeneracy.
Importantly, we find that the numerical results are all consistent with the effective theories in Secs.~\ref{sec:RVB} and \ref{sec:KSL2}.

Section~\ref{sec:anyon} is the most important part, where we discuss our main results on anyon condensation and anyon confinement. We start by articulating the non-abelian ${\rm Ising}\times\overline{\rm Ising}$ topological order of the KSL$\times$KSL state and the abelian $\mathbb{Z}_2$ toric code topological order of the RVB state.
We review the anyon condensation mechanism for the transition between the two topological orders.
Then, we present numerical evidences for the anyon condensation transition in our model. We clarify the condensed anyon by explicitly calculating associated order parameters.
Furthermore, we elucidate the confinement phenomena of Ising anyons by numerically investigating excitation energies of Ising anyons.
We will show a simple picture that enables to intuitively understand the anyon confinement from the condensed anyon.

In Sec.~\ref{sec:general-cases}, we explore generic parameter regimes of the model and show that the emergence of the RVB state or $\mathbb{Z}_2$ spin liquid is determined by the sign structure of the coupling constants of the model.

Section~\ref{sec:discussion} is another important part, where we discuss implications of this work and promising future directions. 
To name a few, we will discuss (i) the Kitaev's sixteenfold way of anyon theories in the perspective of anyon condensation, (ii) vison condensation transition from the RVB state to a VBS state in our Kitaev bilayer system, (iii) a non-Hermitian Kitaev bilayer and dynamical anyon condensation, (iv) generalizations of our model to other lattice geometries, and (v) experimental realizations.

Appendices \ref{app:A} and \ref{app:B} provide details of the sixth order degenerate perturbation theory and the quantum dimer model for the RVB state.

\section{Bilayer model\label{sec:2}}

We place two copies of the Kitaev honeycomb model~\cite{Kitaev2006} on a bilayer geometry of AA stacking as illustrated in Fig.~\ref{fig:1}(a).
In this bilayer setup, our model Hamiltonian consists of three parts:
\begin{eqnarray}
H
&=& K_\sigma \sum_{\langle jk \rangle_\gamma}^{\textup{upper~layer}} \sigma_j^\gamma \sigma_k^\gamma 
+ K_\tau \sum_{\langle jk \rangle_\gamma}^{\textup{lower~layer}} \tau_j^\gamma \tau_k^\gamma
\nonumber\\
&+&
G \sum_{\langle jk \rangle_\gamma}^{\textup{inter-layer}} \sigma_j^\gamma \sigma_k^\gamma \tau_j^\gamma \tau_k^\gamma,
\label{eq:model}
\end{eqnarray}
where $\sigma^\gamma$ and $\tau^\gamma$ ($\gamma=x,y,z$) are Pauli spins on the upper and lower layers coupled by the bond-dependent Kitaev interactions ($K_{\sigma}$, $K_{\tau}$) and also the inter-layer interactions ($G$).
We label upper and lower layer spins with same site-indices ($j,k$), and the inter-layer interactions are nothing but the products of adjacent upper-layer and lower-layer Kitaev interactions, $\sigma_{j}^\gamma\sigma_{k}^\gamma\times\tau_{j}^\gamma\tau_{k}^\gamma$ [Fig.~\ref{fig:1}(b)].
The Hamiltonian commutes with the hexagon plaquette operators defined on the upper and lower layers,
\begin{equation}
\hat{W}_p = 
\sigma_1^z 
\sigma_2^y
\sigma_3^x
\sigma_4^z
\sigma_5^y
\sigma_6^x
~~~~
\&
~~~~
\hat{Z}_p = 
\tau_1^z 
\tau_2^y
\tau_3^x
\tau_4^z
\tau_5^y
\tau_6^x,
\label{eq:plaquette-op}
\end{equation}
{\it i.e.,} $[H,\hat{W}_p]=[H,\hat{Z}_p]=[\hat{W}_p,\hat{Z}_{p'}]=0$; see Fig.~\ref{fig:3}(a) for the site convention within a hexagon plaquette $p$.

In this paper, we mainly focus on the parameter region,
\begin{equation}
K_\sigma=-K_\tau=\cos\theta ~~~\&~~~ G=\sin\theta,
\label{eq:parameter-region}
\end{equation}
where $0\leq \theta \leq \pi/2$.
The model allows two different topological QSLs: (i) KSL$\times$KSL bilayer state in the weak coupling limit, $|K_{\sigma,\tau}| \gg G$ ($\theta \approx 0$), and (ii) RVB state in the strong coupling limit, $|K_{\sigma,\tau}| \ll G$ ($\theta \approx \pi/2$).
We obtain insights on the QSLs by developing analytical theories first for the two limits.

\section{RVB state: strong coupling limit\label{sec:RVB}}

To understand the strong coupling limit in an intuitive way, we employ a dimer mapping to a dual kagome lattice.
Using the dimer mapping, we derive an effective quantum dimer model and the RVB state.

\subsection{Dimer mapping to a dual kagome lattice}

\begin{figure}[b]
\includegraphics[width=\linewidth]{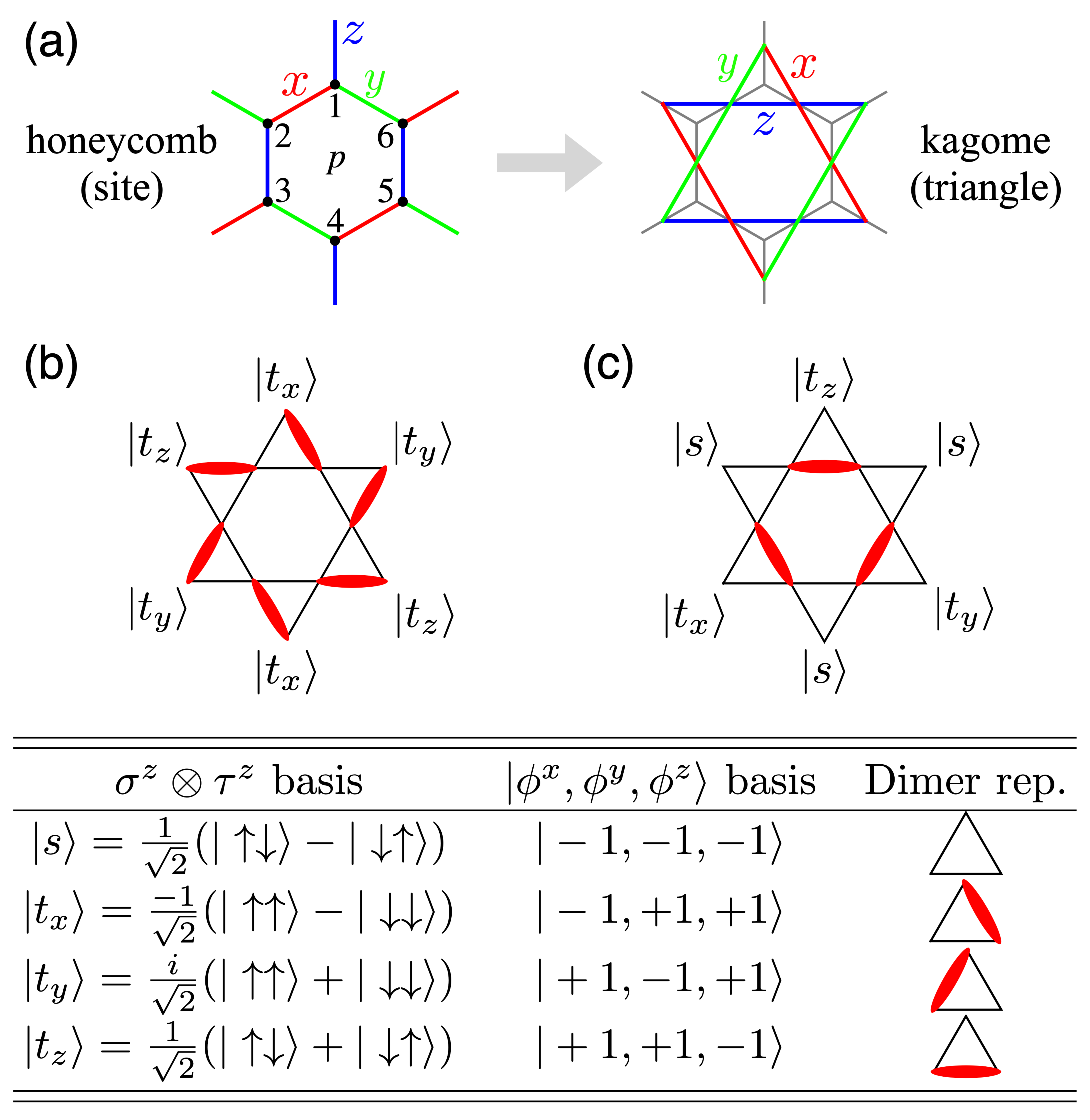}
\caption{Dimer mapping to the dual kagome lattice.
(a) Mapping from the honeycomb lattice to the kagome lattice.
Different colors denote the three bond characters, $x$(red), $y$(green), and $z$(blue).
The numbers ($1\sim6$) indicate the site convention within a hexagon plaquette $p$.
(b),(c) Illustrations of several dimer states.
The four local states $\{|s\rangle,|t_{x}\rangle,|t_{y}\rangle,|t_{z}\rangle\}$ and the dimer mapping are defined in the bottom table.
}
\label{fig:3}
\end{figure}

The bilayer model can be regarded as a single layer honeycomb model with four states per site.
Each site may have either spin-singlet state $|s\rangle$ or spin-triplet state $|t_{x,y,z}\rangle$ as shown in the table of Fig.~\ref{fig:3}.
We take a mapping from the honeycomb lattice to a dual kagome lattice.
In Fig.~\ref{fig:3}(a), the dual kagome lattice is constructed by connecting the mid-points of the bonds of the honeycomb lattice.
Sites of the honeycomb lattice are now replaced by triangles of the kagome lattice.
Interestingly, the kagome lattice has three different bond directions, which are perpendicular to the $x,y,z$-bond directions of the honeycomb lattice.
By using this property, we assign a bond character ($x,y,z$) to each bond of the kagome lattice [denoted by different colors in Fig.~\ref{fig:3}(a)].
In this mapping, the four local states $\{ |s\rangle, |t_x \rangle, |t_y \rangle, |t_z \rangle \} $ on the honeycomb lattice are represented by four dimer states $\{| \parbox{0.4cm}{\includegraphics[width=\linewidth]{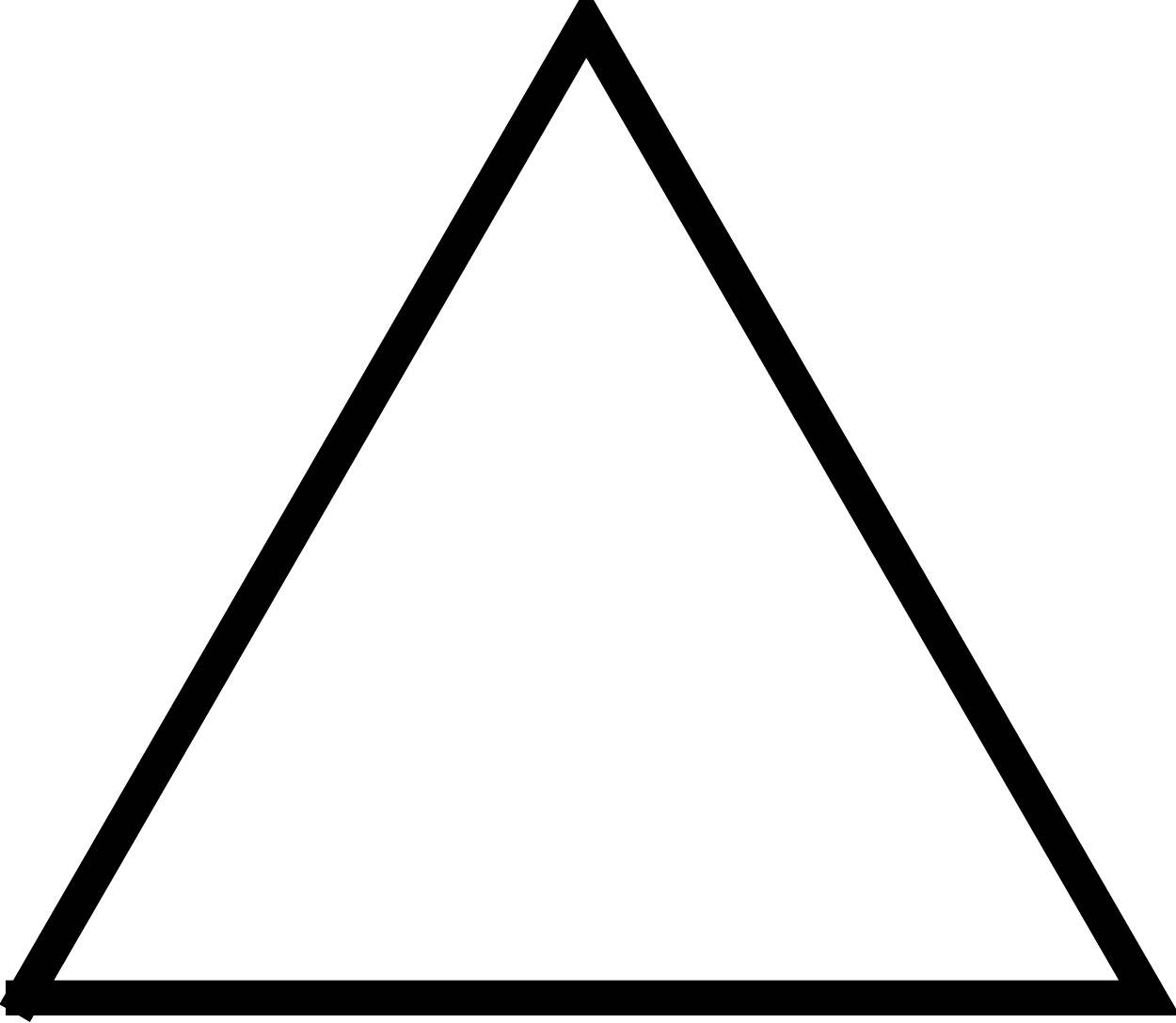}} \rangle,| \parbox{0.4cm}{\includegraphics[width=\linewidth]{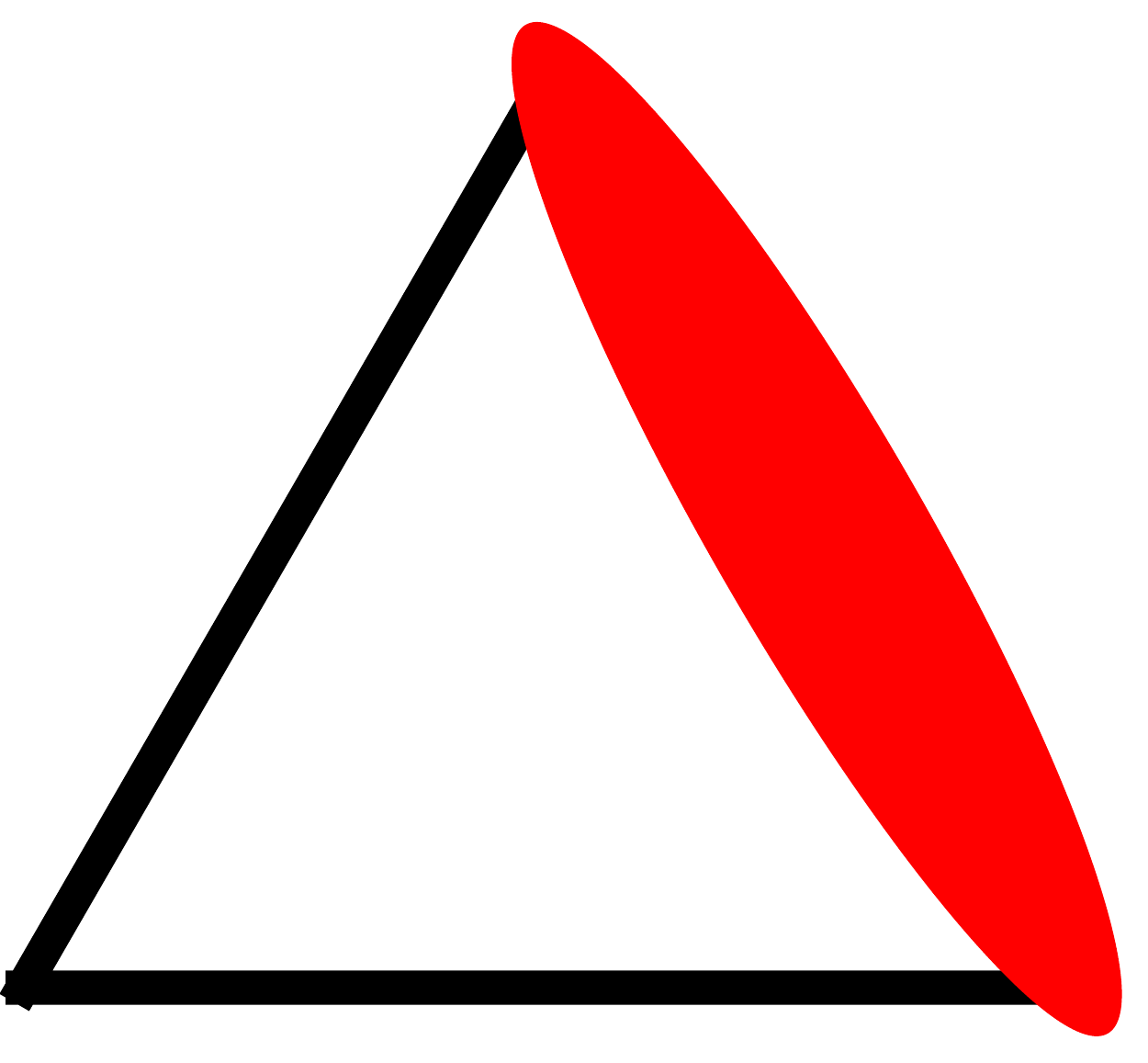}} \rangle,| \parbox{0.4cm}{\includegraphics[width=\linewidth]{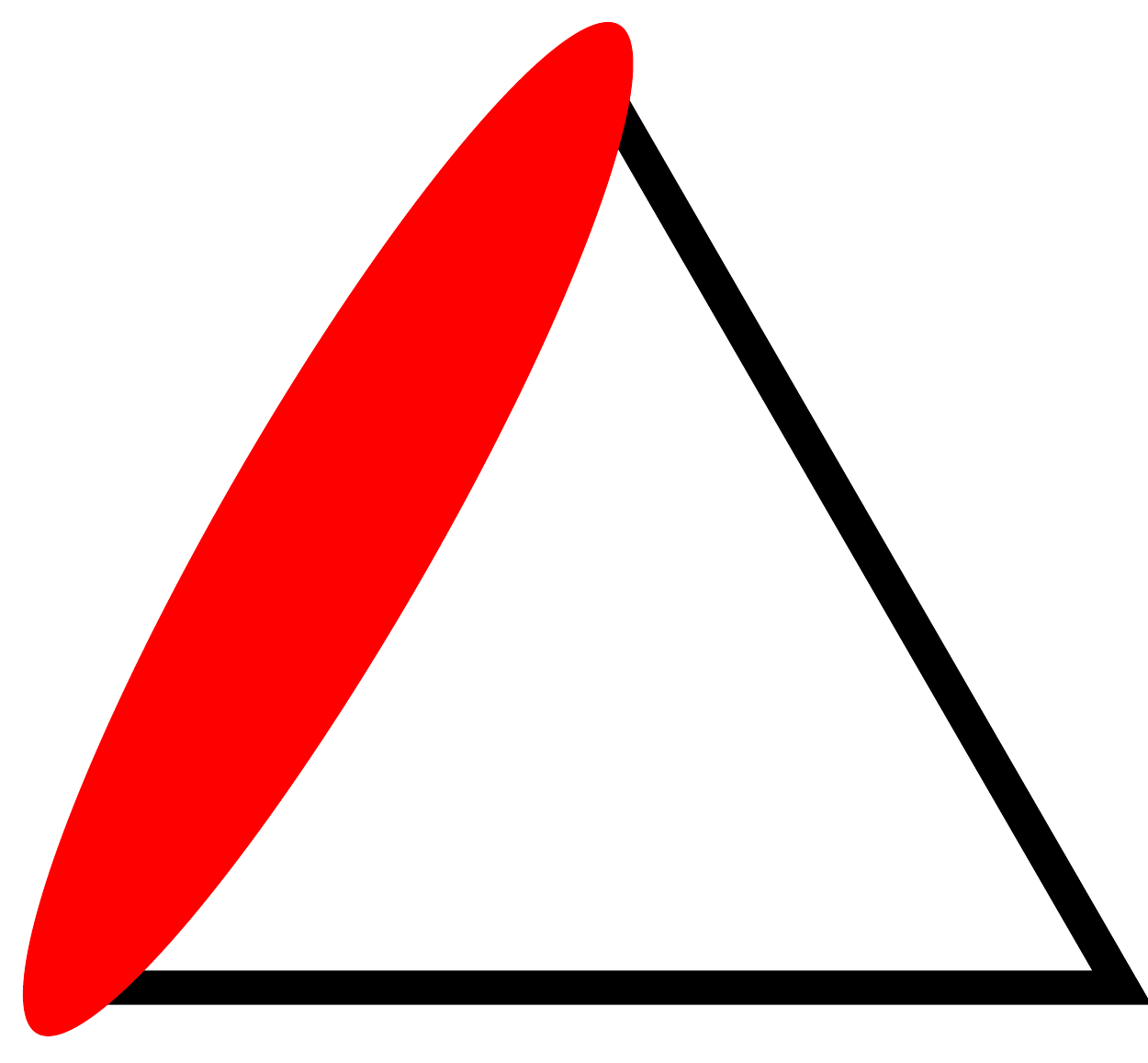}} \rangle,| \parbox{0.4cm}{\includegraphics[width=\linewidth]{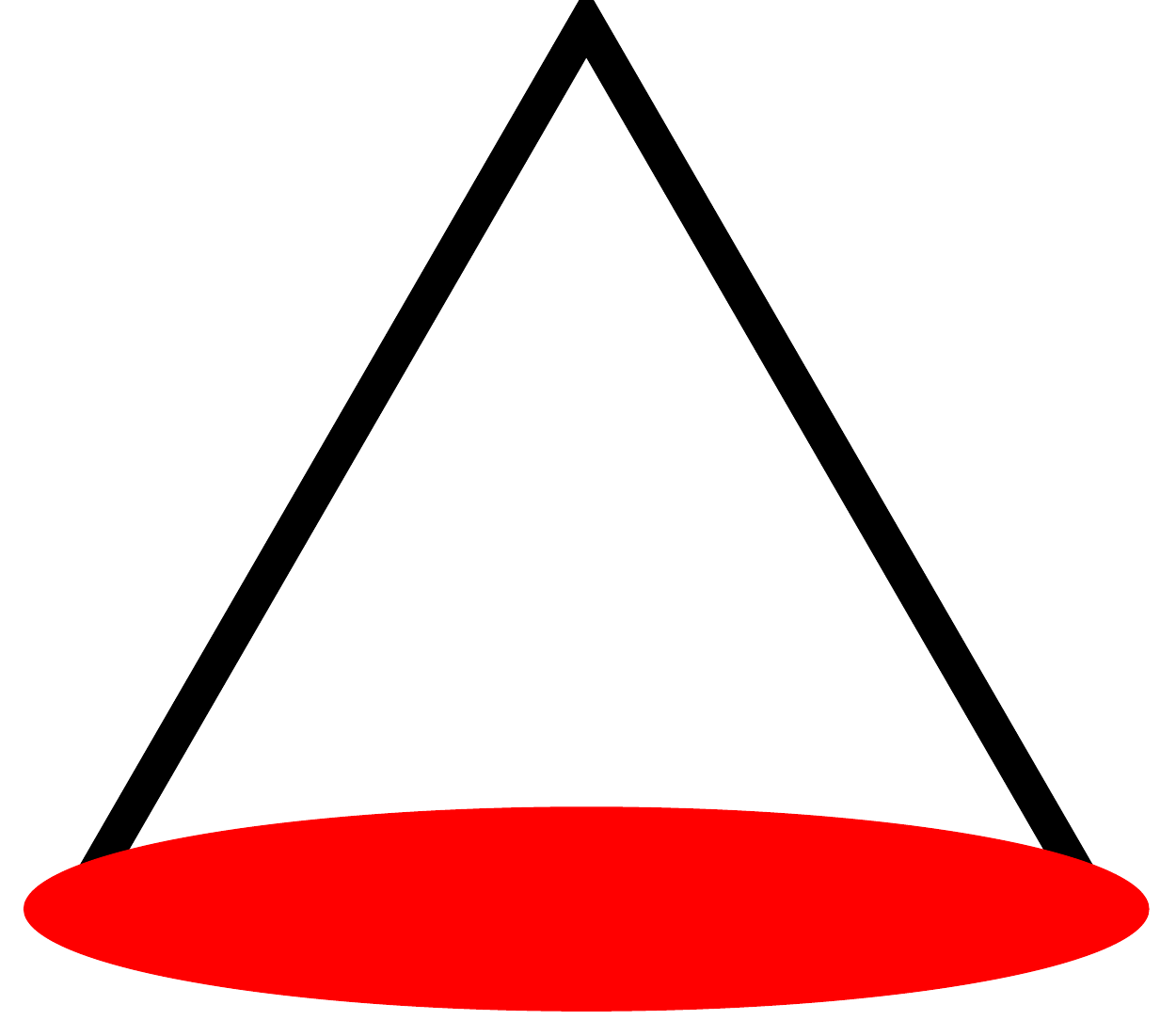}} \rangle\}$ on the dual kagome lattice.
The state $|s\rangle$ is the no-dimer state at the corresponding local triangle of the kagome lattice while the state $|t_{\gamma}\rangle$ is the dimer state occupying the ${\gamma}$-bond of the local triangle; see examples shown in Figs.~\ref{fig:3}(b)~and~\ref{fig:3}(c).
This kind of dimer mapping was introduced in a recent study of a spin-3/2 transverse field Ising model~\cite{Verresen2022}.

By this dimer mapping, spin operators take the following representations.
\begin{eqnarray}
\sigma^z
&=& 
 | s \rangle \langle t_z | 
+ | t_z \rangle \langle s | 
-i | t_x \rangle \langle t_y | 
+i | t_y \rangle \langle t_x | 
\nonumber\\
&=&
| \parbox{0.4cm}{\includegraphics[width=\linewidth]{./dimer-s.pdf}} \rangle
\langle \parbox{0.4cm}{\includegraphics[width=\linewidth]{./dimer-tz.pdf}} |
+
| \parbox{0.4cm}{\includegraphics[width=\linewidth]{./dimer-tz.pdf}} \rangle
\langle \parbox{0.4cm}{\includegraphics[width=\linewidth]{./dimer-s.pdf}} |
-i
| \parbox{0.4cm}{\includegraphics[width=\linewidth]{./dimer-tx.pdf}} \rangle
\langle \parbox{0.4cm}{\includegraphics[width=\linewidth]{./dimer-ty.pdf}} |
+i
| \parbox{0.4cm}{\includegraphics[width=\linewidth]{./dimer-ty.pdf}} \rangle
\langle \parbox{0.4cm}{\includegraphics[width=\linewidth]{./dimer-tx.pdf}} |.
~~~~~
\end{eqnarray}
\begin{eqnarray}
\tau^z
&=& 
- | s \rangle \langle t_z | 
- | t_z \rangle \langle s | 
-i | t_x \rangle \langle t_y | 
+i | t_y \rangle \langle t_x | 
\nonumber\\
&=&
-
| \parbox{0.4cm}{\includegraphics[width=\linewidth]{./dimer-s.pdf}} \rangle
\langle \parbox{0.4cm}{\includegraphics[width=\linewidth]{./dimer-tz.pdf}} |
-
| \parbox{0.4cm}{\includegraphics[width=\linewidth]{./dimer-tz.pdf}} \rangle
\langle \parbox{0.4cm}{\includegraphics[width=\linewidth]{./dimer-s.pdf}} |
-i
| \parbox{0.4cm}{\includegraphics[width=\linewidth]{./dimer-tx.pdf}} \rangle
\langle \parbox{0.4cm}{\includegraphics[width=\linewidth]{./dimer-ty.pdf}} |
+i
| \parbox{0.4cm}{\includegraphics[width=\linewidth]{./dimer-ty.pdf}} \rangle
\langle \parbox{0.4cm}{\includegraphics[width=\linewidth]{./dimer-tx.pdf}} |.
~~~~~
\end{eqnarray}
Other spin operators are obtained by cyclic permutations of ($x,y,z$) in the above.
In this dimer picture, $\sigma^\gamma$ and $\tau^\gamma$ spin operators do monomer pair-creation/annihilation or monomer hopping at the $\gamma$-bond of the kagome lattice ($\gamma=x,y,z$).

\subsection{Emergence of a quantum dimer model and resonating valence bonds}

We develop a degenerate perturbation theory for the strong coupling limit of $H$.
In this case, the unperturbed Hamiltonian is the inter-layer interaction part,
\begin{equation}
H_0=G\sum_{\langle jk \rangle_\gamma} \phi_j^\gamma \phi_k^\gamma.
\end{equation}
Here we employed the composite spin operator,
\begin{equation}
\phi_j^\gamma\equiv\sigma_j^\gamma\tau_j^\gamma, 
\end{equation}
which is $\mathbb{Z}_2$-valued [$(\phi_j^\gamma)^2=1$] and satisfies the commutation relation, $[\phi_j^\gamma,\phi_k^\lambda]=0$, and the local constraint, $\phi_j^x \phi_j^y \phi_j^z = -1$.

Interestingly, the four states $\{ |s\rangle, |t_x \rangle, |t_y \rangle, |t_z \rangle \} $ are the eigenstates of the composite spin operators ($\phi^x, \phi^y, \phi^z$) as summarized in the table of Fig.~\ref{fig:3}.
Hence, the composite spin operators are diagonalized in the dimer representation, {\it e.g.},
\begin{eqnarray}
\phi^z
&=& 
- | s \rangle \langle s | 
+ | t_x \rangle \langle t_x | 
+ | t_y \rangle \langle t_y | 
- | t_z \rangle \langle t_z | 
\nonumber\\
&=&
-
| \parbox{0.4cm}{\includegraphics[width=\linewidth]{./dimer-s.pdf}} \rangle
\langle \parbox{0.4cm}{\includegraphics[width=\linewidth]{./dimer-s.pdf}} |
+
| \parbox{0.4cm}{\includegraphics[width=\linewidth]{./dimer-tx.pdf}} \rangle
\langle \parbox{0.4cm}{\includegraphics[width=\linewidth]{./dimer-tx.pdf}} |
+
| \parbox{0.4cm}{\includegraphics[width=\linewidth]{./dimer-ty.pdf}} \rangle
\langle \parbox{0.4cm}{\includegraphics[width=\linewidth]{./dimer-ty.pdf}} |
-
| \parbox{0.4cm}{\includegraphics[width=\linewidth]{./dimer-tz.pdf}} \rangle
\langle \parbox{0.4cm}{\includegraphics[width=\linewidth]{./dimer-tz.pdf}} |.
~~~~~
\end{eqnarray}
Note that $-\phi^z$ measures the dimer parity across the $x,y$-bonds, and similarly for $-\phi^{x,y}$.

The unperturbed Hamiltonian $H_{0}$ is readily diagonalized in the dimer representation. 
We find that the ground state manifold of $H_{0}$ is extensively degenerate satisfying the condition,
\begin{equation}
\phi_j^\gamma\phi_k^\gamma=-1,
\label{eq:G-only-condition}
\end{equation}
at every bond $\langle jk \rangle_\gamma$ of the honeycomb lattice.
This condition implies the ``hardcore dimer'' constraint on the kagome lattice, i.e., each site of the kagome lattice is occupied by a single dimer.
Specifically, the condition $\phi_j^\gamma\phi_k^\gamma=-1$ only allows the eight configurations,
\begin{equation}
\parbox{7.5cm}{\includegraphics[width=\linewidth]{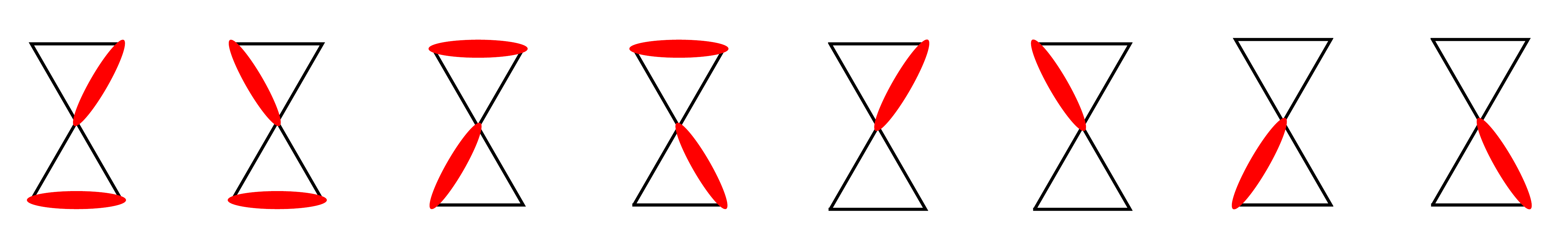}},
\end{equation}
 in every adjacent two triangles.
The central site is always occupied by a single dimer because the condition $\phi_j^\gamma\phi_k^\gamma=-1$ requires the odd dimer parity over the four bonds connected to the central site.
Hence, the ground state manifold of $H_0$ consists of the dimer states that respect the hardcore dimer constraint at every site.
The ground state energy is $E_{0}=-\frac{3N}{2}|G|$, where $N$ is the number of sites of the honeycomb lattice, and the degeneracy is given by $2^{N/2-1}$ in periodic boundary conditions.
It must be noted that the ground state manifold of $H_0$ satisfies the flux condition,
\begin{equation}
W_p Z_p = \prod_{\langle jk \rangle_\gamma \in p} \phi_j^\gamma \phi_k^\gamma = +1,
~~~{\rm i.e.,}~~~
W_p = Z_p .
\end{equation}

The Kitaev terms in $H-H_0$ generate quantum dimer motions in the ground state manifold of $H_0$.
By conducting perturbative expansions within the manifold, we obtain the effective Hamiltonian,
\begin{equation}
\mathcal{H}_{\rm eff}(G \gg K)= -\lambda \sum_p \hat{W}_p,
\label{eq:QDM}
\end{equation}
where $\hat{W}_p$ is the hexagon plaquette operator in Eq.~(\ref{eq:plaquette-op}), and $\lambda\propto K_\sigma^6/G^5$~(Appendix~\ref{app:A}).
It is straightforward to find the ground state of $\mathcal{H}_{\rm eff}(G \gg K)$, 
\begin{equation}
|\Psi \rangle = \mathcal{N} \prod_p \frac{1+\hat{W}_p}{2} | \Phi \rangle,
\label{eq:RVB-wf}
\end{equation}
where $\mathcal{N}$ is a normalization constant and $| \Phi \rangle$ is an arbitrary state in the ground state manifold of $H_0$.
This state is massively quantum superposed with the uniform zero-flux (${W}_p={Z}_p=+1$). Also, it has the property, $\langle\sigma_j^\gamma\sigma_k^\gamma\rangle =\langle\tau_j^\gamma\tau_k^\gamma\rangle=0$, because the Kitaev terms violate the hardcore dimer constraint, Eq.~(\ref{eq:G-only-condition}); see Fig.~\ref{fig:13}.

The ground state $|\Psi \rangle$ is nothing but a resonating valence bond state on the kagome lattice.
To see this, one should understand the effects of $\hat{W}_p$ in the dimer language.
Acting on hardcore dimer states, $\hat{W}_p$ gives rise to dimer resonance motions within the 12-site David star:
\begin{eqnarray}
\hat{W}_p
&=&
\sum_{\mathpzc{D}} f(\mathpzc{D}) | \mathpzc{D} \rangle \langle \bar{\mathpzc{D}} | + {\rm H.c.} 
\nonumber\\
&=&
| \parbox{0.6cm}{\includegraphics[width=\linewidth]{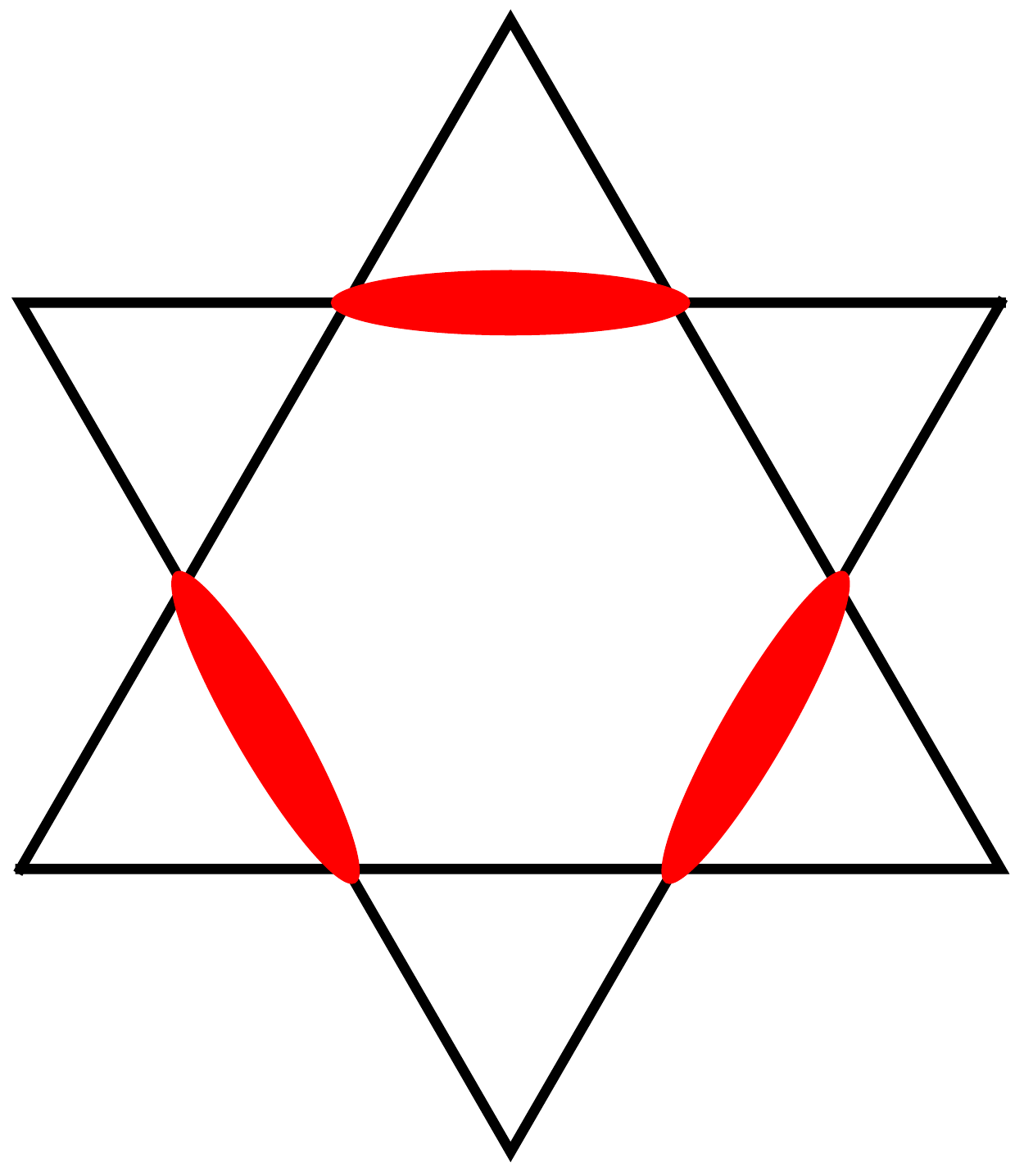}} \rangle
\langle \parbox{0.6cm}{\includegraphics[width=\linewidth]{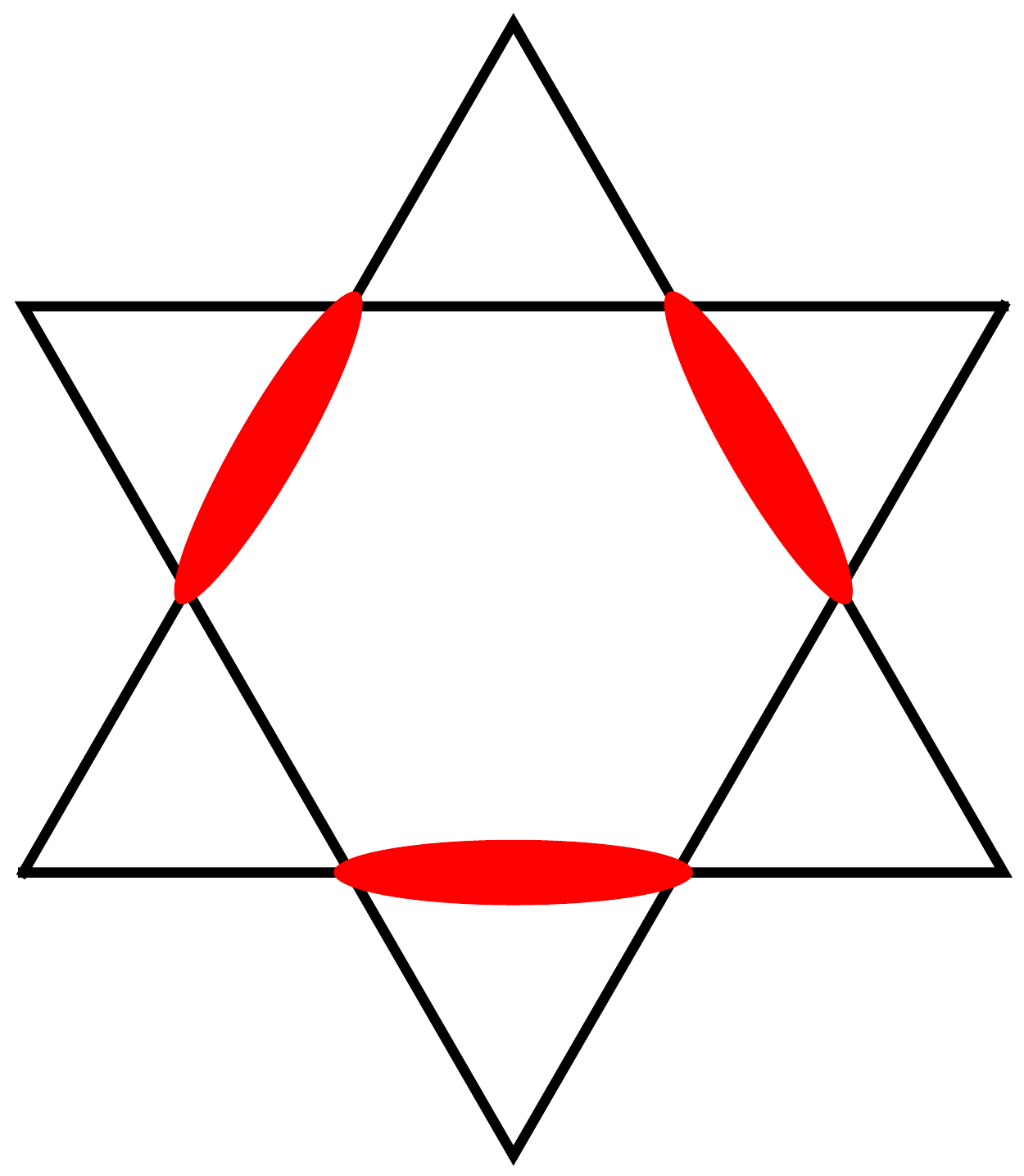}} |
-
| \parbox{0.6cm}{\includegraphics[width=\linewidth]{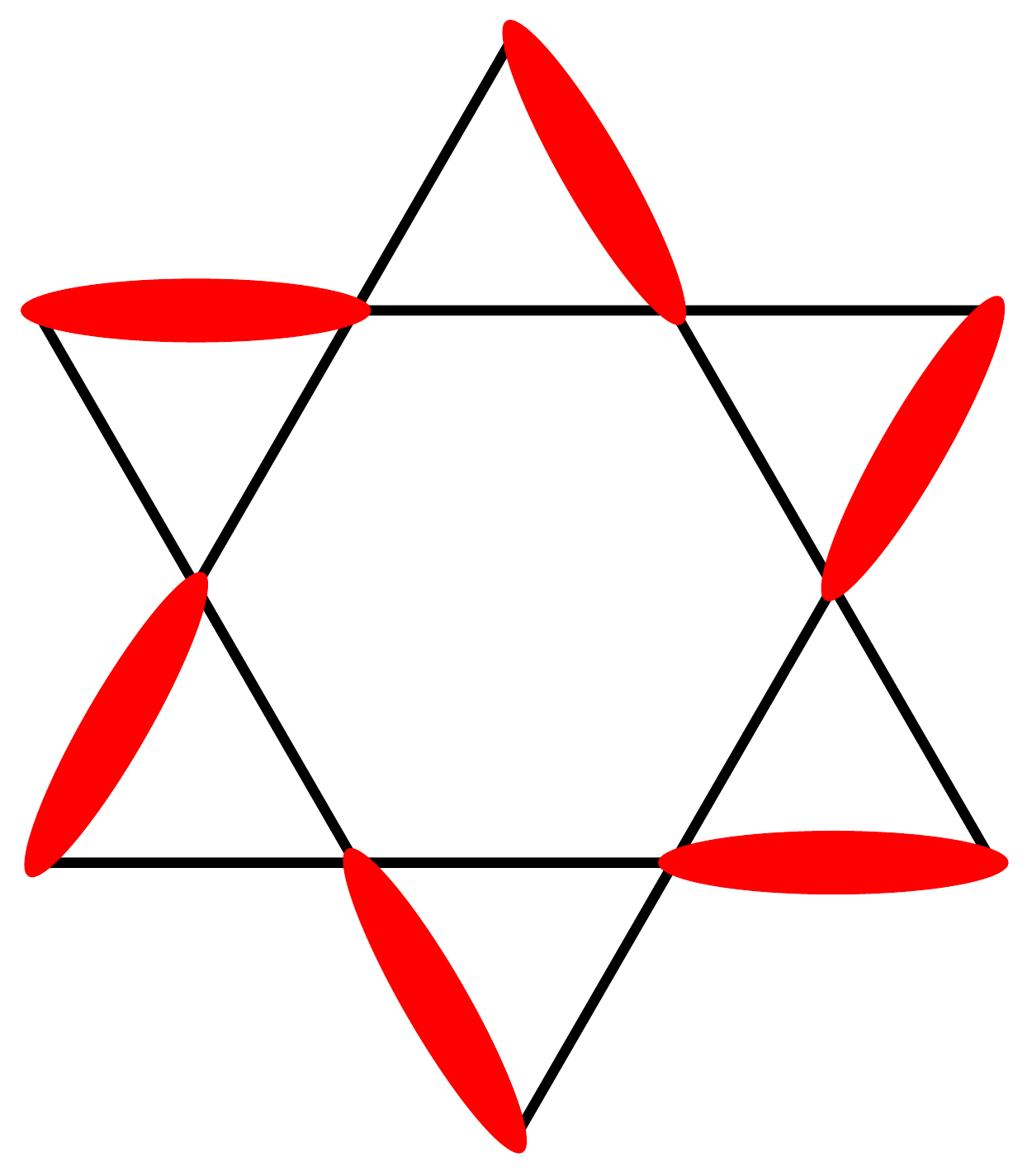}} \rangle
\langle \parbox{0.6cm}{\includegraphics[width=\linewidth]{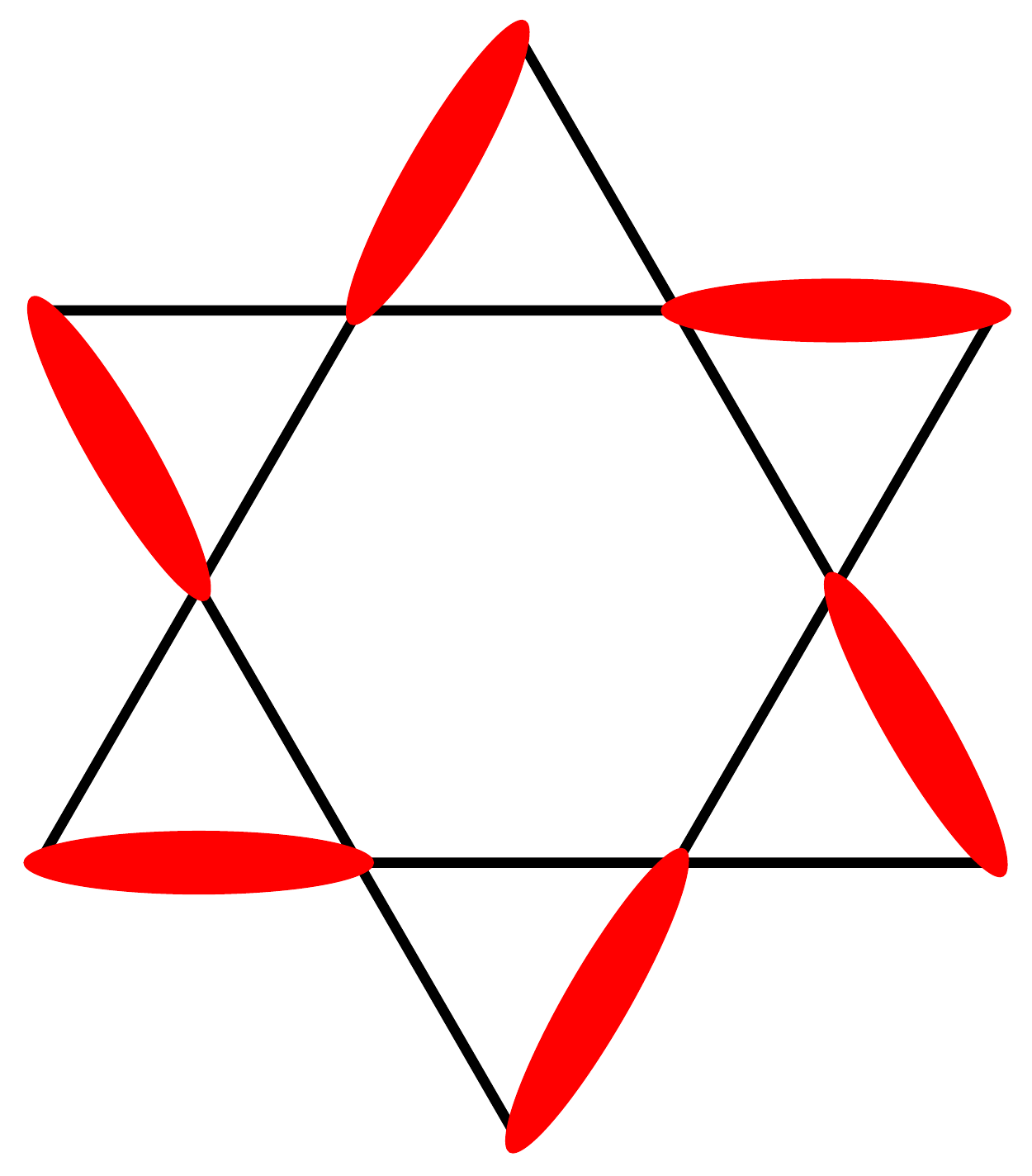}} |
+
\cdots .
\label{eq:dimer-resonance}
\end{eqnarray}
The full list of dimer motions and the associated sign factor, $f(\mathpzc{D})=\pm1$, are provided in Table~\ref{tab:III} (also see Appendix~\ref{app:B}).
We note that the above description (apart from the sign factor) is identical to the quantum dimer model by Misguich, Serban, and Pasquier~\cite{Misguich2002}.
In Eq.~(\ref{eq:RVB-wf}), all possible $2^{N/2-1}$ hardcore dimer configurations are generated by the plaquette operators, and superposed with equal weight.
Therefore, in the strong coupling limit of $H$, we have an effective quantum dimer model $\mathcal{H}_{\rm eff}(G \gg K)$, and the ground state $|\Psi \rangle$ is a resonating valence bond state whose anyon properties are characterized by the $\mathbb{Z}_2$ toric code topological order (discussed later).

\section{KSL$\times$KSL state: weak coupling limit\label{sec:KSL2}}

We construct a degenerate perturbation theory for the weak coupling limit of $H$. Second order perturbations create effective spin interactions that lead to spontaneous symmetry breaking of time-reversal in the KSL$\times$KSL state. We show this by a Majorana mean-field theory.

\subsection{Effective Hamiltonian}

To construct an effective theory for the weak coupling limit, we arrange the Hamiltonian into the form $H=H_0+H_1$ (where $H_0=\sum_{\langle jk\rangle_\gamma}K_\sigma\sigma_j^\gamma\sigma_k^\gamma+K_\tau\tau_j^\gamma\tau_k^\gamma$ and $H_1=G\sum_{\langle jk\rangle_\gamma}\sigma_j^\gamma\sigma_k^\gamma\tau_j^\gamma\tau_k^\gamma$), and conduct a second order degenerate perturbation theory. The resulting effective Hamiltonian is given by
\begin{equation}
\mathcal{H}_{\rm eff}(G \ll K)=H_0+H_1-\frac{1}{\Delta E} H_1 H_1,
\end{equation}
where $\Delta E (\propto |K_\sigma|)$ means the energy difference between the ground state and an intermediate excited state of $H_0$.
Nontrivial effects are generated by the second order term,
\begin{equation}
-\frac{1}{\Delta E} H_1 H_1
\propto
-\frac{G^2}{|K_\sigma|}
\sum_{\langle ij\rangle_\alpha}
\sigma_i^\alpha \sigma_j^\alpha \tau_i^\alpha \tau_j^\alpha
\sum_{\langle kl \rangle_\gamma}
\sigma_k^\gamma \sigma_l^\gamma \tau_k^\gamma \tau_l^\gamma.
\end{equation}
Among the various combinations of spin operators, we are particularly interested in the combinations that are defined on connected two bonds. 
For instance, $(\sigma_1^x \sigma_2^x \tau_1^x \tau_2^x) (\sigma_2^y \sigma_3^y \tau_2^y \tau_3^y)$ on adjacent two bonds, $\langle 12 \rangle_x$ and $\langle 23 \rangle_y$. This eight-spin operator can be simplified to a six-spin operator:
\begin{eqnarray}
(\sigma_1^x \sigma_2^x \tau_1^x \tau_2^x)
(\sigma_2^y \sigma_3^y \tau_2^y \tau_3^y)
&=&
(\sigma_1^x \underbrace{\sigma_2^x \sigma_2^y}_{i\sigma_2^z} \sigma_3^y) 
(\tau_1^x \underbrace{\tau_2^x \tau_2^y}_{i\tau_2^z} \tau_3^y)
\nonumber\\
&=&
-
( \sigma_1^x \sigma_2^z \sigma_3^y )
( \tau_1^x \tau_2^z \tau_3^y ).
\end{eqnarray}
By collecting this kind of terms, we arrange the effective Hamiltonian into the following form.
\begin{equation}
\mathcal{H}_{\rm eff}(G \ll K)=H_0+\eta \sum_{\langle ij\rangle_\alpha \langle jk\rangle_\gamma} \sigma_i^\alpha \sigma_j^\beta \sigma_k^\gamma  \tau_i^\alpha \tau_j^\beta \tau_k^\gamma + \cdots,
\label{eq:H-eff-weak}
\end{equation}
where $\eta (\propto G^2/|K_\sigma|)$ is a positive constant, the summation is for all connected two bonds $\{\langle ij\rangle_\alpha \langle jk\rangle_\gamma\}$, and the three spin indices are all different ($\alpha\ne\beta\ne\gamma$).

\begin{figure}[tb]
\includegraphics[width=\linewidth]{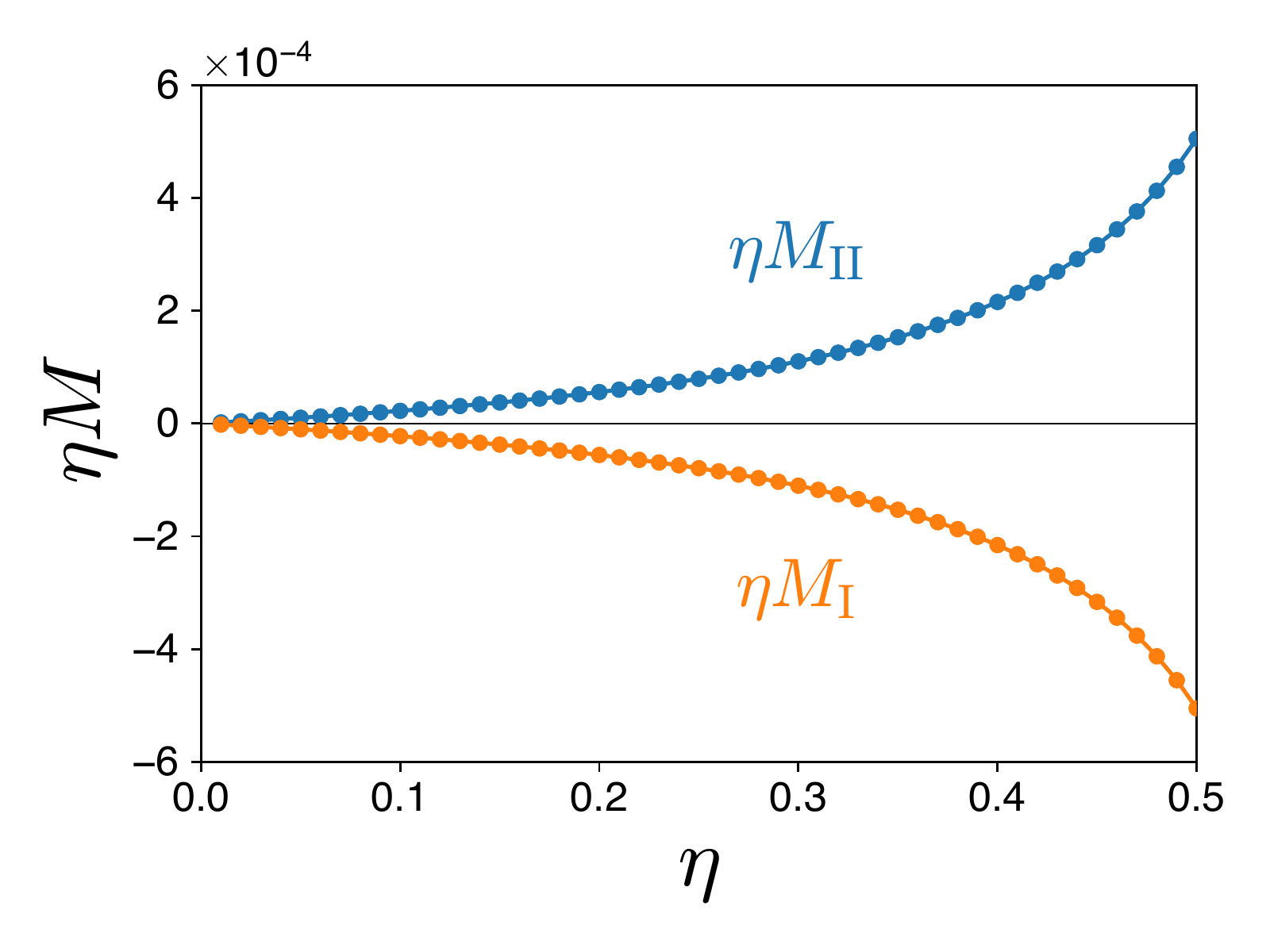}
\caption{Result of the Majorana mean-field theory.
The mean-field parameters ($M_{\rm I,II}$) are obtained by solving Eq.~(\ref{eq:MF-eq}) self-consistently. The Kitaev term is fixed by $|K_\sigma|=|K_\tau|=1$. The result does not depend on the relative sign of the coupling constants.
}
\label{fig:4}
\end{figure}

\subsection{Majorana mean-field theory}

The effective Hamiltonian is solved by a mean-field theory. To the six-spin operators, we apply the mean-field decoupling,
\begin{eqnarray}
\sigma_i^\alpha \sigma_j^\beta \sigma_k^\gamma  \tau_i^\alpha \tau_j^\beta \tau_k^\gamma
&\rightarrow&
\sigma_i^\alpha \sigma_j^\beta \sigma_k^\gamma \langle \tau_i^\alpha \tau_j^\beta \tau_k^\gamma \rangle
+
\langle \sigma_i^\alpha \sigma_j^\beta \sigma_k^\gamma \rangle \tau_i^\alpha \tau_j^\beta \tau_k^\gamma
\nonumber\\
&&-
\langle \sigma_i^\alpha \sigma_j^\beta \sigma_k^\gamma \rangle \langle \tau_i^\alpha \tau_j^\beta \tau_k^\gamma \rangle .
\end{eqnarray}
Similar mean-field decoupling schemes may be applied to the other terms that are not explicitly shown in Eq.~(\ref{eq:H-eff-weak}).
Yet, those terms merely renormalize the Kitaev term $H_0$.
Hence, we ignore those terms and just focus on the mean-field Hamiltonian,
\begin{eqnarray}
\mathcal{H}_{\rm MF} &=& 
H_0
+
\eta \sum_{\langle ij\rangle_\alpha \langle jk\rangle_\gamma}
\left[
\sigma_i^\alpha \sigma_j^\beta \sigma_k^\gamma \langle \tau_i^\alpha \tau_j^\beta \tau_k^\gamma \rangle \right.
\nonumber\\
&+&
\left.
\langle \sigma_i^\alpha \sigma_j^\beta \sigma_k^\gamma \rangle \tau_i^\alpha \tau_j^\beta \tau_k^\gamma
-
\langle \sigma_i^\alpha \sigma_j^\beta \sigma_k^\gamma \rangle \langle \tau_i^\alpha \tau_j^\beta \tau_k^\gamma \rangle 
\right].~~~~~~~
\label{eq:weak-coupling-H_MF}
\end{eqnarray}
The three-spin terms, $\sigma_i^\alpha\sigma_j^\beta\sigma_k^\gamma$ \& $\tau_i^\alpha\tau_j^\beta\tau_k^\gamma$, break time-reversal symmetry and create a finite energy gap in the fermion excitations of the Kitaev spin liquid state on each layer~\cite{Kitaev2006}.

This can be shown by using the Majorana representation for spin-1/2 operator, 
\begin{equation}
\sigma_i^\alpha=ib_{{\rm I},i}^\alpha c_{{\rm I},i}~~~\&~~~\tau_i^\alpha=ib_{{\rm II},i}^\alpha c_{{\rm II},i},
\end{equation}
where the subscript (I,II) means the upper/lower layer.
This representation leads to the Majorana mean-field Hamiltonian,
\begin{eqnarray}
\mathcal{H}_{\rm MF}
&=&\sum_{\langle ij \rangle} \left[ (-K_\sigma) ic_{{\rm I},i}c_{{\rm I},j} + (-K_\tau) ic_{{\rm II},i}c_{{\rm II},j} \right]
\\
&+&\sum_{\langle\langle ik \rangle\rangle} \left[ (\eta M_{\rm II}) ic_{{\rm I},i}c_{{\rm I},k} + (\eta M_{\rm I}) ic_{{\rm II},i}c_{{\rm II},k} - \eta M_{\rm I} M_{\rm II} \right], \nonumber
\end{eqnarray}
where we have chosen the simplest gauge for the uniform zero-flux ($u_{{\rm I},ij}=ib_{{\rm I},i}^\alpha b_{{\rm I},j}^\alpha=+1$ and $u_{{\rm II},ij}=ib_{{\rm II},i}^\alpha b_{{\rm II},j}^\alpha=+1$ with site $i$ in A sublattice and $j$ in B sublattice), and introduced the mean-field parameters,
\begin{equation}
M_{\rm I}=\langle i c_{{\rm I},i} c_{{\rm I},k} \rangle~~~\&~~~M_{\rm II}=\langle i c_{{\rm II},i} c_{{\rm II},k} \rangle.
\label{eq:MF-eq}
\end{equation}
In this mean-field theory, the two layers are coupled only by the mean-field parameters, $M_{\rm I}$ and $M_{\rm II}$.

Figure~\ref{fig:4} shows the result of the self-consistent mean-field calculations.
We find that the mean-field parameters have a same magnitude but opposite signs ($M_{\rm I}=-M_{\rm II}$).
This implies that the KSL state of each layer has a finite energy gap ($\Delta$) and a nonzero Chern number ($\nu$)~\cite{Kitaev2006}: 
\begin{eqnarray}
\Delta_{\rm I} = 6\sqrt{3}|\eta M_{\rm II}|
~~~& \& &~~~
\nu_{\rm I} = {\rm sgn}(\eta M_{\rm II}) ,
\\
\Delta_{\rm II} = 6\sqrt{3}|\eta M_{\rm I}|
~~~& \& &~~~
\nu_{\rm II} = {\rm sgn}(\eta M_{\rm I}) .
\end{eqnarray}
More importantly, the two layers have opposite signs for the Chern number (e.g., $\nu_{\rm I}=+1$ \& $\nu_{\rm II}=-1$).
This theory indicates that the KSL$\times$KSL state has the non-abelian ${\rm Ising}\times\overline{\rm Ising}$ topological order~\cite{Kitaev2006,Pachos2012,Burnell2018}.

On the time-reversal symmetry breaking in the KSL$\times$KSL state, we provide further evidences from our numerical calculations shown in the next section.
We will discuss the spontaneous symmetry breaking and the gap-openning problem more carefully by comparing our results with a field theory argument about gapless Dirac fermions in Sec.~\ref{sec:discussion}.

\section{Exact diagonalization\label{sec:ED}}

Now we investigate the phase diagram of $H$ by exact diagonalization (ED) on the (24+24)-site bilayer cluster in Fig.~\ref{fig:1}(a).
We impose periodic boundary conditions and utilize the flux quantum numbers, $W_p=\pm1$ \& $Z_p=\pm1$, to reduce the size of the Hilbert space in our ED calculations.

Figure~\ref{fig:1}(c) displays the resulting phase diagram.
We find that the ground state appears in the uniform zero-flux sector ($W_p=Z_p=+1$).
The KSL$\times$KSL and RVB states, which we considered in the weak and strong coupling limits, are stabilized over large regions separated by a single transition at $\theta_{\rm c}\simeq0.26\pi$ (shown by a small peak in the derivative of the ground state energy, $-\partial^2 E_{\rm gs}/\partial \theta^2$).
We characterize the two QSL states by investigating entanglement entropy, chirality structure factor, hardcore dimer constraint, and topological degeneracy below.

\begin{figure}[tb]
\includegraphics[width=\linewidth]{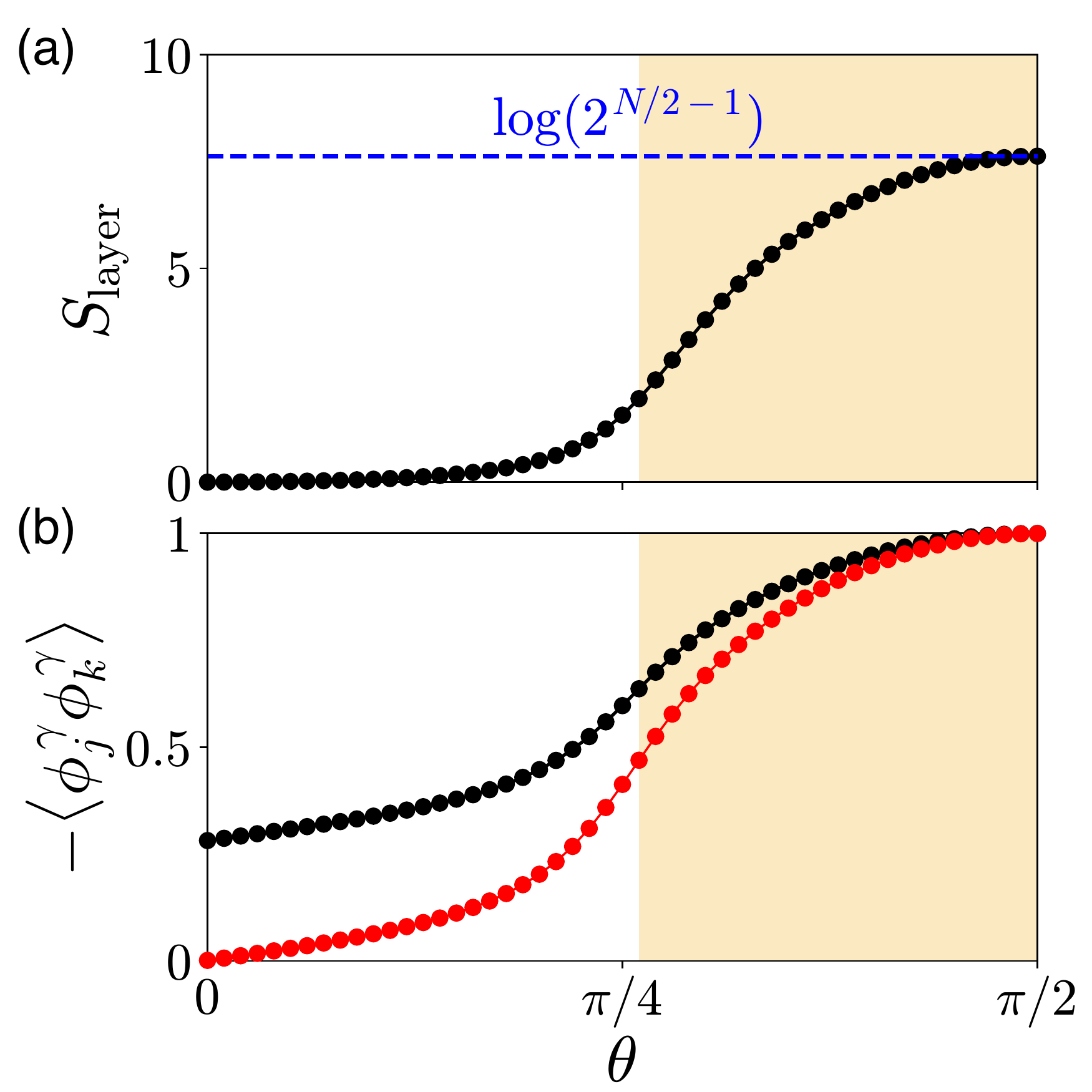}
\caption{Entanglement entropy and four-spin correlator for the hardcore dimer constraint.
(a) The entanglement entropy $S_{\rm layer}$. Dashed line marks the value of $\log(2^{N/2-1})$, where $N=24$.
(b) Four-spin correlators to check the hardcore dimer constraint. Black: $-\langle \phi_j^\gamma \phi_k^\gamma \rangle$. Red: $-\langle \phi_j^\gamma\phi_k^\gamma \rangle+\langle\sigma_j^\gamma\sigma_k^\gamma\rangle \langle\tau_j^\gamma\tau_k^\gamma\rangle$.
}
\label{fig:5}
\end{figure}

\subsection{Entanglement entropy}

Quantum entanglement between the two layers can be measured by the entanglement entropy,
\begin{equation}
S_{\rm layer}=-\log ({\rm Tr}_{\{\sigma\}} \rho_{\rm layer}^2),
\end{equation}
where $\rho_{\rm layer}={\rm Tr}_{\{\tau\}}|\Psi_{\rm gs}\rangle \langle \Psi_{\rm gs} |$ is the reduced density matrix obtained by tracing out $\tau$-spins in the ground state wave function $|\Psi_{\rm gs}\rangle$.
We find that the KSL$\times$KSL state has negligible entanglement ($S_{\rm layer}\approx0$), indicating that it is basically a product state of two layers of KSL. By contrast, the RVB state shows strong interlayer entanglement and the dimension of the dimer Hilbert space manifests through the entanglement entropy $S_{\rm layer}={\rm log}(2^{N/2-1})$ in the strong coupling limit [Fig.~\ref{fig:5}(a)].

\begin{figure*}[tb]
\includegraphics[width=\linewidth]{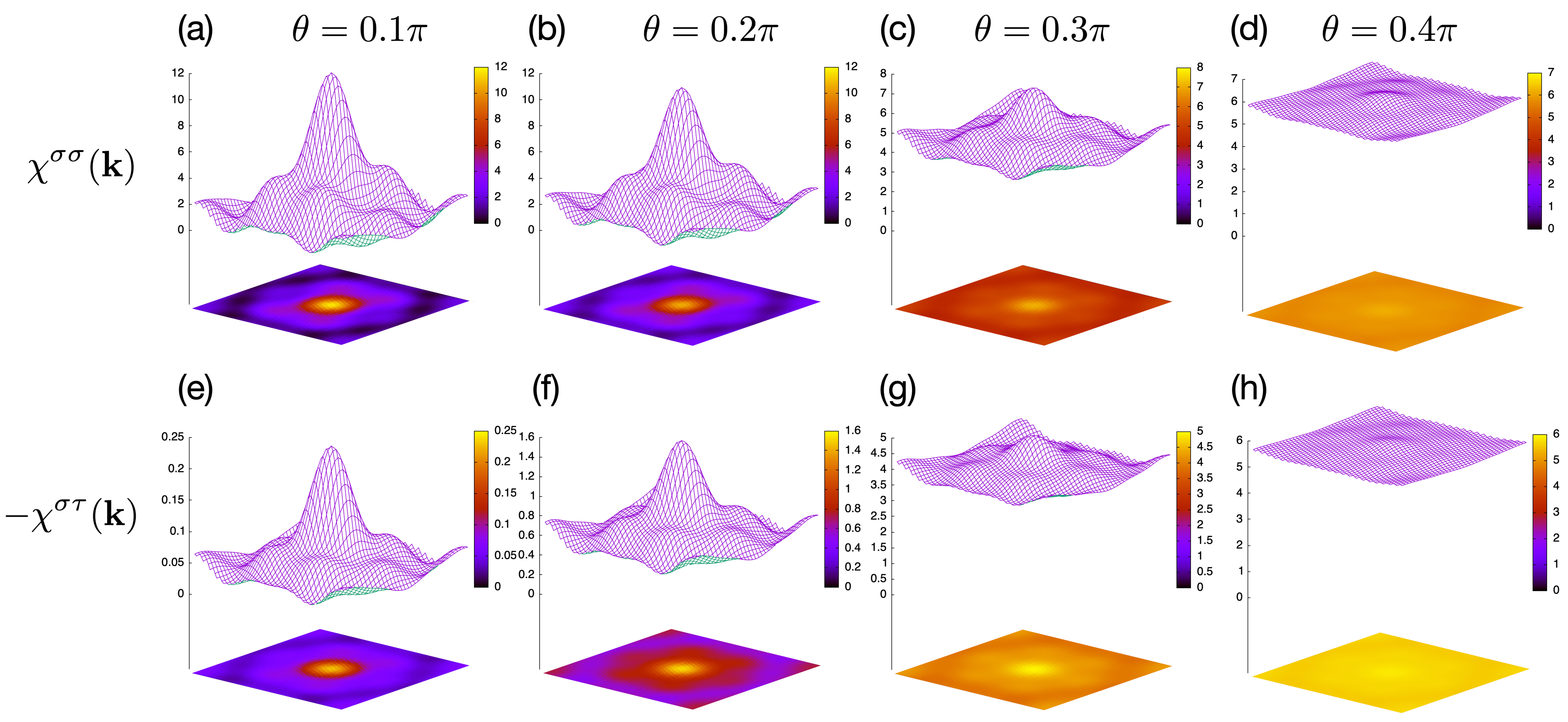}
\caption{Chirality structure factor.
(a),(b),(c),(d) The upper panels depict $\chi^{\sigma\sigma}({\bf k})=\chi^{\tau\tau}({\bf k})$ at four representative points of the phase diagram ($\theta/\pi=0.1,0.2,0.3,0.4$). (e),(f),(g),(h) The lower panels illustrate $-\chi^{\sigma\tau}({\bf k})$ for the same parameters ($\theta/\pi=0.1,0.2,0.3,0.4$). 
In each case, the chirality structure factor is depicted in two different fashions, 3D plot (purple surface) and 2D plot (color map at the bottom).
The center of the $xy$ plane corresponds to the zero momentum point (${\bf k=0}$).
The KSL$\times$KSL state exhibits a well-defined peak at ${\bf k=0}$, suggesting a long-range order in the chirality in the thermodynamic limit [(a),(b),(e),(f)].
The RVB state has short-ranged correlations in the chirality as shown by relatively flat chirality structure factors [(c),(d),(g),(h)].
}
\label{fig:6}
\end{figure*}

\subsection{Chirality order}

Time-reversal symmetry breaking and the opposite Chern numbers discussed in our mean-field theory for the KSL$\times$KSL state can be checked in our ED via the chirality structure factor,
\begin{equation}
\chi^{IJ}({\bf k}) 
= 
\frac{1}{N/2} \sum_{p,q} \langle \hat{\chi}_p^I \hat{\chi}_q^J  \rangle e^{i{\bf k}\cdot({\bf r}_p-{\bf r}_q)},
\label{eq:CSF}
\end{equation}
where $I,J(=\sigma,\tau)$ are layer indices and $p,q$ are plaquette indices.
The chiral spin operators, $\hat{\chi}_p^{\sigma}$ \& $\hat{\chi}_p^{\tau}$, are defined at each plaquette $p$ as 
\begin{equation}
\hat{\chi}_p^\sigma = \sum_{\langle ij \rangle_\alpha\langle jk \rangle_\gamma} \sigma_i^{\alpha} \sigma_j^{\beta} \sigma_k^{\gamma}
~~\&~~
\hat{\chi}_p^\tau = \sum_{\langle ij \rangle_\alpha\langle jk \rangle_\gamma} \tau_i^{\alpha} \tau_j^{\beta} \tau_k^{\gamma}
\label{eq:chiral-spin-op}
\end{equation}
with $\alpha\ne\beta\ne\gamma$.
To be specific, the chiral spin operators can be written as
\begin{eqnarray}
\hat{\chi}_p^\sigma &=& 
\sigma_1^{x} \sigma_2^{y} \sigma_3^{z}
+
\sigma_2^{z} \sigma_3^{x} \sigma_4^{y}
+
\sigma_3^{y} \sigma_4^{z} \sigma_5^{x}
\nonumber\\
&+&
\sigma_4^{x} \sigma_5^{y} \sigma_6^{z}
+
\sigma_5^{z} \sigma_6^{x} \sigma_1^{y}
+
\sigma_6^{y} \sigma_1^{z} \sigma_2^{x} ,
\\
\hat{\chi}_p^\tau &=& 
\tau_1^{x} \tau_2^{y} \tau_3^{z}
+
\tau_2^{z} \tau_3^{x} \tau_4^{y}
+
\tau_3^{y} \tau_4^{z} \tau_5^{x}
\nonumber\\
&+&
\tau_4^{x} \tau_5^{y} \tau_6^{z}
+
\tau_5^{z} \tau_6^{x} \tau_1^{y}
+
\tau_6^{y} \tau_1^{z} \tau_2^{x} ,
\label{eq:chirality-op}
\end{eqnarray}
where we followed the site convention shown in Fig.~\ref{fig:3}(a). 
The operators $\hat{\chi}_p^\sigma~\&~\hat{\chi}_p^\tau$ are nothing but the (time-reversal odd) gap-opening terms considered by Kitaev~\cite{Kitaev2006}, which we already have seen in Eq.~(\ref{eq:weak-coupling-H_MF}).

Figure~\ref{fig:6} presents the calculated $\chi^{IJ}({\bf k})$.
The two QSLs show distinguished behaviors.
First, the RVB state shows broad features rather than sharp peaks, implying no long-range order in the chirality $\langle \hat{\chi} \rangle$.
At $\theta=\pi/2$, one can exactly show that
\begin{equation}
\langle \hat{\chi}_p^I \hat{\chi}_q^J \rangle=\pm6\delta_{p,q}~~~{\rm and}~~~\chi^{IJ}({\bf k})=\pm6,
\end{equation}
where the sign is positive (negative) when $I=J$ ($I\ne J$).
Flat structure factors shown in the RVB state indicate short-ranged correlations but essentially no long-range order in the chirality.
See  Figs.~\ref{fig:6}(c),~\ref{fig:6}(d),~\ref{fig:6}(g),~and~\ref{fig:6}(h).

In sharp contrast, the ${\rm KSL}\times{\rm KSL}$ state exhibits a sharp peak at ${\bf k}={\bf 0}$, indicating a long-range order in the chirality.
The chirality order is characterized by
\begin{eqnarray}
\chi^{\sigma\sigma}({\bf 0})=\chi^{\tau\tau}({\bf 0})>0&:&\textup{intra-layer~ferro-chirality},
\nonumber\\
\chi^{\sigma\tau}({\bf 0})<0&:&\textup{inter-layer~antiferro-chirality}.
\nonumber
\end{eqnarray}
Namely, in thermodynamics limit, $\sigma$- and $\tau$-layers have a ${\bf k=0}$ order of opposite chiralities, e.g., $\langle \hat{\chi}^\sigma \rangle>0$ and $\langle \hat{\chi}^\tau \rangle<0$.
See Figs.~\ref{fig:6}(a),~\ref{fig:6}(b),~\ref{fig:6}(e),~and~\ref{fig:6}(f).

This result, together with the entanglement entropy in Fig.~\ref{fig:5}(a), suggests that in thermodynamic limit the two layers of the KSL$\times$KSL state can be individually described by the effective Hamiltonians~\cite{Hwang2022},
\begin{eqnarray}
\mathcal{H}_{\rm eff}^\sigma &=& K_\sigma \sum_{\langle jk \rangle_\gamma} \sigma_j^\gamma \sigma_k^\gamma -\lambda_\sigma \sum_p \hat{\chi}_p^\sigma + \cdots,
\\
\mathcal{H}_{\rm eff}^\tau &=& K_\tau \sum_{\langle jk \rangle_\gamma} \tau_j^\gamma \tau_k^\gamma -\lambda_\tau \sum_p \hat{\chi}_p^\tau + \cdots,
\end{eqnarray}
where the coupling constants $\lambda_\sigma ~\&~ \lambda_\tau$ have the opposite signs ($\lambda_\sigma \lambda_\tau < 0$) as required by the inter-layer antiferro-chirality.
Notice that the above effective Hamiltonians are exactly same as what we have derived in our Majorana mean-field theory~[Eq.~(\ref{eq:weak-coupling-H_MF})].
Therefore, time-reversal symmetry breaking in the KSL$\times$KSL state is unequivocally shown not only from the mean-field theory but also from the numerical calculations of chirality structure factor.

Our approach of using the chirality-chirality correlation to detect the broken time-reversal in the KSL$\times$KSL state is analogous to the conventional approach of using the spin-spin correlation to distinguish between magnetic orders and spin liquids (see Table~\ref{tab:I}).

\begin{figure*}[tb]
\includegraphics[width=\linewidth]{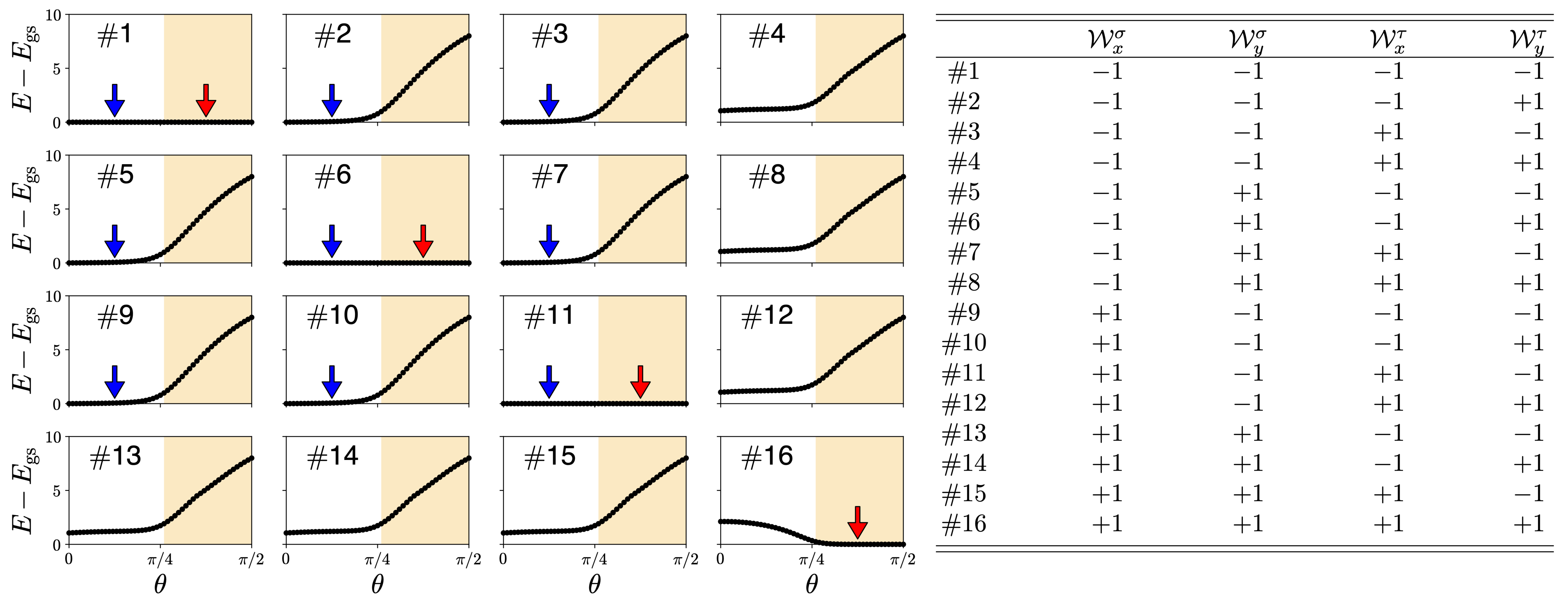}
\caption{Topological degeneracy.
Energy spectrum in the zero-flux sector ($W_p=Z_p=1$) where the ground state appears.
The sixteen panels represent different topological sectors defined by the four Wilson loop fluxes, $\{\mathcal{W}_x^\sigma=\pm1,\mathcal{W}_y^\sigma=\pm1,\mathcal{W}_x^\tau=\pm1,\mathcal{W}_y^\tau=\pm1\}$.
The table lists the flux values of the sixteen topological sectors ($\#1\sim16$).
In each topological sector, only the lowest energy level is displayed to avoid complication.
The KSL$\times$KSL state exhibits ninefold degeneracy in $\#1,2,3,5,6,7,9,10,11$ (marked by blue arrows).
The RVB state shows fourfold degeneracy in $\#1,6,11,16$ (marked by red arrows).
}
\label{fig:7}
\end{figure*}

\begin{table}[b]
\begin{ruledtabular}
\begin{tabular}{c|c|c}
 &  {\bf KSL$\times$KSL} & {\bf RVB}
 \\
\hline
Chirality correlation & Long-ranged & Short-ranged
 \\
 $\langle \hat{\chi}_p^I \hat{\chi}_q^J \rangle$ & &
 \\
\hline
Time-reversal & Broken & Unbroken
 \\
& in each layer &
\\
\hline
\hline
 & {\bf Magnetic order} & {\bf Spin liquid}
 \\
\hline
Spin correlation & Long-ranged & Short-ranged
 \\
 $\langle \boldsymbol{\sigma}_i \cdot \boldsymbol{\sigma}_j \rangle$ & &
 \\
\hline
Time-reversal & Broken & Unbroken
\end{tabular}
\end{ruledtabular}
\caption{Similarities of the chirality and spin correlations in detecting phases of broken time-reversal.}
\label{tab:I}
\end{table}

\subsection{Hardcore dimer constraint}

The hardcore dimer constraint in Eq.~(\ref{eq:G-only-condition}) is another good measure to distinguish between the KSL$\times$KSL and RVB states.
Based on our analytical approaches, the two states are expected to show distinct behaviors: $\langle \phi_j^\gamma\phi_k^\gamma \rangle\approx\langle\sigma_j^\gamma\sigma_k^\gamma\rangle \langle\tau_j^\gamma\tau_k^\gamma\rangle$ in the KSL$\times$KSL state near $\theta=0$ and $\langle \phi_j^\gamma\phi_k^\gamma \rangle\approx-1$ in the RVB state around $\theta=\pi/2$.
We confirm the distinct behaviors,
\begin{equation}
-\langle \phi_j^\gamma\phi_k^\gamma \rangle+\langle\sigma_j^\gamma\sigma_k^\gamma\rangle \langle\tau_j^\gamma\tau_k^\gamma\rangle
\approx
\left\{
\begin{array}{cc}
0 & ({\rm KSL\times KSL})
\\
1 & ({\rm RVB})
\end{array}
\right. ,
\end{equation}
in our ED results [Fig.~\ref{fig:5}(b)].

\begin{figure*}[tb]
\includegraphics[width=\linewidth]{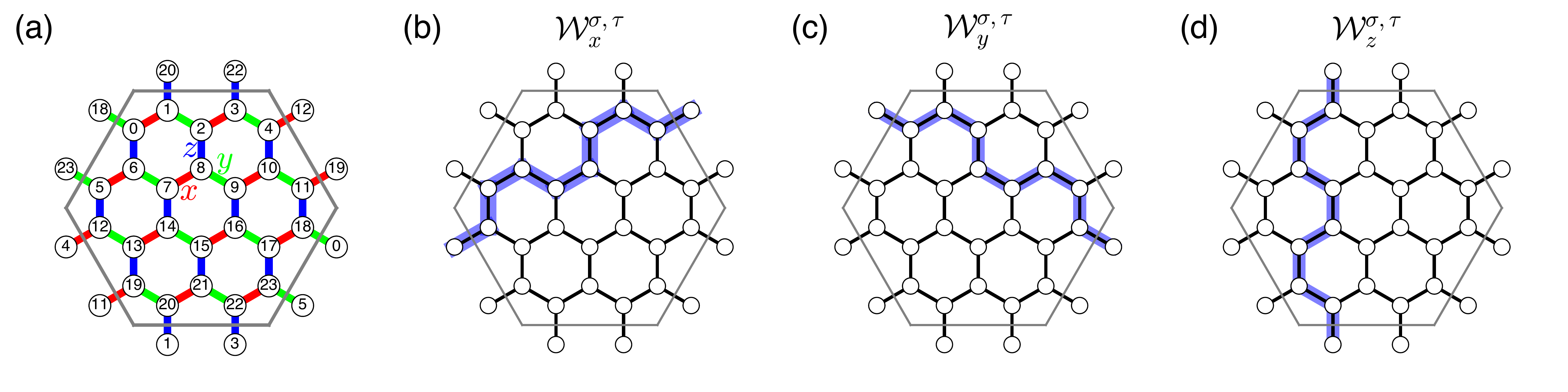}
\caption{Wilson loop operators on the 24-site cluster.
(a) The 24-site cluster with four states per site and periodic boundary conditions. (b),(c),(d) Visualizations of the Wilson loop operators, $\mathcal{W}_x^{\sigma,\tau}$, $\mathcal{W}_y^{\sigma,\tau}$, $\mathcal{W}_z^{\sigma,\tau}$. 
}
\label{fig:8}
\end{figure*}

\subsection{Topological degeneracy}

Topological degeneracy, i.e., the ground state degeneracy on torus geometry, is a useful probe to detect a topological order in the system. Namely, topological degeneracy tells about the number of anyon types allowed in the system~\cite{Wen1990,WenNiu1990}.
We identify ``ninefold'' degeneracy in the KSL$\times$KSL state and ``fourfold'' degeneracy in the RVB state (see Fig.~\ref{fig:7}).

For the analysis of topological degeneracy, we employ four Wilson loop operators, $\{\mathcal{W}_{x}^\sigma,\mathcal{W}_{y}^\sigma,\mathcal{W}_{x}^\tau,\mathcal{W}_{y}^\tau\}$, commuting with the Hamiltonian $H$.
The Hilbert space can then be partitioned into sixteen topological sectors distinguished by the eigenvalues of the Wilson loop operators ($\mathcal{W}_{x,y}^{\sigma,\tau}=\pm 1$).

The Wilson loop operators are defined by generalizing the hexagon plaquette operators ($\hat{W}_p$,$\hat{Z}_p$) to non-contractible loops of the system.
Figure~\ref{fig:8}(a) shows the cluster used in our exact diagonalization.
Periodic boundary conditions are imposed at the cluster boundary (gray hexagon).
Along the non-contractible loop in Fig.~\ref{fig:8}(b), we define the Wilson operator $\mathcal{W}_x^\sigma$ as follows.
\begin{eqnarray}
\mathcal{W}_x^\sigma 
&=&
 (\sigma_{12}^z \sigma_5^z) (\sigma_5^x \sigma_6^x) (\sigma_6^y \sigma_7^y) (\sigma_7^x \sigma_8^x) 
\nonumber\\ 
 &&
 (\sigma_8^z \sigma_2^z) (\sigma_2^x \sigma_3^x) (\sigma_3^y \sigma_4^y) (\sigma_4^x \sigma_{12}^x)
\nonumber\\
&=&
- \sigma_{12}^y \sigma_5^y \sigma_6^z \sigma_7^z
 \sigma_8^y \sigma_2^y \sigma_3^z \sigma_4^z .
\end{eqnarray}
This is nothing but the product of the Kitaev terms of $\sigma$-spins along the non-contractible loop.
Repeating for other possible non-contractible loops [Figs.~\ref{fig:8}(c)~and~\ref{fig:8}(d)],
we can define other Wilson loop operators:
\begin{eqnarray}
\mathcal{W}_y^\sigma 
=
- \sigma_0^z \sigma_1^z \sigma_2^x \sigma_8^x
 \sigma_9^z \sigma_{10}^z \sigma_{11}^x \sigma_{18}^x , 
\end{eqnarray}
\begin{eqnarray}
\mathcal{W}_z^\sigma 
=
- \sigma_1^y \sigma_0^y \sigma_6^x \sigma_7^x
 \sigma_{14}^y \sigma_{13}^y \sigma_{19}^x \sigma_{20}^x .
\end{eqnarray}
Note that the three operators satisfy the relationship, $\mathcal{W}_x^\sigma \mathcal{W}_y^\sigma \mathcal{W}_z^\sigma = +1$, in the ground state manifold due to the uniform zero-flux property ($W_p=Z_p=+1$).
Among the three operators, we use $\mathcal{W}_x^\sigma$ \& $\mathcal{W}_y^\sigma$.
Similarly, we repeat the same procedure for $\tau$-spins and obtain the following Wilson loop operators.
\begin{eqnarray}
\mathcal{W}_x^\tau 
=- \tau_{12}^y \tau_5^y \tau_6^z \tau_7^z
 \tau_8^y \tau_2^y \tau_3^z \tau_4^z ,
\end{eqnarray}
\begin{eqnarray}
\mathcal{W}_y^\tau 
=
- \tau_0^z \tau_1^z \tau_2^x \tau_8^x
 \tau_9^z \tau_{10}^z \tau_{11}^x \tau_{18}^x ,
\end{eqnarray}
\begin{eqnarray}
\mathcal{W}_z^\tau 
=
- \tau_1^y \tau_0^y \tau_6^x \tau_7^x
 \tau_{14}^y \tau_{13}^y \tau_{19}^x \tau_{20}^x .
\end{eqnarray}
Again, we have the relationship, $\mathcal{W}_x^\tau \mathcal{W}_y^\tau \mathcal{W}_z^\tau = +1$, and only use $\mathcal{W}_x^\tau$ \& $\mathcal{W}_y^\tau$.
The Wilson loop operators, $\{\mathcal{W}_x^\sigma,\mathcal{W}_y^\sigma,\mathcal{W}_x^\tau,\mathcal{W}_y^\tau\}$, commute with themselves, all the hexagon plaquette operators, and the Hamiltonian $H$.
The sixteen topological sectors of the Wilson loop operators ($\# 1\sim 16$) are specified in the table of Fig.~\ref{fig:7}.

In the KSL$\times$KSL state, the ground states appear in the topological sectors having $-1$ in two of $\mathcal{W}_{x,y,z}^\sigma$ and also in two of $\mathcal{W}_{x,y,z}^\tau$ (ninefold degeneracy).
In the RVB state, the ground states stay in the topological sectors satisfying the condition, $\mathcal{W}_{l}^\sigma=\mathcal{W}_{l}^\tau~(l=x,y,z)$ (fourfold degeneracy).

\section{Topological transition by anyon condensation and confinement\label{sec:anyon}}

The topological degeneracies provide useful hints about the underlying topological orders of the two QSLs. The ninefold degeneracy in the KSL$\times$KSL state is consistent with the ${\rm Ising}\times\overline{\rm Ising}$ topological order with the nine different anyon-pairs, $\{1_{\rm I},\sigma_{\rm I},\psi_{\rm I}\}\boxtimes \{1_{\rm II},\sigma_{\rm II},\psi_{\rm II}\}$. 
Here the subscript means the layer index (I: upper layer, II: lower layer), and each layer has the three anyon sectors: trivial boson ($1$), Ising anyon ($\sigma$), and fermion ($\psi$). Note that there is nontrivial braiding between $\sigma$ and $\psi$; moving a $\psi$-particle around a $\sigma$-particle changes the overall sign of the wave function ($|\Psi\rangle \rightarrow e^{i\pi}|\Psi\rangle$). These anyons satisfy the fusion rules,
\begin{equation}
\psi \times \psi = 1,~~~\sigma \times \psi = \sigma,~~~\sigma\times\sigma=1+\psi,
\end{equation}
where the fusion outcome of two $\sigma$-particles has two possibilities due to the non-abelian nature of the Ising anyon $\sigma$ (quantum dimension: $\sqrt{2}$)~\cite{Kitaev2006,Nayak2008,Bombin2010,Pachos2012,Simon2023book}.
This Ising anyon topological order emerges in each layer of the KSL$\times$KSL state, hence the ${\rm Ising}\times\overline{\rm Ising}$ topological order for the whole system (bar indicates that the two layers are time-reversal partners)~\cite{Kitaev2006,Bais2009,Burnell2018,Pachos2012,Simon2023book}.

On the other hand, the fourfold degeneracy in the RVB state implies the $\mathbb{Z}_2$ toric code topological order with the four different anyons, $\{1,e,m,\epsilon\}$; the $e$- and $m$-particles are self-bosons with mutual statistics, and the $\epsilon$-particle is a self-fermion composed of the $e$- and $m$-particles. All these particles are abelian anyons satisfying the fusion rules,
\begin{equation}
e\times e = m \times m = \epsilon \times \epsilon = 1,~~~e \times m = \epsilon,
\label{eq:TC-fusion}
\end{equation}
which define the $\mathbb{Z}_2$ toric code topological order~\cite{Kitaev2003,Kitaev2006}.

The transition between the non-abelian ${\rm Ising}\times\overline{\rm Ising}$ topological order and the abelian $\mathbb{Z}_2$ topological order can be understood by the mechanism of anyon condensation transition~\cite{Bais2009,Burnell2011,Burnell2012,Burnell2018}.
Suppose we condense the fermion-pair $\psi_{\rm I}\boxtimes\psi_{\rm II}$.
Then, $\psi_{\rm I}\boxtimes1_{\rm II}$ and $1_{\rm I}\boxtimes\psi_{\rm II}$ become indistinguishable and {\it identified} as a same type of anyons in the condensed phase: $\psi_{\rm I}\boxtimes1_{\rm II}=1_{\rm I}\boxtimes\psi_{\rm II}$.
Moreover, anyons having nontrivial braiding with the fermion-pair are {\it confined}, thus only $\psi_{\rm I}\boxtimes1_{\rm II}$, $1_{\rm I}\boxtimes\psi_{\rm II}$, and $\sigma_{\rm I}\boxtimes\sigma_{\rm II}$ remain deconfined. The following table highlights the identical anyons (colored) and the confined anyons (crossed) resulting from the anyon condensation.
\begin{equation}
\begin{array}{|c|c|c|c|}
\hline
 & 1_{\rm II} & \sigma_{\rm II} & \psi_{\rm II}
 \\
 \hline
1_{\rm I} & \cellcolor{yellow!25} 1_{\rm I}\boxtimes1_{\rm II} &  \mynot{purple}{1_{\rm I}\boxtimes\sigma_{\rm II}} & \cellcolor{blue!25} 1_{\rm I}\boxtimes\psi_{\rm II}
 \\
 \hline
\sigma_{\rm I} & \mynot{purple}{\sigma_{\rm I}\boxtimes1_{\rm II}} & \cellcolor{red!25} \sigma_{\rm I}\boxtimes\sigma_{\rm II} & \mynot{purple}{\sigma_{\rm I}\boxtimes\psi_{\rm II}}
 \\
 \hline
\psi_{\rm I} & \cellcolor{blue!25} \psi_{\rm I}\boxtimes1_{\rm II} & \mynot{purple}{\psi_{\rm I}\boxtimes\sigma_{\rm II}} & \cellcolor{yellow!25} \psi_{\rm I}\boxtimes\psi_{\rm II}
\\
\hline
\end{array}
\end{equation}
We note that
\begin{eqnarray}
&&
(\sigma_{\rm I}\boxtimes\sigma_{\rm II}) \times (\sigma_{\rm I}\boxtimes\sigma_{\rm II}) 
\nonumber\\
&=& (1_{\rm I}\boxtimes1_{\rm II}) + (\psi_{\rm I}\boxtimes1_{\rm II}) + (1_{\rm I}\boxtimes\psi_{\rm II}) + (\psi_{\rm I}\boxtimes\psi_{\rm II}),~~~~~~~
\label{eq:bilayer-fusion}
\end{eqnarray} 
where we have two trivial bosons ($1_{\rm I}\boxtimes1_{\rm II}$,$\psi_{\rm I}\boxtimes\psi_{\rm II}$) and two fermions ($\psi_{\rm I}\boxtimes1_{\rm II}$,$1_{\rm I}\boxtimes\psi_{\rm II}$).
In order to reproduce the abelian $\mathbb{Z}_2$ topological order in the condensed phase, $\sigma_{\rm I}\boxtimes\sigma_{\rm II}$ (which has the quantum dimension 2) should be {\it split} into two abelian anyons.
Therefore, we may assume that 
\begin{eqnarray}
&&
1_{\rm I}\boxtimes1_{\rm II} = \psi_{\rm I}\boxtimes\psi_{\rm II} = 1,
\nonumber\\
&&
\psi_{\rm I}\boxtimes1_{\rm II} = 1_{\rm I}\boxtimes\psi_{\rm II} = \epsilon,
\nonumber\\
&&\sigma_{\rm I}\boxtimes\sigma_{\rm II}= e+m.
\end{eqnarray}
Then, Eq.~(\ref{eq:bilayer-fusion}) becomes identical to the toric code's fusion rules shown in Eq.~(\ref{eq:TC-fusion})~\cite{Bais2009,Burnell2018,Simon2023book}.

\subsection{Anyon condensation}

\begin{figure}[tb]
\includegraphics[width=\linewidth]{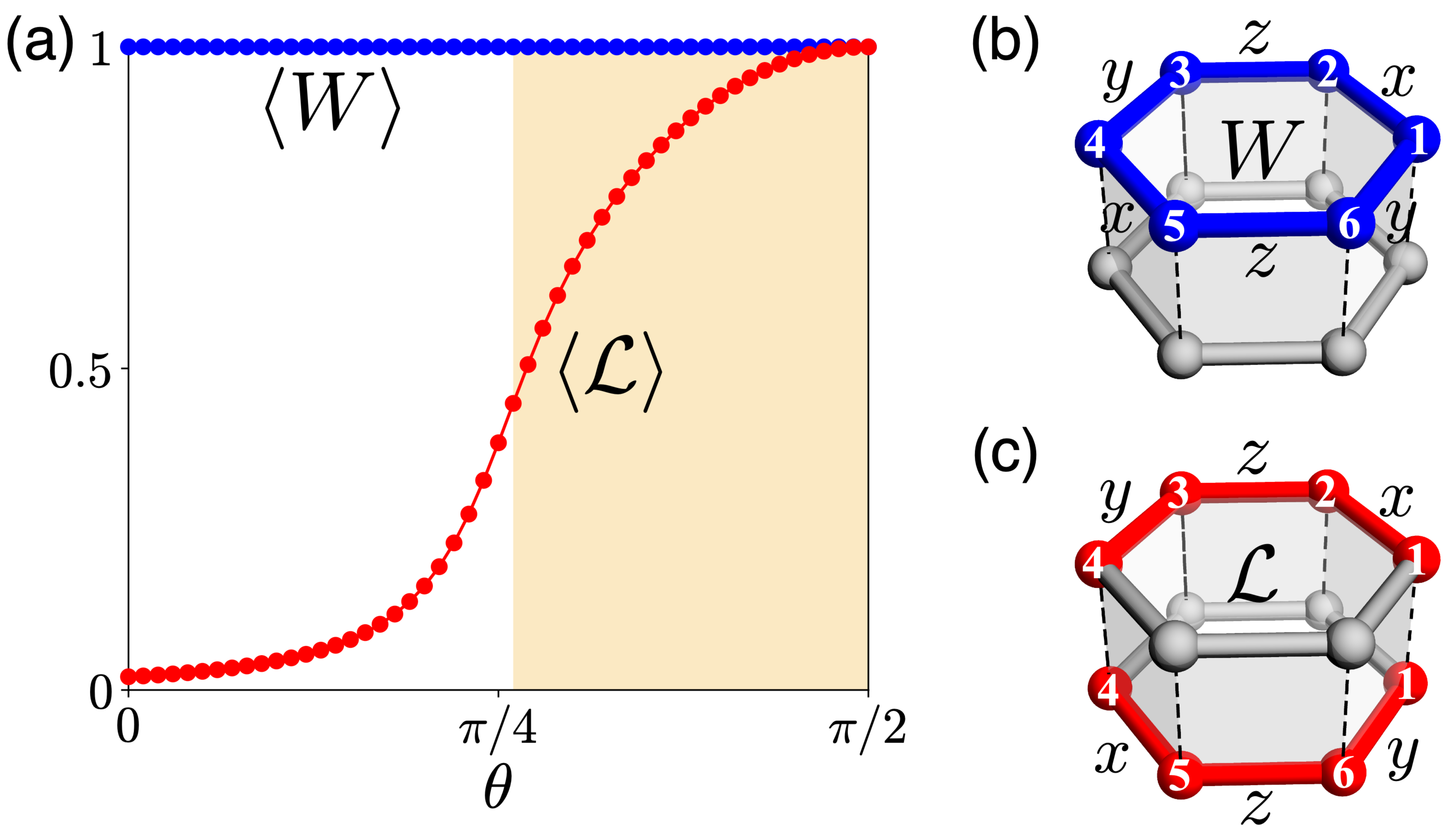}
\caption{Condensation of the fermion-pair.
(a) The expectation values of the loop operators, $\langle W \rangle$ and $\langle \mathcal{L} \rangle$.
In the RVB state, the substantially large values of $\langle \mathcal{L} \rangle$ indicate the condensation of the fermion-pair, $\psi_{\rm I}\boxtimes\psi_{\rm II}$.
(b),(c) Visualizations of the $W$ and $\mathcal{L}$ operators.
}
\label{fig:9}
\end{figure}

Now we confirm the mechanism of the anyon condensation induced transition in our microscopic model.
First, we check if $\psi_{\rm I}\boxtimes1_{\rm II}$ and $1_{\rm I}\boxtimes\psi_{\rm II}$ become indeed indistinguishable identical anyons ($\psi_{\rm I}\boxtimes1_{\rm II} = 1_{\rm I}\boxtimes\psi_{\rm II}$).
To this end, we consider two different loop operators.
The first one is the usual hexagon plaquette operator,
\begin{equation}
W=(\sigma_1^y\sigma_6^y)(\sigma_6^z\sigma_5^z)(\sigma_5^x\sigma_4^x)(\sigma_4^y\sigma_3^y)(\sigma_3^z\sigma_2^z)(\sigma_2^x\sigma_1^x) ,
\end{equation}
which is defined within a single layer [Fig.~\ref{fig:9}(b)].
If we apply the Majorana representation, $\sigma_j^\gamma\sigma_k^\gamma=-u_{jk} (i c_j c_k)$ where $u_{jk}=i b_j^\gamma b_k^\gamma$,
the action of $W$ is to create a pair of $c$-fermions, move one of the fermions along the hexagon, and finally annihilate the fermion-pair~\cite{Kitaev2006}.
Notice that the $c$-fermions correspond to $\psi_{\rm I}$ fermions.
The second one is defined over the two layers:
\begin{equation}
\mathcal{L}=-(\tau_1^y\tau_6^y)(\tau_6^z\tau_5^z)(\tau_5^x\tau_4^x)(\sigma_4^y\sigma_3^y)(\sigma_3^z\sigma_2^z)(\sigma_2^x\sigma_1^x) .
\label{eq:loop-op-2}
\end{equation}
This operator moves a fermion ($\psi_{\rm I}$) around an upper ``half'' hexagon, and moves another fermion ($\psi_{\rm II}$) around a lower ``half'' hexagon [Fig.~\ref{fig:9}(c)].
This type of loop operator has been considered in a recent study on anyon condensation in kagome quantum spin liquids~\cite{Meng2021}.

Figure~\ref{fig:9}(a) shows the expectation value $\langle \mathcal{L}\rangle$.
We find that  $\langle \mathcal{L}\rangle$ is small in the KSL$\times$KSL phase (close to zero near $\theta=0$), meaning that $\psi_{\rm I}$ and $\psi_{\rm II}$ cannot move across the two layers because they are distinct anyons supported on different layers.
By contrast, substantially large values of $\langle \mathcal{L}\rangle$ are observed in the RVB phase (reaching one at $\theta=\pi/2$).
This implies that $\psi_{\rm I}$ and $\psi_{\rm II}$ become identical particles, thus the fermions ($\psi_{\rm I}=\psi_{\rm II}$) can complete the hexagon loop motion.
It is only possible when we have the condensation of $\psi_{\rm I}\boxtimes\psi_{\rm II}$.
Therefore, $\langle \mathcal{L}\rangle$ plays a role of an order parameter for the anyon condensation.

\begin{figure*}[tb]
\includegraphics[width=\linewidth]{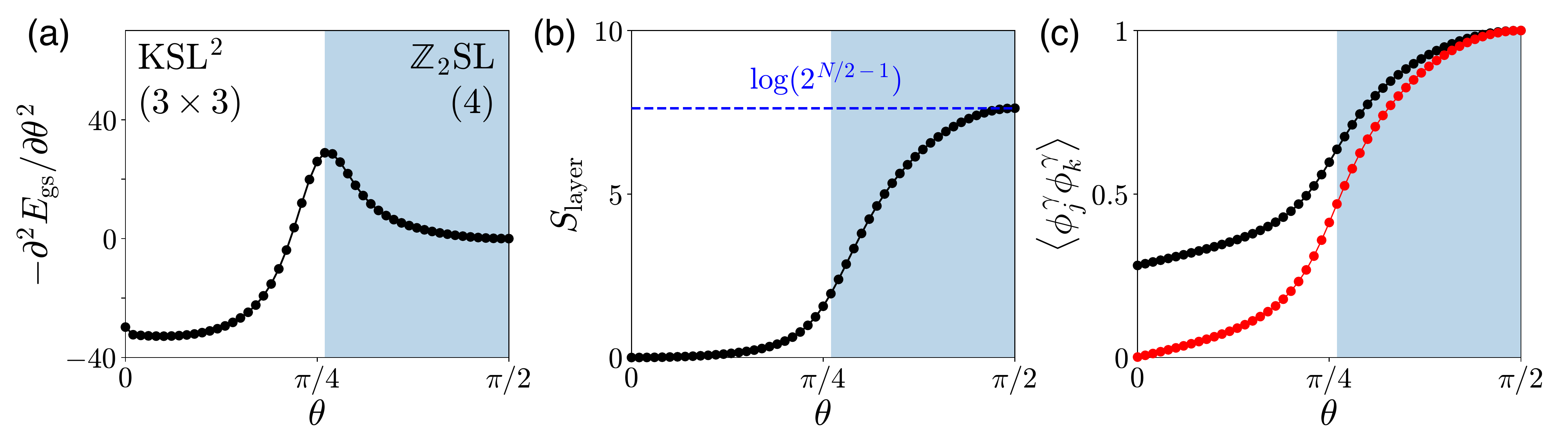}
\caption{
ED results for the cases \#5 and \#6.
(a) Phase diagram of the model. The KSL$\times$KSL and $\mathbb{Z}_2{\rm SL}$ states are connected by a continuous transition at $\theta_c \simeq 0.26 \pi$ as indicated by a small peak in the derivative of the ground state energy, $-{\partial^2 E_{\rm gs}}/{\partial \theta^2}$. The numbers in the parentheses indicate topological degeneracies.
(b) The entanglement entropy $S_{\rm layer}$. The dashed line marks the value of $\log(2^{N/2-1})$, where $N=24$.
(c) The four-spin correlators, $\langle \phi_j^\gamma \phi_k^\gamma \rangle$ (black) and $\langle \phi_j^\gamma\phi_k^\gamma \rangle-\langle\sigma_j^\gamma\sigma_k^\gamma\rangle \langle\tau_j^\gamma\tau_k^\gamma\rangle$ (red).
Exactly same results of $E_{\rm gs}$, $S_{\rm layer}$, $\langle \phi_j^\gamma \phi_k^\gamma \rangle$, $\langle \phi_j^\gamma\phi_k^\gamma \rangle-\langle\sigma_j^\gamma\sigma_k^\gamma\rangle \langle\tau_j^\gamma\tau_k^\gamma\rangle$ are obtained for both cases (\#5,6).
}
\label{fig:10}
\end{figure*}

\subsection{Anyon confinement}

Next, we investigate the confinement of the Ising anyons, $\sigma_{\rm I}\boxtimes1_{\rm II}~(=\sigma_{\rm I}\boxtimes\psi_{\rm II})$ \& $1_{\rm I}\boxtimes\sigma_{\rm II}~(=\psi_{\rm I}\boxtimes\sigma_{\rm II})$.
To allow Ising anyon excitations, we consider flux sectors with $W=-1$ or $Z=-1$. Under broken time-reversal symmetry, flux excitations with $W=-1$ or $Z=-1$ trap localized Majorana zero modes, behaving as the non-abelian Ising anyons~\cite{Kitaev2006,Pachos2012,Pachos2020}.
By comparing energy costs of different flux patterns, we can identify in which flux patterns the confinement occurs.

Figure~\ref{fig:2} shows the lowest excitation energy profiles in various flux patterns.
Depending on the phase, flux excitations have different behaviors in the energy cost.
In the KSL$\times$KSL phase, the excitation energy with respect to the (zero-flux) ground state is roughly proportional to the number of excited fluxes.
In sharp contrast, in the RVB phase, such simple counting does not work.
To be specific, when fluxes are excited only on a single layer, the excitation energy dramatically increases across the transition point $\theta_c\simeq0.26\pi$ [see the flux pattern \#1 in Figs.~\ref{fig:2}(a) and~\ref{fig:2}(e)].
In general, flux excitation energy significantly increases if any flux mismatch (i.e., unpaired flux) exists between the upper layer and lower layer as shown in the flux patterns \#1,3,4 [Figs.~\ref{fig:2}(a),~\ref{fig:2}(c),~\ref{fig:2}(d),~and~\ref{fig:2}(e)].
However, if fluxes are all paired up between the two layers at same locations ($W_p=Z_p=-1$), the energy cost vanishingly decreases.
See the flux pattern \#2 in Figs.~\ref{fig:2}(b)~and~\ref{fig:2}(e) and the flux patterns \#5-8 in Figs.~\ref{fig:2}(f)-(j).

The large energy costs due to unpaired fluxes, shown in the flux patterns \#1,3,4, can be explained by the hardcore dimer constraint of the RVB phase ($\langle \phi_j^\gamma \phi_k^\gamma \rangle \approx-1$).
Any unpaired flux ($W_p Z_p = -1$) necessarily violates the hardcore dimer constraint leading to an energy increase in the order of $G$.
In contrast, paired fluxes between the two layers at same locations ($W_p=Z_p=-1$) can be consistent with the hardcore dimer constraint.
The lowest excitations of such paired fluxes are captured by the effective quantum dimer model which we derived in large-$G$ limit [Eq.~(\ref{eq:QDM})].
Notice that the excitation energy scale is $\lambda~(\propto K_\sigma^6/G^5 = \cos^6 \theta / \sin^5 \theta)$, which explains the vanishingly decreasing excitation energies of the flux patterns \#2,5-8 [Figs.~\ref{fig:2}(e)~and~\ref{fig:2}(f)].
Remarkably, unpaired fluxes occur at an extremely high energy scale ($G$) compare to paired fluxes described by the quantum dimer model ($\lambda$).
In the low energy physics of the RVB phase, unpaired fluxes never appear, essentially confined due to the extremely large energy cost.

The paired fluxes in Figs.~\ref{fig:2}(b)~and~\ref{fig:2}(g)-(j) correspond to the paired Ising anyons, $\sigma_{\rm I}\boxtimes\sigma_{\rm II}$.
The unpaired fluxes in Figs.~\ref{fig:2}(a),~\ref{fig:2}(c),~and~\ref{fig:2}(d) correspond to the Ising anyons, $\sigma_{\rm I}\boxtimes1_{\rm II}$ \& $1_{\rm I}\boxtimes\sigma_{\rm II}$.
The energy profiles in Figs.~\ref{fig:2}(e)~and~\ref{fig:2}(f) elucidate the confinement of $\sigma_{\rm I}\boxtimes1_{\rm II}$ \& $1_{\rm I}\boxtimes\sigma_{\rm II}$ and reveal the explicit form of $\sigma_{\rm I}\boxtimes\sigma_{\rm II}$ in our model (flux-pair at a same location).

The fact that only paired fluxes appear as deconfined anyons in the RVB phase can be understood in a more intuitive way.
In order for $\psi_{\rm I}$ \& $\psi_{\rm II}$ to be identical particles, $\psi_{\rm I}$ \& $\psi_{\rm II}$ have to see a same flux pattern on the upper and lower layers.
Otherwise, $\psi_{\rm I}$ \& $\psi_{\rm II}$ cannot be identical.
This simple fact results in the confinement of unpaired fluxes ($\sigma_{\rm I}$ \& $\sigma_{\rm II}$), only allowing paired fluxes ($\sigma_{\rm I}\boxtimes\sigma_{\rm II}$) as deconfined anyons.

\subsection{Topological degeneracy revisited}

The fourfold topological degeneracy of the RVB state is also easily understood by the same picture. Fermions, $\psi_{\rm I}$ \& $\psi_{\rm II}$, must see a same pattern of Wilson loop fluxes on the upper and lower layers.
In other words, the topological sectors allowed by the $\psi_{\rm I}\boxtimes\psi_{\rm II}$-condensation must satisfy the condition, $\mathcal{W}_{l}^\sigma=\mathcal{W}_{l}^\tau~(l=x,y,z)$, which leads to exactly the four sectors ($\#1,6,11,16$) in Fig.~\ref{fig:7}.

\section{General cases\label{sec:general-cases}}

We investigate other parameter regions of the model beyond Eq.~(\ref{eq:parameter-region}).
We still constrain ourselves into the cases of $|K_\sigma|=|K_\tau|$ since more generic cases can be easily understood from our discussion here.
We find that the signs of the coupling constants $\{G,K_{\sigma},K_{\tau}\}$ determine the emergence of a RVB-type $\mathbb{Z}_2$ spin liquid. Table \ref{tab:II} lists six possible cases for the sign structure. Three cases (\#1,5,6) stabilize a $\mathbb{Z}_2$ spin liquid ($\mathbb{Z}_2$SL) whereas the other three cases do not. The case \#1 corresponds to the parameter choice in Eq.~(\ref{eq:parameter-region}).

\begin{table}[b]
\begin{ruledtabular}
\begin{tabular}{ccccc}
Case  &  $G$  &  $K_{\sigma}$  &  $K_{\tau}$  & $\mathbb{Z}_2$ spin liquid
\\
\hline
\#1  &  $+$  & $+$  &  $-$  &  \checkmark
\\
\#2  &  $+$  & $+$  &  $+$  &  
\\
\#3  &  $+$  & $-$  &  $-$  &  
\\
\hline
\#4  &  $-$  & $+$  &  $-$  &  
\\
\#5  &  $-$  & $+$  &  $+$  &  \checkmark
\\
\#6  &  $-$  & $-$  &  $-$  &  \checkmark
\\
\end{tabular}
\caption{
Possible sign structures of the coupling constants $\{G,K_{\sigma},K_{\tau}\}$ and the emergence of a RVB-type $\mathbb{Z}_2$ spin liquid state.}
\label{tab:II}
\end{ruledtabular}
\end{table}

Here we discuss the cases \#5 and \#6 ($G<0,~K_\sigma=K_\tau$). Figure~\ref{fig:10} shows ED results for the two cases with the parametrization,
\begin{equation}
G=-\sin\theta
~~~\&~~~
K_{\sigma}=K_{\tau}=
\left\{
\begin{array}{cc}
+\cos\theta & ({\rm Case~\#5})
\\
-\cos\theta & ({\rm Case~\#6})
\end{array}
\right\}.
\end{equation}
The results are almost same as in the case \#1 in terms of the ground state energy $E_{\rm gs}$, the topological degeneracy, the entanglement entropy $S_{\rm layer}$, and the four-spin correlator $\langle \phi_j^\gamma\phi_k^\gamma \rangle$. 
Yet, there is a difference between \#5,6 and  \#1 in the sign of $\langle \phi_j^\gamma\phi_k^\gamma \rangle$, which is positive in the former cases but negative in the latter case.
In the strong coupling limit ($|G| \gg |K_{\sigma,\tau}|$), we see distinct behaviors in the associated $\mathbb{Z}_2$ spin liquids:
\begin{equation}
\langle \phi_j^\gamma\phi_k^\gamma \rangle \rightarrow
\left\{
\begin{array}{cc}
+1 & ({\rm Cases~\#5,6})
\\
-1 & ({\rm Case~\#1})
\end{array}
\right\}.
\end{equation}
The property $\phi_j^\gamma\phi_k^\gamma =-1$ of \#1 allows us to construct the effective quantum dimer model on the dual kagome lattice as we have seen in Sec.~\ref{sec:RVB}. However, the property $\phi_j^\gamma\phi_k^\gamma =+1$ of \#5,6 leads to states violating the hardcore dimer constraint, so the dimer representation is no longer useful for the two cases.

Instead, a $\mathbb{Z}_2$ gauge theory can be more conveniently constructed on the honeycomb lattice for the strong coupling limits of the cases \#5,6.
Utilizing the property $\phi_j^\gamma\phi_k^\gamma =+1$ at each bond $\langle jk \rangle_\gamma$,
we may define a $\mathbb{Z}_2$ link variable,
\begin{equation}
X_{jk}={\rm sgn} (\phi_j^\gamma)={\rm sgn} (\phi_k^\gamma)=\pm 1.
\end{equation}
Such $X$-variables are subject to the Gauss law constraint,
\begin{equation}
X_{jk} X_{jl} X_{jm} = -1 ,
\end{equation}
which is simply the product of $X$-variables on the three links sharing site $j$.
The Gauss law constraint arises as a combined effect of (i) the property $\phi_j^\gamma\phi_k^\gamma =+1$ at each bond $\langle jk \rangle_\gamma$ and (ii) the constraint $\phi_j^x\phi_j^y\phi_j^z=-1$ at each site $j$.
In other words, the property $\phi_j^\gamma\phi_k^\gamma =+1$ defines the Hilbert space of the $\mathbb{Z}_2$ link variables $\{X_{jk}\}$ subject to the Gauss law constraint.

We repeat a sixth order degenerate perturbation theory for the cases \#5,6 and obtain an effective Hamiltonian that is exactly in the same form of Eq.~(\ref{eq:QDM}):
\begin{equation}
\mathcal{H}_{\rm eff}= - \lambda \sum_p \hat{W}_p ~~~{\rm where}~~~\hat{W}_p \propto \prod_{jk \in p} Z_{jk} .
\end{equation}
The plaquette operator $\hat{W}_p$ flips the signs of $X$-variables at plaquette $p$ as denoted by the Pauli $Z$-operators in the above.
Here we see a $\mathbb{Z}_2$ gauge theory on the honeycomb lattice.

In the other cases (\#2,3,4), the effects of the $K_{\sigma,\tau}$ terms completely cancel each other, failing to create any $\mathbb{Z}_2$ spin liquid state in the strong coupling limit. For instance, in the cases \#2,3 ($G>0$ and $K_\sigma=K_\tau$), we have the relationship,
\begin{equation}
\sigma_j^\gamma \sigma_k^\gamma | \Phi \rangle 
= 
- \tau_j^\gamma \tau_k^\gamma | \Phi \rangle,
\end{equation}
where $| \Phi \rangle$ is an arbitrary ground state ($\phi_j^\gamma \phi_k^\gamma=-1$) of the interlayer interactions.
Then, one can easily see that
\begin{equation}
( K_\sigma \sigma_j^\gamma \sigma_k^\gamma + K_\tau \tau_j^\gamma \tau_k^\gamma ) | \Phi \rangle 
= 
( K_\sigma - K_\tau ) \sigma_j^\gamma \sigma_k^\gamma | \Phi \rangle 
=0 .
\end{equation}
Due to this cancellation, the Kitaev interactions do not create any effect on the dimer Hilbert space and any $\mathbb{Z}_2$ spin liquid.

\section{Discussion and outlook\label{sec:discussion}}

The anyon condensation transition between the non-abelian ${\rm KSL}\times{\rm KSL}$ state and the ${\rm RVB}$ state was identified in our model via the loop operator $\mathcal{L}$.
Interestingly, there is an intimate relationship between the anyon condensation and the hardcore dimer constraint [Eq.~(\ref{eq:G-only-condition})].
To see this, we consider the constraint in a slightly different fashion: $\sigma_j^\gamma \sigma_k^\gamma=-\tau_j^\gamma \tau_k^\gamma$.
This tells us that moving the fermions, $\psi_{\rm I}$ \& $\psi_{\rm II}$, has a same effect, meaning that the RVB state does not distinguish between $\psi_{\rm I}$ and $\psi_{\rm II}$; they are essentially same particles.
To be more precise, if we apply the hardcore dimer constraint to Eq.~(\ref{eq:loop-op-2}), we obtain the relationship $\mathcal{L}=W=Z$.
Therefore, the hardcore dimer constraint itself means the $\psi_{\rm I}\boxtimes\psi_{\rm II}$ anyon condensation.

We emphasize that the inter-layer interactions ($\sigma_j^\gamma \sigma_k^\gamma\tau_j^\gamma \tau_k^\gamma=\phi_j^\gamma\phi_k^\gamma$) are crucial for realizing the two topological spin liquids in the bilayer system and also for understanding the anyon condensation transition.
In the weak coupling regime, a chirality order is developed in each layer due to the inter-layer interactions, yielding the non-abelian KSL$\times$KSL state.
In the strong coupling regime, the inter-layer interactions construct the Hilbert space of hardcore dimers for the RVB state.
The hardcore dimer constraint established by the inter-layer interactions plays a role of an order parameter for the anyon condensation transition between the two topological spin liquids.
Furthermore, it was due to the inter-layer interactions conserving the flux quantum number that we could easily trace the Ising anyons and their confinement across the transition (Fig.~\ref{fig:2}).

\begin{figure}[tb]
\includegraphics[width=\linewidth]{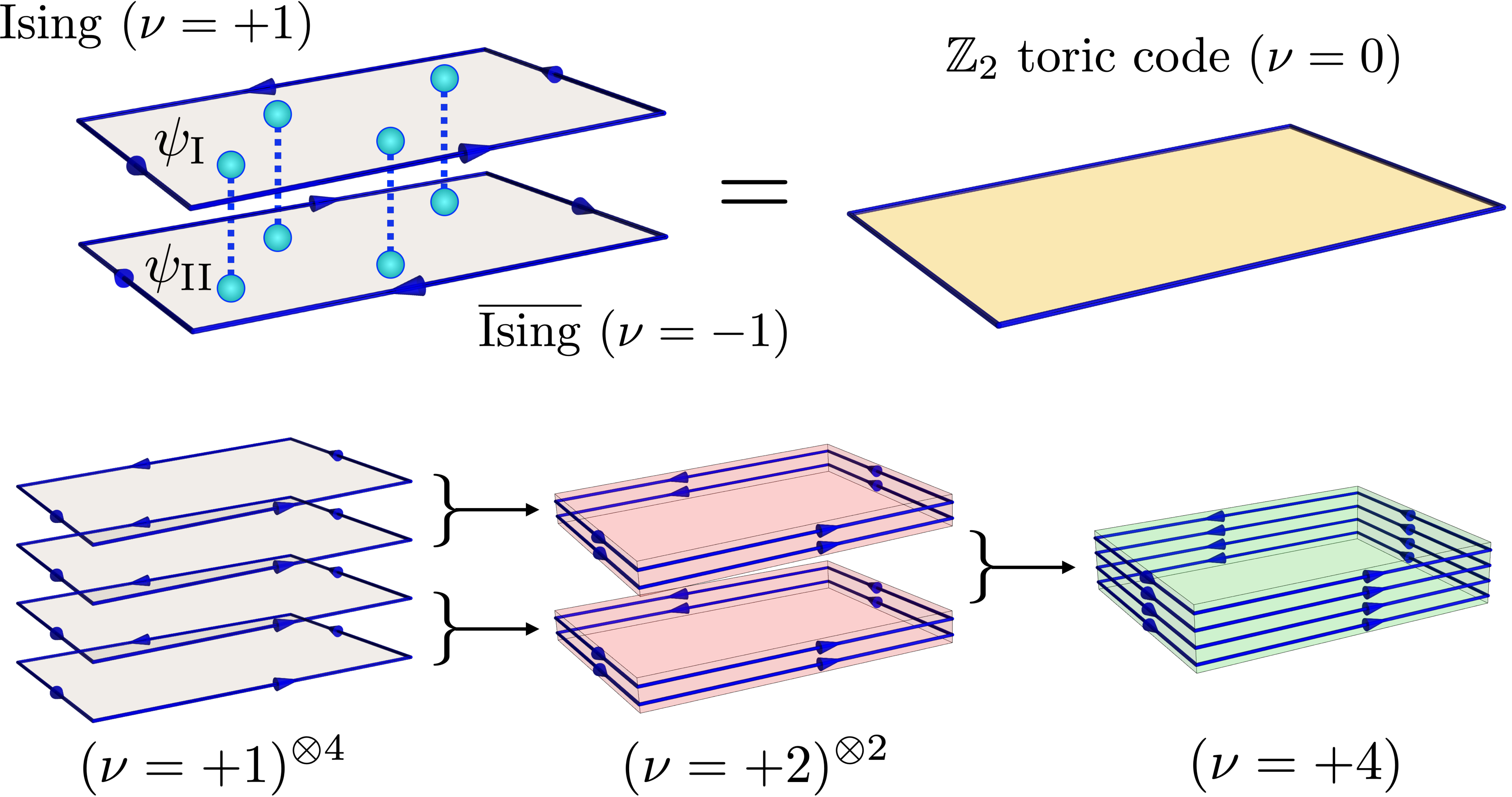}
\caption{Anyon-condensed multilayer constructions of the Kitaev's sixteenfold way of anyon theories. 
Upper: creation of the toric code topological order ($\nu=0$) by attaching two layers of Ising topological orders ($\nu=\pm1$) with the glue of anyon condensation ($\langle\psi_{\rm I}\boxtimes\psi_{\rm II}\rangle\ne0$).
Lower: Consecutive anyon condensation transitions from four layers of $\nu=+1$ topological order, to two layers of $\nu=+2$ topological order, finally to a single layer of $\nu=+4$ topological order.
}
\label{fig:11}
\end{figure}

This work demonstrates an anyon-condensed multilayer construction for the Kitaev's sixteenfold way of anyon theories.
In his original work, Kitaev provided a simple idea to construct sixteen different topological orders~\cite{Kitaev2006}. Namely, in the system of Majorana fermions coupled with $\mathbb{Z}_2$ fluxes, the spectral Chern number ($\nu$) of the Majorana fermions determines the anyon properties of the $\mathbb{Z}_2$ fluxes. Changing the Chern number results in sixteen distinct types of topological orders determined by $\nu~{\rm mod}~16$.
Such sixteenfold way constructions have been discussed in several solvable models recently~\cite{Zhang2020,Vidal2020,Tu2020,Zhou2023}.
Here we discuss a different approach using anyon condensation in mutilayer systems.
First, we note that the non-abelian KSL with $\nu=1$ realizes the Ising anyon topological order.
Then, the bilayer system of KSLs with the opposite Chern numbers $\nu_{\rm I}=+1~\&~\nu_{\rm II}=-1$ (having zero for the net Chern number: $\nu = \nu_{\rm I}+\nu_{\rm II}=0$) may construct the $\mathbb{Z}_2$ toric code topological order.
Our bilayer model demonstrates this via the RVB state that emerges from the KSL$\times$KSL state by anyon condensation.
Generalizing this idea, all the sixteen topological orders could be constructed by stacking non-abelian KSLs with $\nu=\pm1$ and inducing anyon condensation over the multilayers, i.e., anyon-condensed multilayer constructions of the Kitaev's sixteenfold way~(Fig.~\ref{fig:11}).

In our study, the KSL$\times$KSL state was shown to open a finite energy gap by developing the chirality order that breaks time-reversal symmetry. From renormalization group analyses for interacting Dirac fermions, any short-ranged four-fermion interaction is, however, perturbatively irrelevant in the systems with time-reversal and lattice symmetries. Hence, the gaplessness of the Dirac fermions is maintained until some finite strengths of four-fermion interactions~\cite{Herbut2006}. The same physics applies to the Kitaev spin liquid where the low energy physics is described by the Majorana version of the field theory. Such gaplessness is expected around the weak coupling limit of our system since the inter-layer interactions are essentially four-fermion interactions in the Majorana representation [$\sigma_j^\gamma\sigma_k^\gamma\tau_j^\gamma\tau_k^\gamma=u_{{\rm I},jk}u_{{\rm II},jk}(ic_{{\rm I},j}c_{{\rm I},k})(ic_{{\rm II},j}c_{{\rm II},k})$, where I \& II are layer indices]. Nonetheless, our mean-field theory and finite-size ED calculations suggest that the bilayer system can spontaneously break time-reversal symmetry and open a finite energy gap before the transition to the RVB state. Combining our results with the field theory argument, we anticipate that in the thermodynamic limit the bilayer system has a gapless region from the $\theta=0$ point, then opens up a gap at some finite value $\theta_{\rm gap}$, and finally undergoes the anyon condensation transition to the RVB state at $\theta_{\rm c}$ $(0<\theta_{\rm gap}<\theta_{\rm c})$.

Low energy field theory for the anyon condensation transition (at $\theta_{\rm c}$) is an interesting problem lying beyond the Landau-Ginzburg-Wilson paradigm. To our best knowledge, a general framework for such transitions is currently unknown. Nonetheless, the 3D Ising universality has been proposed for the $\psi_{\rm I}\boxtimes\psi_{\rm II}$ condensate induced transition between the ${\rm Ising}\times\overline{\rm Ising}$ and $\mathbb{Z}_2$ topological orders by Burnell,~Simon,~and~Slingerland~\cite{Burnell2012,Burnell2011}. Our system is expected to be in the same 3D Ising universality class.

We compare our work with the recent spin-$3/2$ transverse field Ising model by Verresen and Vishwanath~\cite{Verresen2022}.
Both cases lead to emergent quantum dimer models with similar structures.
The hardcore dimer constraint for the dimer Hilbert space is energetically implemented, and the dimer resonance is induced by anyon fluctuations---by the Kitaev interactions in our case, and by the transverse field in their case.
Despite the similarities, the two works have different interests.
Ref.~\cite{Verresen2022} focuses on the possible quantum liquids induced by different types of anyon fluctuations.
In our study, we are mainly interested in the anyon condensation transition and its identification.

\begin{table*}[tb]
\begin{ruledtabular}
\begin{tabular}{c|cccccccc}
\makecell{Dimer state $|\mathpzc{D}\rangle$ \\ \& motion graph}
& \parbox{5em}{\includegraphics[width=0.6\linewidth]{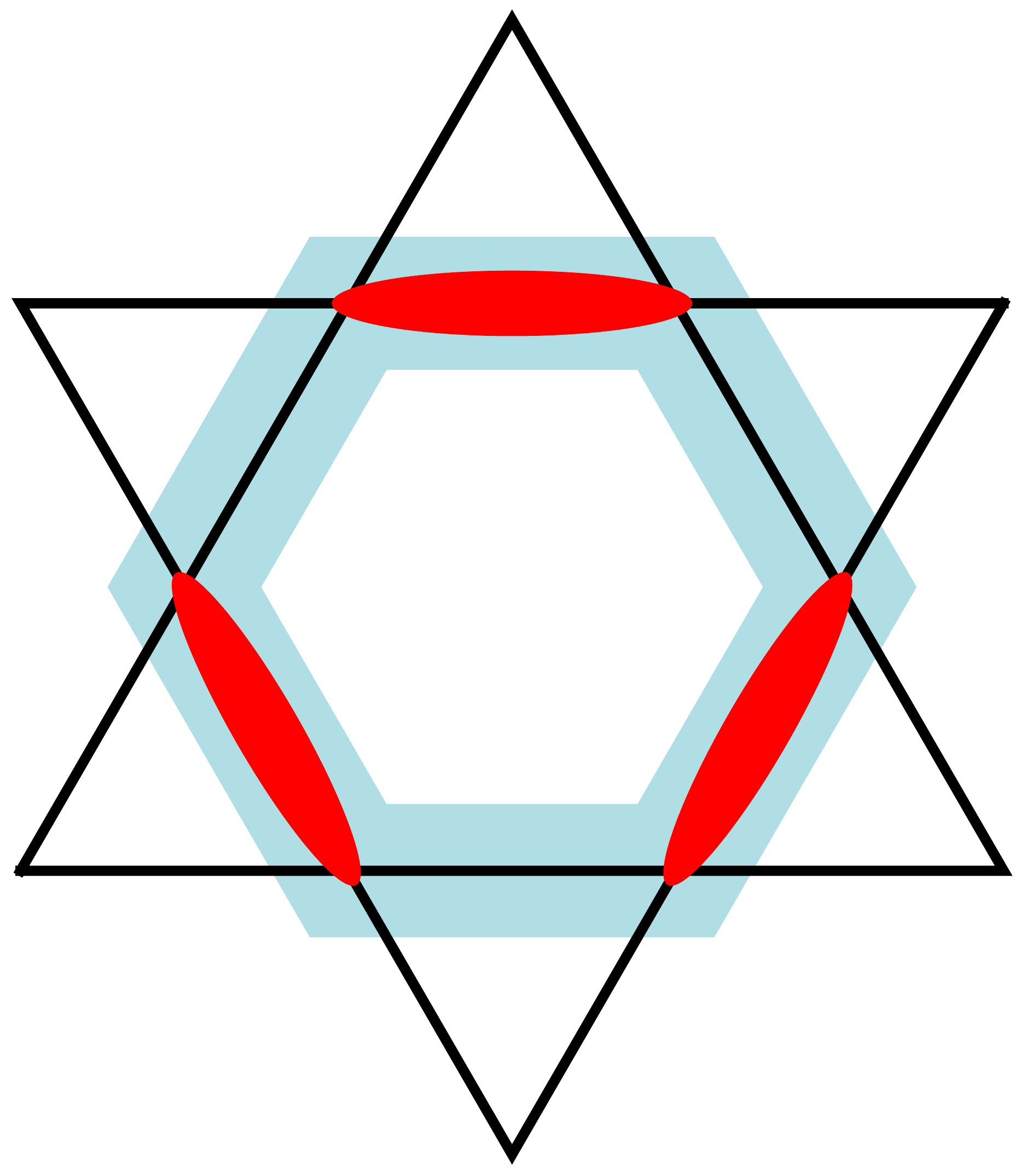}}  
& \parbox{5em}{\includegraphics[width=0.6\linewidth]{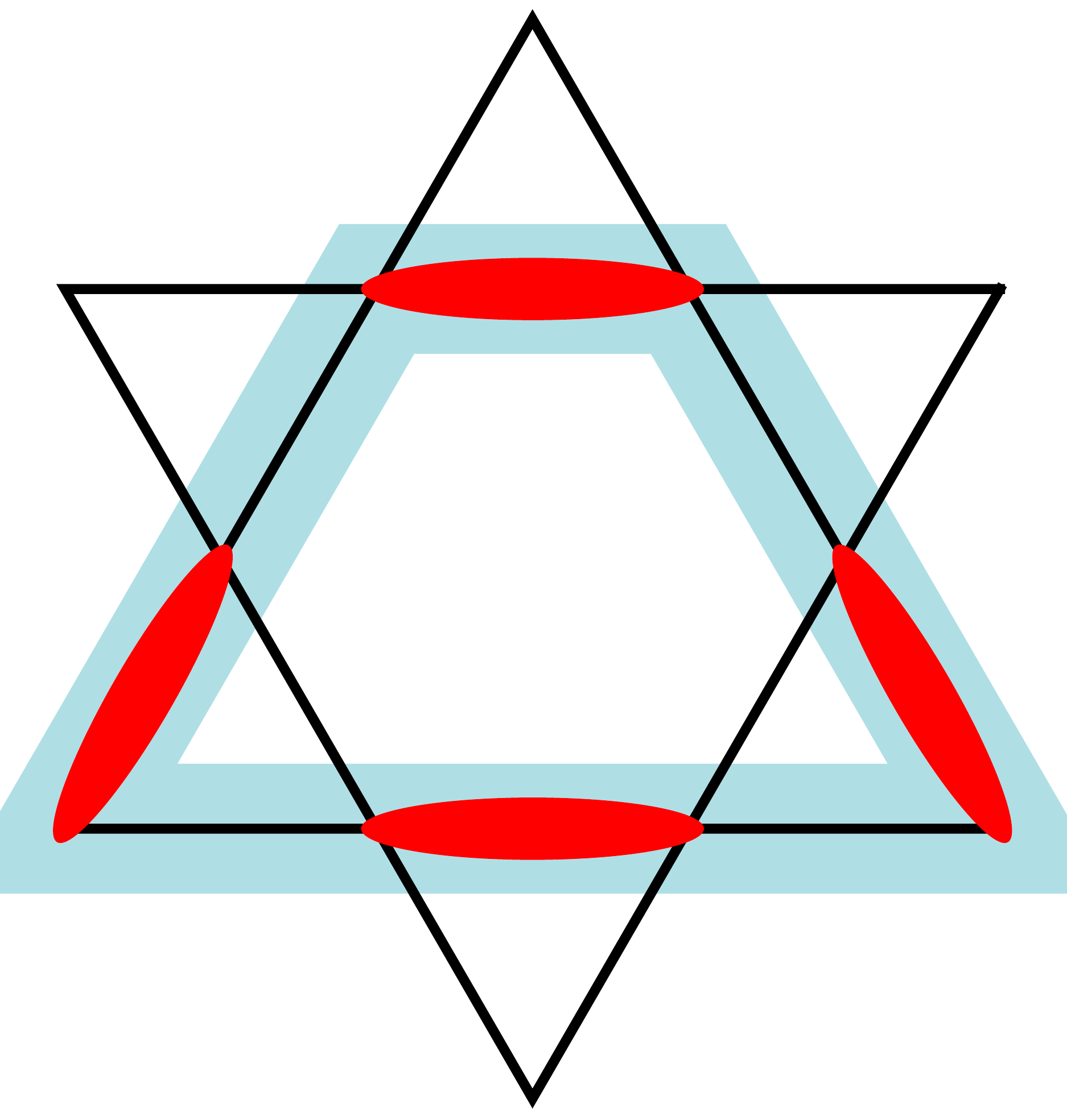}} 
& \parbox{5em}{\includegraphics[width=0.6\linewidth]{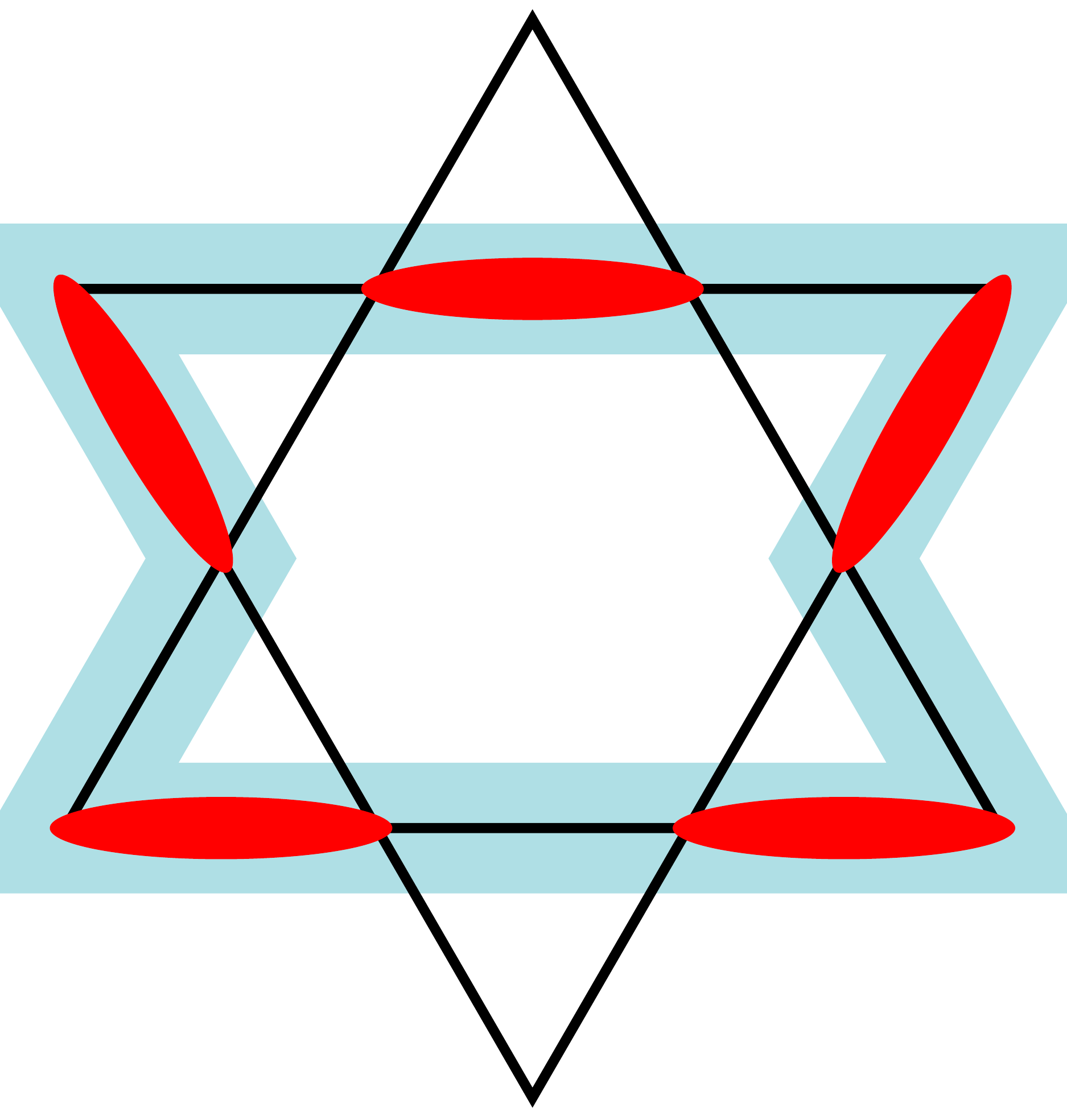}} 
& \parbox{5em}{\includegraphics[width=0.6\linewidth]{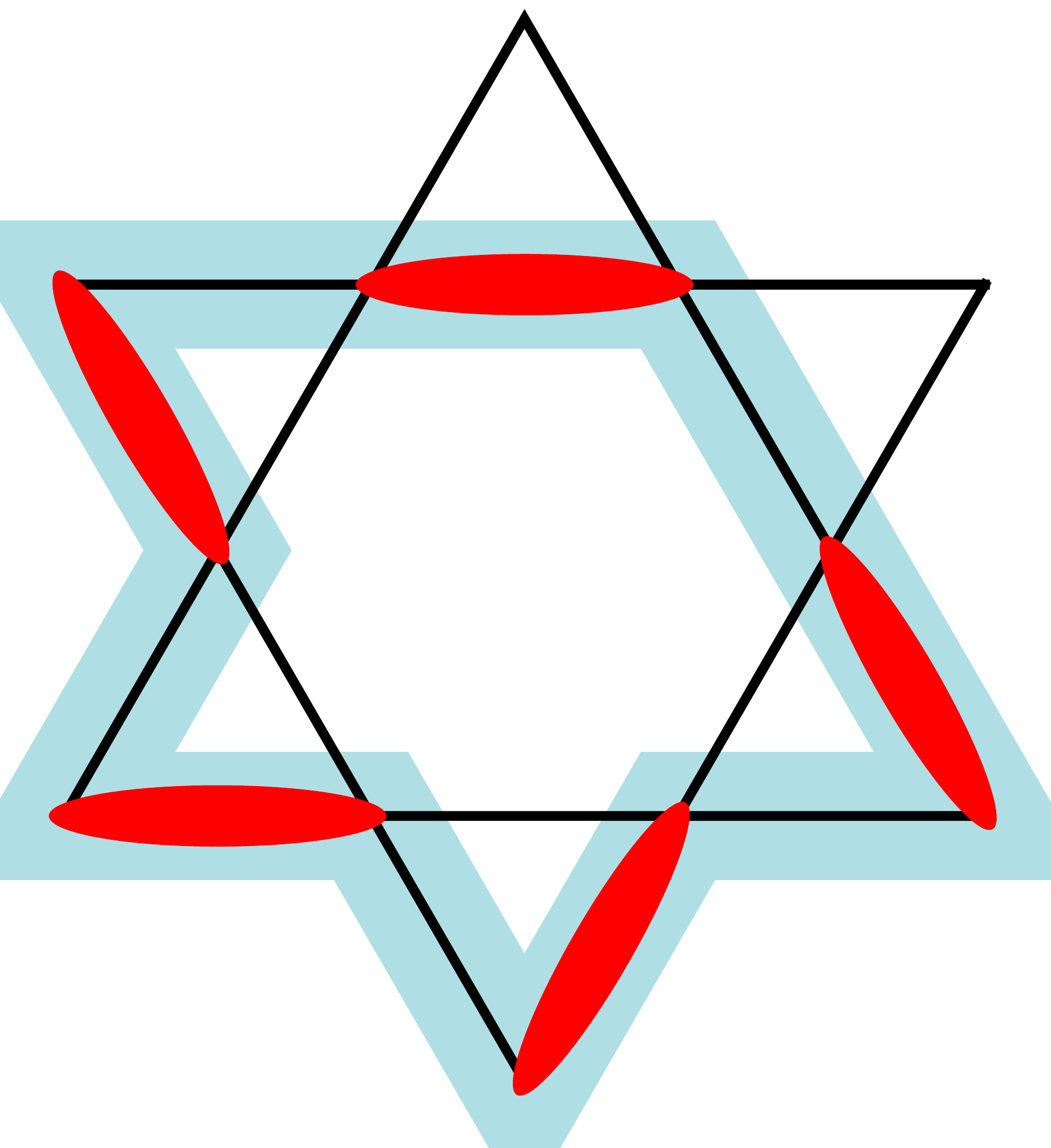}} 
& \parbox{5em}{\includegraphics[width=0.6\linewidth]{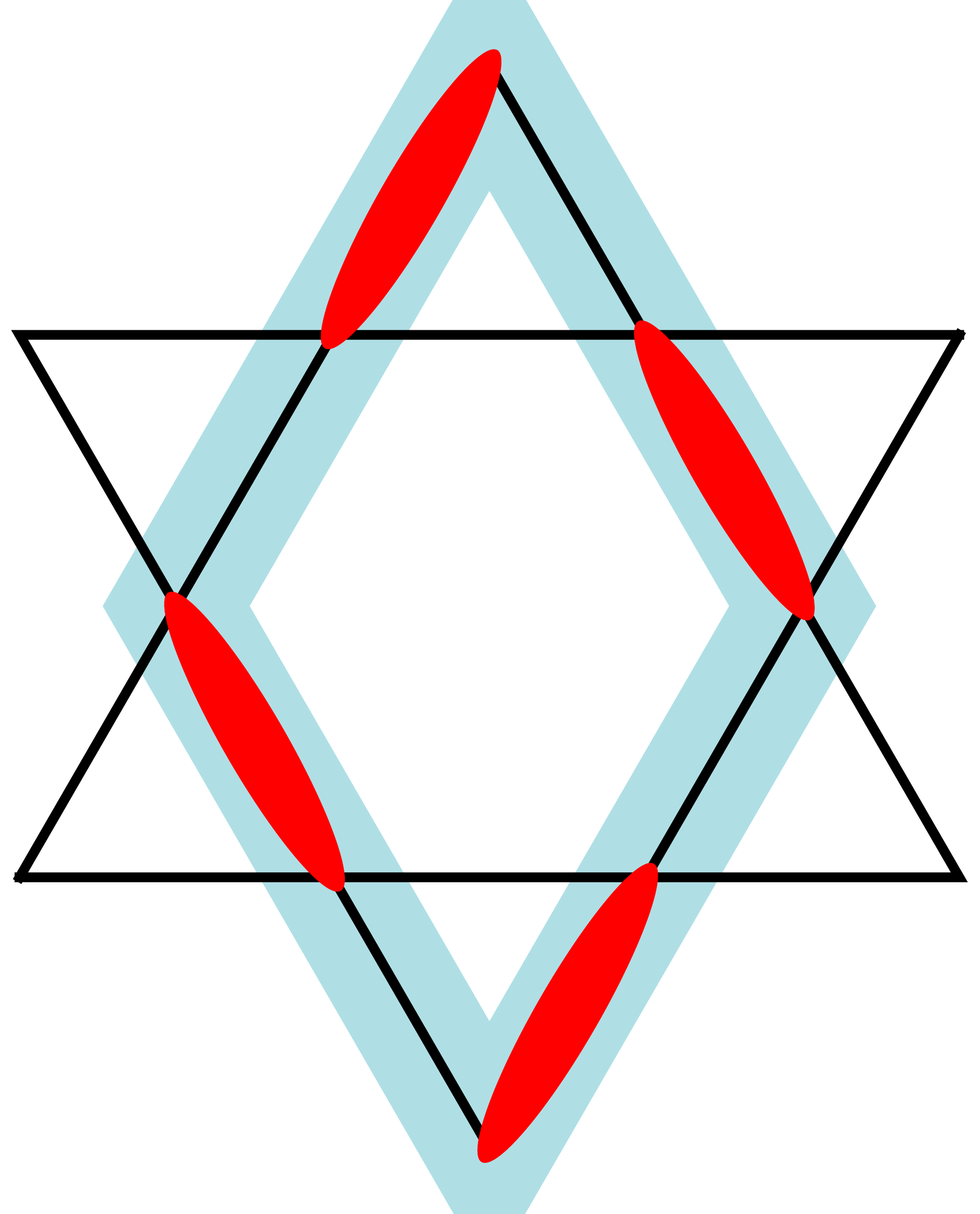}} 
& \parbox{5em}{\includegraphics[width=0.6\linewidth]{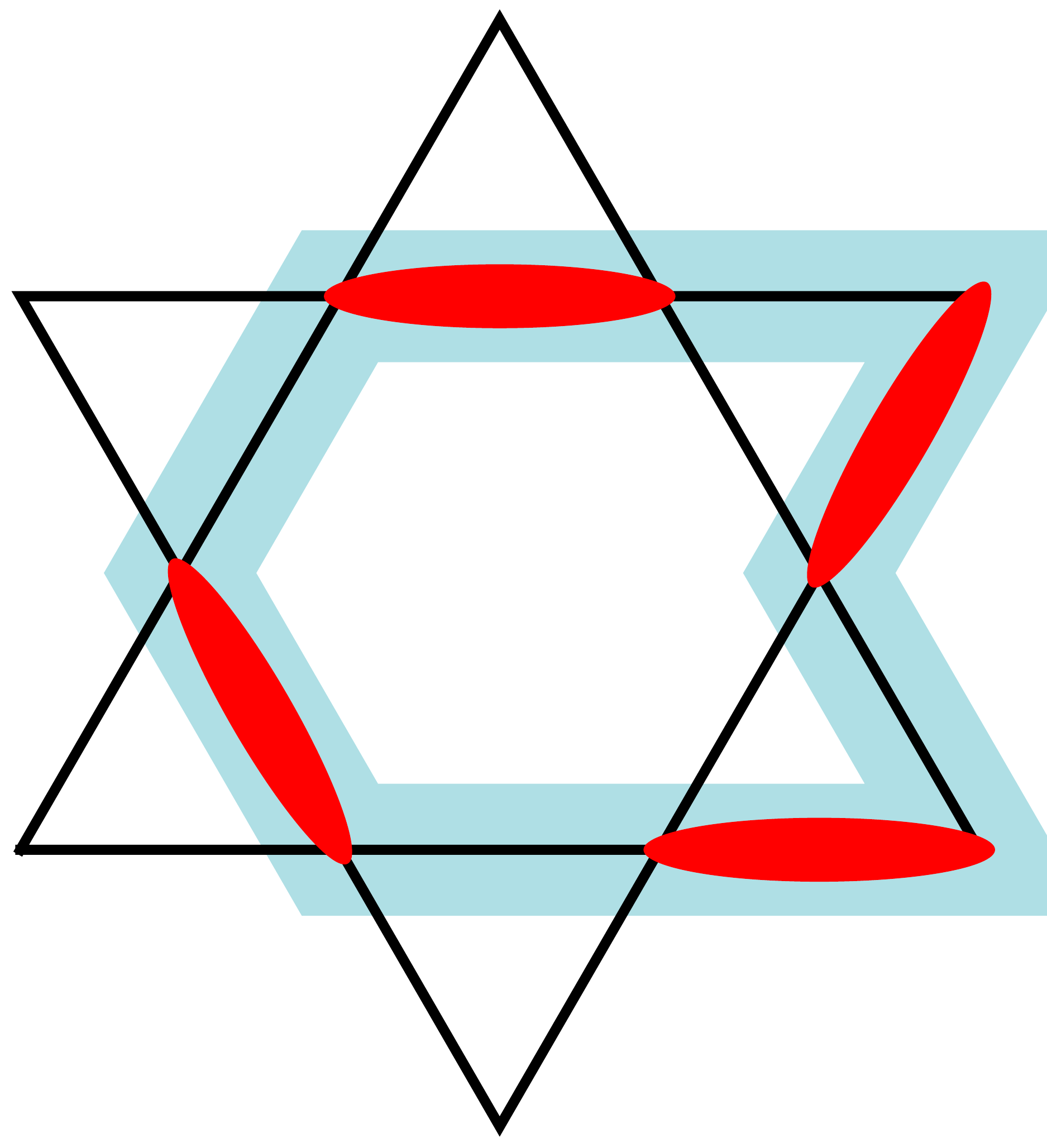}} 
& \parbox{5em}{\includegraphics[width=0.6\linewidth]{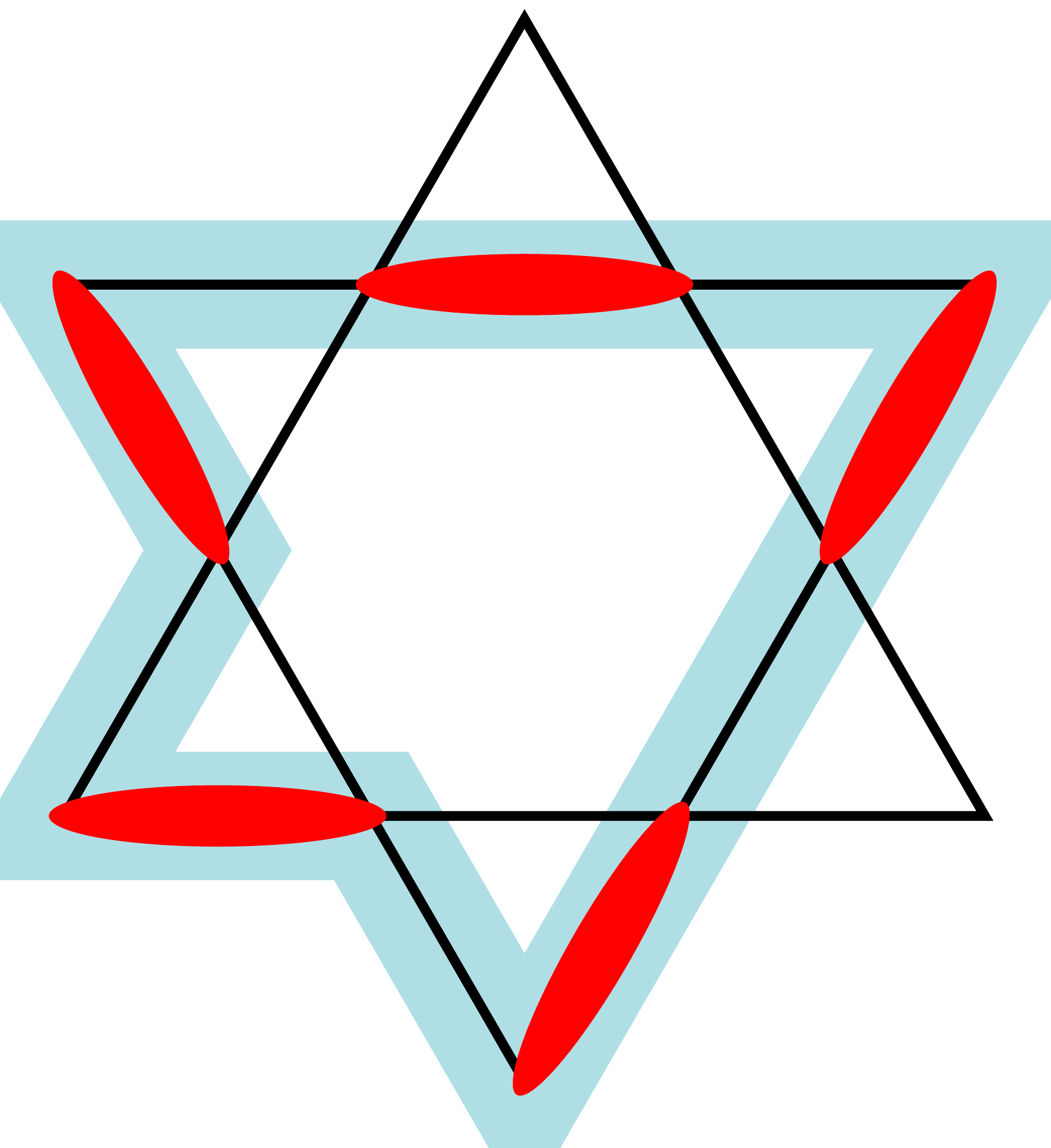}} 
& \parbox{5em}{\includegraphics[width=0.6\linewidth]{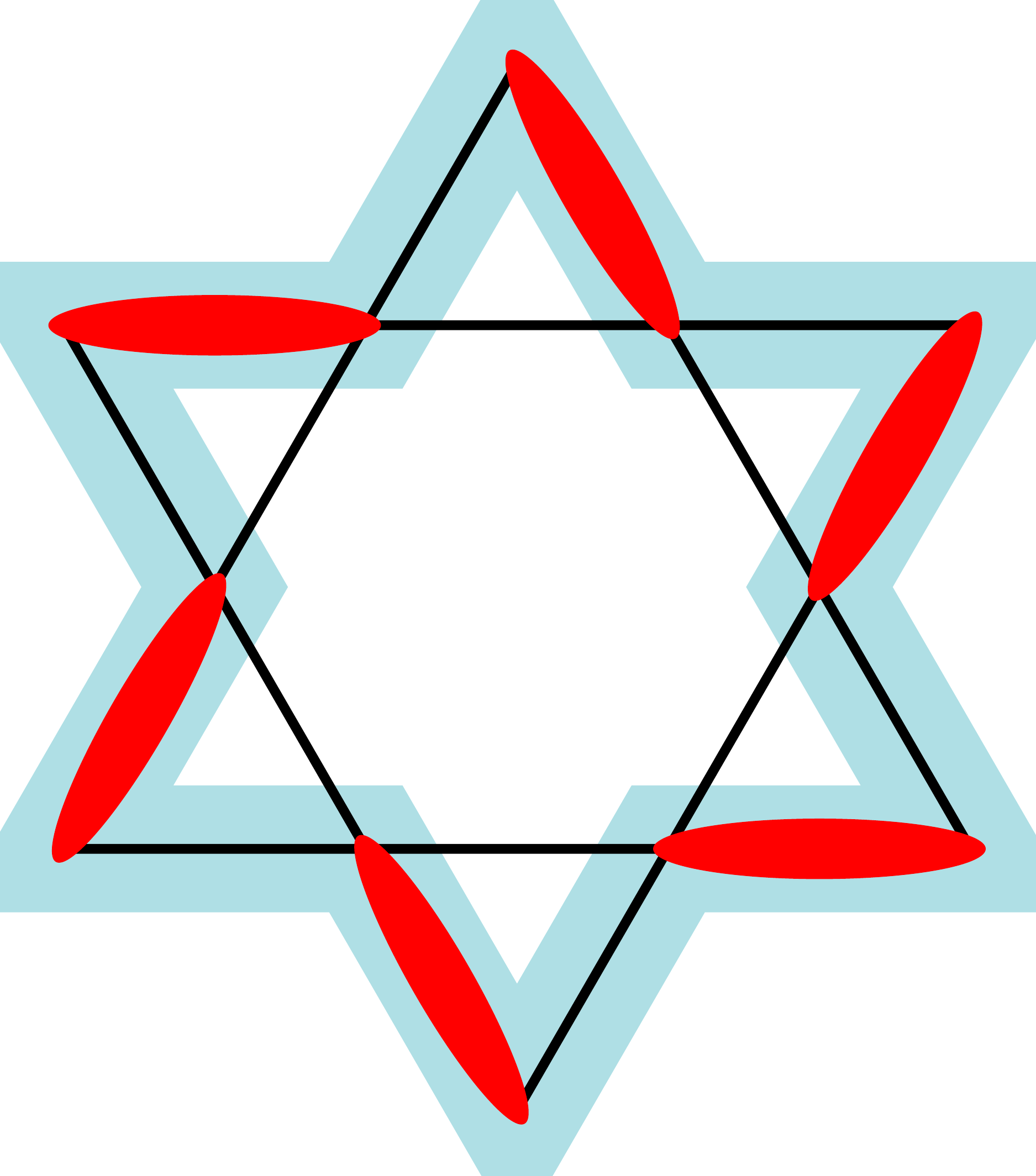}} 
\\
\hline
$f(\mathpzc{D})$ & 1 & 1 & 1 & 1 & -1 & -1 & -1 & -1
\\
\hline
$E_\mathpzc{D}$ & 6 & -2 & 6 & -2 & -6 & 2 & 2 & -6
\end{tabular}
\end{ruledtabular}
\caption{Dimer motion graphs and dimer interaction energies.
Top: each graph depicts a dimer state (red) and the closed path (light blue) of the dimer motion by $\hat{W}_p$.
Middle: the sign factor $f(\mathpzc{D})$ in Eq.~(\ref{eq:dimer-resonance}).
Bottom: the energy coefficient $E_\mathpzc{D}$ of $\hat{V}_p$ [Eq.~(\ref{eq:dimer-int})].
Other cases related by symmetry are dropped for simplicity.}
\label{tab:III}
\end{table*}

\begin{figure}[tb]
\includegraphics[width=0.9\linewidth]{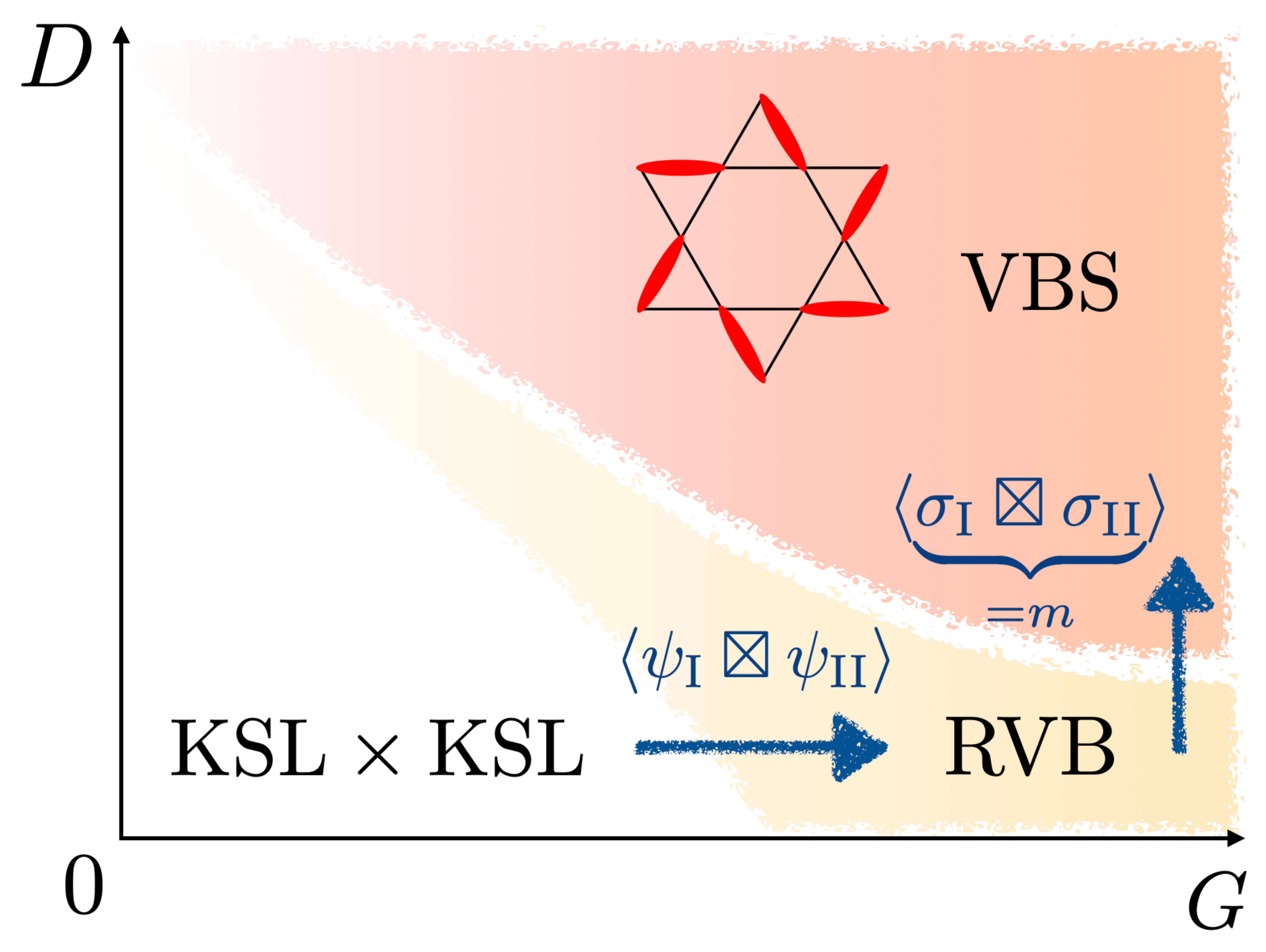}
\caption{A schematic phase diagram of the bilayer model [Eq.~(\ref{eq:model})] extended by the interactions in Eq.~(\ref{eq:NNN-phi-phi}).
Here we assume $K_\sigma=1~\&~K_\tau=-1$.
In large-$G$ limit, it is expected that the RVB-to-VBS transition occurs around $D\sim\lambda(\propto K_\sigma^6/G^5)$ and the resulting VBS state is the 12-site pinwheel VBS state that is well known from kagome antiferromagnetism~\cite{Norman2016,Broholm2020}.
}
\label{fig:12}
\end{figure}

Another anyon condensation transition can be generated in our bilayer system if we add the interactions, 
\begin{equation}
D \sum_{\langle ij \rangle_\alpha \langle jk \rangle_\beta} \phi_i^\gamma \phi_k^\gamma ,
\label{eq:NNN-phi-phi}
\end{equation}
where next-nearest-neighbor composite spins are coupled in a bond-dependent way.
If the composite spins are connected by $\alpha$- and $\beta$-bonds, then only the $\gamma$-components are coupled ($\alpha\ne\gamma\ne\beta$).
In the dimer Hilbert space (appearing in large-$G$ limit), such interactions become the dimer interactions,
\begin{equation}
D \sum_p \hat{V}_p, ~~~{\rm where}~~~
\hat{V}_p
=
\sum_{\mathpzc{D}} E_\mathpzc{D} | \mathpzc{D} \rangle \langle \mathpzc{D} |
\label{eq:dimer-int}
\end{equation}
and the values of the energy coefficient $E_\mathpzc{D}$ are listed in Table~\ref{tab:III}.
The emergence of the dimer interactions implies that there should be a transition from the RVB state to a valence bond solid (VBS) state, i.e., a crystalline order of dimers.
The RVB-to-VBS transition is driven by the condensation of $m$-particles (or visons) of the RVB state~\cite{Hwang2015,Wan2013}.
By the vison condensation transition, the topological spin liquid (RVB state) becomes trivial loosing all nontrivial anyons. Namely, $e$- and $\epsilon$-particles are confined due to their nontrival braiding with $m$-particles.
An anticipated phase diagram is schematically drawn in Fig.~\ref{fig:12}.

Our model can be generalized to other tri-coordinated lattices beyond the honeycomb lattice as in the original Kitaev model~\cite{Yang2007}. 
\begin{equation}
\parbox{8cm}{\includegraphics[width=\linewidth]{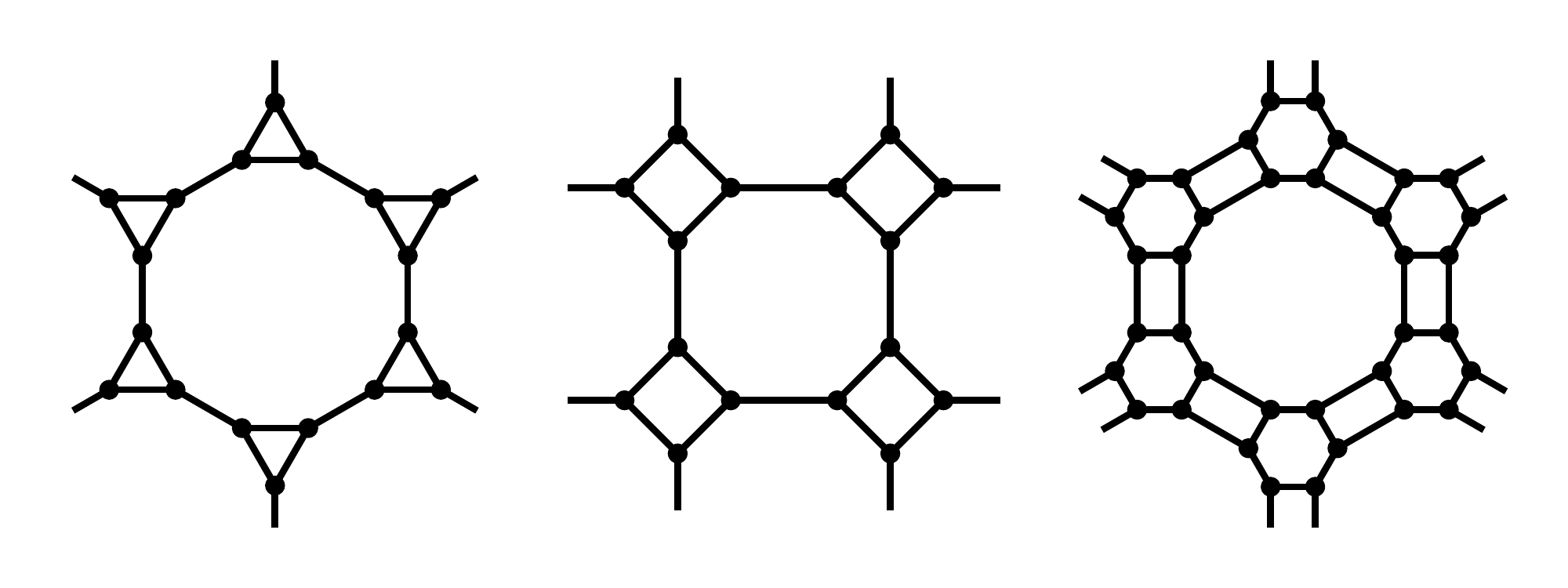}}
\nonumber
\end{equation}
In particular, the star lattice (left figure) would be an interesting case where the pure Kitaev model itself allows a chiral spin liquid state as shown by Yao and Kivelson~\cite{Yao2007}.
Such spontaneous time-reversal breaking may have some novel effects on the RVB side and the supported anyon types. Generalization of our work to other lattice geometries would be an interesting direction for future studies.

The bilayer model has an interesting connection to a non-Hermitian Kitaev system. Motivated by inevitable environment effects such as decoherence and dissipation on quantum many-body states prepared in quantum devices, there has been growing interest in open quantum systems recently~\cite{Katsura2019,Bergholtz2021,Altman2023,Lee2023,Bao2023mixedstate,Hwang2023mixedstate}.
Such environment effects can be investigated by using the so called Choi-Jamiołkowski isomorphism, which transforms the system's density matrix ($\hat{\rho}$) into a doubled state vector ($|\rho\rangle\!\rangle$) that is governed by a non-Hermitian Schr\"odinger equation:
\begin{eqnarray}
\hat{\rho}
=
\sum_{m,n} \rho_{mn} |m\rangle \langle n|
~~\Rightarrow~~
|\rho\rangle\!\rangle
=
\sum_{m,n} \rho_{mn} |m\rangle \otimes |n\rangle .
\end{eqnarray}
The problem of a Kitaev spin liquid coupled to an environment becomes identical to a bilayer Kitaev system featured with non-Hermitian inter-layer interactions, i.e., a non-Hermitian analog of our bilayer model~\cite{Hwang2023mixedstate}.
More interestingly, anyon condensation phenomena dynamically occur in the non-Hermitian Kitaev system by the non-Hermitian inter-layer interactions~\cite{Hwang2023mixedstate}.
Quantum spin liquids coupled to environments (open quantum spin liquids) are an interesting setup which allows novel anyon physics via non-unitary dynamics.

Lastly, we comment on experimental realizations of our system.
There have been several concrete proposals upon realizing the Kitaev's honeycomb model and toric code model by using ultracold atoms or Rydberg atoms together with Floquet engineering~\cite{Duan2003,Pan2017,Bukov2023,Lukin2023,Liu2023}.
Moreover, non-abelian topological orders and anyons have been successfully simulated in superconducting-qubit and trapped-ion quantum processors recently~\cite{Google2023,Deng2023,Iqbal2023}.
We expect that such quantum simulators with high controllability, especially the reconfigurable atom arrays~\cite{Lukin2024}, could realize and further extend the rich anyon physics of our bilayer system in future.

\begin{figure}[b]
\includegraphics[width=\linewidth]{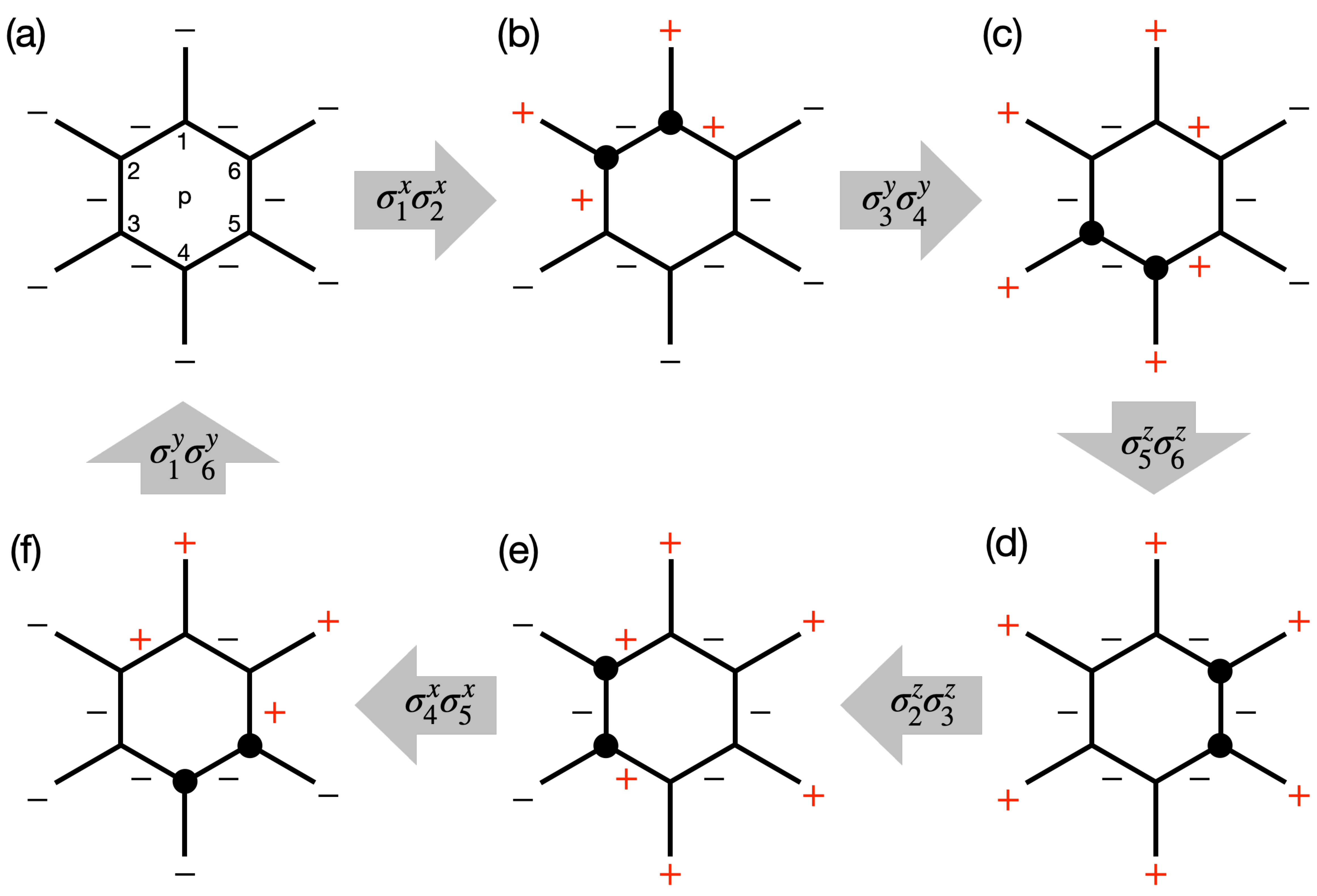}
\caption{A sixth order perturbation. The signs indicate the value of $\phi_j^\gamma\phi_k^\gamma(=\pm1)$ at each bond.
(a) An initial ground state $|\Phi'\rangle$.
(b) The intermediate state $| \nu \rangle = (K_{\sigma}\sigma_1^x\sigma_2^x+K_{\tau}\tau_1^x\tau_2^x) | \Phi' \rangle= 2K_{\sigma}\sigma_1^x\sigma_2^x | \Phi' \rangle$.
(c) The next intermediate state $| \mu \rangle = (K_{\sigma}\sigma_3^y\sigma_4^y+K_{\tau}\tau_3^y\tau_4^y) | \nu \rangle= 2K_{\sigma}\sigma_3^y\sigma_4^y | \nu \rangle$.
Similarly for other intermediate states $| \rho \rangle,|\beta\rangle,|\alpha\rangle$ in (d),(e),(f).
After all the Kitaev interactions around the plaquette $p$, the state comes back to some other ground state $|\Phi\rangle$.
}
\label{fig:13}
\end{figure}

\acknowledgements

I am grateful to Eun-Gook Moon, Hae-Young Kee, Nandini Trivedi, Oleg Tchernyshyov, Onur Erten, Suk Bum Chung, Jong Yeon Lee, Jung Hoon Han, and Gil Young Cho for helpful discussions.
Computations were performed on clusters at the Center for Advanced Computation (CAC) of Korea Institute for Advanced Study (KIAS).
This work was supported by KIAS Individual Grants (No.~PG071402~\&~PG071403).

\appendix

\begin{widetext}

\section{Perturbation theory for the strong coupling limit\label{app:A}}

To investigate the strong coupling limit, we separate the Hamiltonian into the two parts, $H=H_0+H'$, where $H_0=\sum_{\langle jk \rangle_{\gamma}} G \phi_j^\gamma\phi_k^\gamma$, and $H'=\sum_{\langle jk \rangle_{\gamma}} K_{\sigma}\sigma_j^\gamma\sigma_k^\gamma+K_{\tau}\tau_j^\gamma\tau_k^\gamma$.
Remember that we focus on the parameter region of $G>0~\&~K_{\tau}=- K_{\sigma}$.
The ground state manifold of $H_0$ is extensively degenerate and characterized by the constraint, $\phi_j^\gamma\phi_k^\gamma=-1$, at each bond.
Note that the eigenstates of the composite spin operators $\{\phi_j^\gamma\}$ are automatically the energy eigenstates of $H_0$.
In this basis, we find that the action of each Kitaev term of $H'$ is determined by the value of $\phi_j^\gamma\phi_k^\gamma$:
\begin{eqnarray}
( K_{\sigma}\sigma_j^\gamma\sigma_k^\gamma+K_{\tau}\tau_j^\gamma\tau_k^\gamma ) | \Phi \rangle
=
\left\{
\begin{array}{cc}
2K_{\sigma}\sigma_j^\gamma\sigma_k^\gamma | \Phi \rangle & ({\rm when}~~\phi_j^\gamma\phi_k^\gamma=-1)
\\
0 & ({\rm when}~~\phi_j^\gamma\phi_k^\gamma=+1)
\end{array}
\right\} ,
\label{eq:Kitaev-action}
\end{eqnarray}
where $| \Phi \rangle$ is an arbitrary state of $\{\phi_j^\gamma\}$.

The effective Hamiltonian $\mathcal{H}_{\rm eff}$ for the ground state manifold is constructed by a degenerate perturbation theory~\cite{Winkler2003}.
Nontrivial interaction terms (other than constant energy shifts) arise at the sixth order perturbations:
\begin{eqnarray}
\langle \Phi |\mathcal{H}_{\rm eff} | \Phi' \rangle 
&=& 
\frac{1}{2}
\sum_{\alpha,\beta,\rho,\mu,\nu} 
\frac{
\langle \Phi | H' | \alpha \rangle \langle \alpha | H' | \beta \rangle \langle \beta | H' | \rho \rangle \langle \rho | H' | \mu \rangle  \langle \mu | H' | \nu \rangle  \langle \nu | H' | \Phi' \rangle
}
{
(E_{\Phi}^{(0)}-E_{\alpha}^{(0)})(E_{\Phi}^{(0)}-E_{\beta}^{(0)})(E_{\Phi}^{(0)}-E_{\rho}^{(0)})(E_{\Phi}^{(0)}-E_{\mu}^{(0)})(E_{\Phi}^{(0)}-E_{\nu}^{(0)})
}
\nonumber\\
&+& 
\frac{1}{2}
\sum_{\alpha,\beta,\rho,\mu,\nu} 
\frac{
\langle \Phi | H' | \alpha \rangle \langle \alpha | H' | \beta \rangle \langle \beta | H' | \rho \rangle \langle \rho | H' | \mu \rangle  \langle \mu | H' | \nu \rangle  \langle \nu | H' | \Phi' \rangle
}
{
(E_{\Phi'}^{(0)}-E_{\alpha}^{(0)})(E_{\Phi'}^{(0)}-E_{\beta}^{(0)})(E_{\Phi'}^{(0)}-E_{\rho}^{(0)})(E_{\Phi'}^{(0)}-E_{\mu}^{(0)})(E_{\Phi'}^{(0)}-E_{\nu}^{(0)}),
}
\label{eq:Heff}
\end{eqnarray}
where $|\Phi \rangle$ and $| \Phi' \rangle$ are states in the ground state manifold, $H_0 | \Phi \rangle = E_{\Phi}^{(0)} | \Phi \rangle$, $H_0 | \Phi' \rangle = E_{\Phi'}^{(0)} | \Phi' \rangle$, and similarly for the intermediate states, $|\alpha\rangle$, $|\beta\rangle$, $|\rho\rangle$, $|\mu\rangle$, $|\nu\rangle$.
Figure~\ref{fig:13} illustrates one example (among various possible sixth order processes): we act the Kitaev terms, $K_{\sigma}\sigma_j^\gamma\sigma_k^\gamma+K_{\tau}\tau_j^\gamma\tau_k^\gamma$, along a plaquette $p$ using Eq.~(\ref{eq:Kitaev-action}). Its contribution to $\mathcal{H}_{\rm eff}$ is given by the expression,
\begin{equation}
\frac{
(2K_\sigma)^6 \langle \Phi |  \sigma_1^y \sigma_6^y | \alpha \rangle \langle \alpha | \sigma_4^x \sigma_5^x | \beta \rangle \langle \beta | \sigma_2^z \sigma_3^z | \rho \rangle \langle \rho | \sigma_5^z \sigma_6^z | \mu \rangle  \langle \mu | \sigma_3^y \sigma_4^y | \nu \rangle  \langle \nu | \sigma_1^x \sigma_2^x | \Phi' \rangle
}
{
(-4G)(-6G)(-6G)(-6G)(-4G)
}
=
- \frac{1}{54}\frac{K_\sigma^6}{G^5} 
\langle \Phi |  \hat{W}_p | \Phi' \rangle .
\end{equation}
Collecting all the sixth order contributions, we obtain the effective Hamiltonian,
\begin{equation}
\mathcal{H}_{\rm eff} = - \lambda \sum_p \hat{W}_p,
\end{equation}
where $\lambda \propto K_{\sigma}^6/G^5$.

\section{Quantum dimer model\label{app:B}}

In the dimer representation, the effective Hamiltonian $\mathcal{H}_{\rm eff}$ has the interpretation of dimer resonance motions.
The dimer Hilbert space is provided by the ground states of $H_0$ respecting the hardcore dimer constraint.
Acting on the hardcore dimer states, a plaquette operator $\hat{W}_p$ generates dimer motions along closed paths around the plaquette $p$.
For example, pinwheel dimers are moved along the path of 12-site David star:
\begin{eqnarray}
 \parbox{1.3cm}{\includegraphics[width=\linewidth]{./pinwheel.pdf}} 
 &=&
  |t_x \rangle_1 \otimes |t_z \rangle_2 \otimes |t_y \rangle_3 \otimes |t_x \rangle_4 \otimes |t_z \rangle_5 \otimes |t_y \rangle_6
 \nonumber\\
 &\xrightarrow{\hat{W}_p}&
 \sigma_1^z |t_x \rangle_1 \otimes \sigma_2^y |t_z \rangle_2 \otimes \sigma_3^x |t_y \rangle_3 \otimes \sigma_4^z |t_x \rangle_4 \otimes \sigma_5^y |t_z \rangle_5 \otimes \sigma_6^x |t_y \rangle_6
 \nonumber\\
&=&
 i |t_y \rangle_1 \otimes i |t_x \rangle_2 \otimes i |t_z \rangle_3 \otimes i |t_y \rangle_4 \otimes i |t_x \rangle_5 \otimes i |t_z \rangle_6
 =
 (-1)\parbox{1.3cm}{\includegraphics[width=\linewidth]{./pinwheel2.pdf}} .
\end{eqnarray}
Notice that the resulting dimer state is accompanied with an extra minus sign.
On the other hand, hexagon dimers are shifted along the path of 6-site hexagon:
\begin{eqnarray}
 \parbox{1.3cm}{\includegraphics[width=\linewidth]{./hexagon.pdf}} 
 &=&
  |t_z \rangle_1 \otimes |s \rangle_2 \otimes |t_x \rangle_3 \otimes |s \rangle_4 \otimes |t_y \rangle_5 \otimes |s \rangle_6
 \nonumber\\
 &\xrightarrow{\hat{W}_p}&
 \sigma_1^z |t_z \rangle_1 \otimes \sigma_2^y |s \rangle_2 \otimes \sigma_3^x |t_x \rangle_3 \otimes \sigma_4^z |s \rangle_4 \otimes \sigma_5^y |t_y \rangle_5 \otimes \sigma_6^x |s \rangle_6
 \nonumber\\
&=&
 |s \rangle_1 \otimes |t_y \rangle_2 \otimes |s \rangle_3 \otimes |t_z \rangle_4 \otimes |s \rangle_5 \otimes |t_x \rangle_6
 =
 \parbox{1.3cm}{\includegraphics[width=\linewidth]{./hexagon2.pdf}} .
\end{eqnarray}
In this case, there is no sign change.
Repeating the same calculations for other dimer configurations, we obtain the dimer model,
\begin{equation}
\hat{W}_p = \sum_{\mathpzc{D}} f(\mathpzc{D}) | \mathpzc{D} \rangle \langle \bar{\mathpzc{D}} | + {\rm H.c.},
\end{equation}
where $\mathpzc{D}$ runs over 32 distinct dimer configurations around the local plaquette $p$, and $|\bar{\mathpzc{D}}\rangle$ means the conjugate dimer configuration of $|\mathpzc{D}\rangle$ connected by $\hat{W}_p$.
The full list of the dimer motions and the associated sign factor $f(\mathpzc{D})$ are described in Table~\ref{tab:III}.

\end{widetext}


\begin{thebibliography}{123}%
\makeatletter
\providecommand \@ifxundefined [1]{%
 \@ifx{#1\undefined}
}%
\providecommand \@ifnum [1]{%
 \ifnum #1\expandafter \@firstoftwo
 \else \expandafter \@secondoftwo
 \fi
}%
\providecommand \@ifx [1]{%
 \ifx #1\expandafter \@firstoftwo
 \else \expandafter \@secondoftwo
 \fi
}%
\providecommand \natexlab [1]{#1}%
\providecommand \enquote  [1]{``#1''}%
\providecommand \bibnamefont  [1]{#1}%
\providecommand \bibfnamefont [1]{#1}%
\providecommand \citenamefont [1]{#1}%
\providecommand \href@noop [0]{\@secondoftwo}%
\providecommand \href [0]{\begingroup \@sanitize@url \@href}%
\providecommand \@href[1]{\@@startlink{#1}\@@href}%
\providecommand \@@href[1]{\endgroup#1\@@endlink}%
\providecommand \@sanitize@url [0]{\catcode `\\12\catcode `\$12\catcode
  `\&12\catcode `\#12\catcode `\^12\catcode `\_12\catcode `\%12\relax}%
\providecommand \@@startlink[1]{}%
\providecommand \@@endlink[0]{}%
\providecommand \url  [0]{\begingroup\@sanitize@url \@url }%
\providecommand \@url [1]{\endgroup\@href {#1}{\urlprefix }}%
\providecommand \urlprefix  [0]{URL }%
\providecommand \Eprint [0]{\href }%
\providecommand \doibase [0]{https://doi.org/}%
\providecommand \selectlanguage [0]{\@gobble}%
\providecommand \bibinfo  [0]{\@secondoftwo}%
\providecommand \bibfield  [0]{\@secondoftwo}%
\providecommand \translation [1]{[#1]}%
\providecommand \BibitemOpen [0]{}%
\providecommand \bibitemStop [0]{}%
\providecommand \bibitemNoStop [0]{.\EOS\space}%
\providecommand \EOS [0]{\spacefactor3000\relax}%
\providecommand \BibitemShut  [1]{\csname bibitem#1\endcsname}%
\let\auto@bib@innerbib\@empty
\bibitem [{\citenamefont {Wilczek}(1982)}]{Wilczek1982}%
  \BibitemOpen
  \bibfield  {author} {\bibinfo {author} {\bibfnamefont {F.}~\bibnamefont
  {Wilczek}},\ }\bibfield  {title} {\bibinfo {title} {{Quantum Mechanics of
  Fractional-Spin Particles}},\ }\href
  {https://doi.org/10.1103/PhysRevLett.49.957} {\bibfield  {journal} {\bibinfo
  {journal} {Phys. Rev. Lett.}\ }\textbf {\bibinfo {volume} {49}},\ \bibinfo
  {pages} {957} (\bibinfo {year} {1982})}\BibitemShut {NoStop}%
\bibitem [{\citenamefont {Wen}(2004)}]{WenBook}%
  \BibitemOpen
  \bibfield  {author} {\bibinfo {author} {\bibfnamefont {X.~G.}\ \bibnamefont
  {Wen}},\ }\href@noop {} {\emph {\bibinfo {title} {{Quantum Field Theory of
  Many-Body Systems}}}}\ (\bibinfo  {publisher} {Oxford University Press},\
  \bibinfo {year} {2004})\BibitemShut {NoStop}%
\bibitem [{\citenamefont {Fradkin}(2013)}]{FradkinBook}%
  \BibitemOpen
  \bibfield  {author} {\bibinfo {author} {\bibfnamefont {E.}~\bibnamefont
  {Fradkin}},\ }\href@noop {} {\emph {\bibinfo {title} {{Field Theories of
  Condensed Matter Physics}}}}\ (\bibinfo  {publisher} {Cambridge University
  Press},\ \bibinfo {year} {2013})\BibitemShut {NoStop}%
\bibitem [{\citenamefont {Sachdev}(2023)}]{SachdevBook}%
  \BibitemOpen
  \bibfield  {author} {\bibinfo {author} {\bibfnamefont {S.}~\bibnamefont
  {Sachdev}},\ }\href@noop {} {\emph {\bibinfo {title} {{Quantum Phases of
  Matter}}}}\ (\bibinfo  {publisher} {Cambridge University Press},\ \bibinfo
  {year} {2023})\BibitemShut {NoStop}%
\bibitem [{\citenamefont {Kitaev}(2003)}]{Kitaev2003}%
  \BibitemOpen
  \bibfield  {author} {\bibinfo {author} {\bibfnamefont {A.}~\bibnamefont
  {Kitaev}},\ }\bibfield  {title} {\bibinfo {title} {{Fault-tolerant quantum
  computation by anyons}},\ }\href
  {https://doi.org/https://doi.org/10.1016/S0003-4916(02)00018-0} {\bibfield
  {journal} {\bibinfo  {journal} {Ann. Phys.}\ }\textbf {\bibinfo {volume}
  {303}},\ \bibinfo {pages} {2} (\bibinfo {year} {2003})}\BibitemShut {NoStop}%
\bibitem [{\citenamefont {Kitaev}(2006)}]{Kitaev2006}%
  \BibitemOpen
  \bibfield  {author} {\bibinfo {author} {\bibfnamefont {A.}~\bibnamefont
  {Kitaev}},\ }\bibfield  {title} {\bibinfo {title} {{Anyons in an exactly
  solved model and beyond}},\ }\href
  {https://doi.org/https://doi.org/10.1016/j.aop.2005.10.005} {\bibfield
  {journal} {\bibinfo  {journal} {Ann. Phys.}\ }\textbf {\bibinfo {volume}
  {321}},\ \bibinfo {pages} {2} (\bibinfo {year} {2006})}\BibitemShut {NoStop}%
\bibitem [{\citenamefont {Levin}\ and\ \citenamefont
  {Wen}(2005)}]{LevinWen2005}%
  \BibitemOpen
  \bibfield  {author} {\bibinfo {author} {\bibfnamefont {M.~A.}\ \bibnamefont
  {Levin}}\ and\ \bibinfo {author} {\bibfnamefont {X.-G.}\ \bibnamefont
  {Wen}},\ }\bibfield  {title} {\bibinfo {title} {{String-net condensation: A
  physical mechanism for topological phases}},\ }\href
  {https://doi.org/10.1103/PhysRevB.71.045110} {\bibfield  {journal} {\bibinfo
  {journal} {Phys. Rev. B}\ }\textbf {\bibinfo {volume} {71}},\ \bibinfo
  {pages} {045110} (\bibinfo {year} {2005})}\BibitemShut {NoStop}%
\bibitem [{\citenamefont {Savary}\ and\ \citenamefont
  {Balents}(2016)}]{Savary2016}%
  \BibitemOpen
  \bibfield  {author} {\bibinfo {author} {\bibfnamefont {L.}~\bibnamefont
  {Savary}}\ and\ \bibinfo {author} {\bibfnamefont {L.}~\bibnamefont
  {Balents}},\ }\bibfield  {title} {\bibinfo {title} {{Quantum spin liquids: a
  review}},\ }\href {https://doi.org/10.1088/0034-4885/80/1/016502} {\bibfield
  {journal} {\bibinfo  {journal} {Rep. Prog. Phys.}\ }\textbf {\bibinfo
  {volume} {80}},\ \bibinfo {pages} {016502} (\bibinfo {year}
  {2016})}\BibitemShut {NoStop}%
\bibitem [{\citenamefont {Norman}(2016)}]{Norman2016}%
  \BibitemOpen
  \bibfield  {author} {\bibinfo {author} {\bibfnamefont {M.~R.}\ \bibnamefont
  {Norman}},\ }\bibfield  {title} {\bibinfo {title} {{Colloquium:
  Herbertsmithite and the search for the quantum spin liquid}},\ }\href
  {https://doi.org/10.1103/RevModPhys.88.041002} {\bibfield  {journal}
  {\bibinfo  {journal} {Rev. Mod. Phys.}\ }\textbf {\bibinfo {volume} {88}},\
  \bibinfo {pages} {041002} (\bibinfo {year} {2016})}\BibitemShut {NoStop}%
\bibitem [{\citenamefont {Zhou}\ \emph {et~al.}(2017)\citenamefont {Zhou},
  \citenamefont {Kanoda},\ and\ \citenamefont {Ng}}]{Ng2017}%
  \BibitemOpen
  \bibfield  {author} {\bibinfo {author} {\bibfnamefont {Y.}~\bibnamefont
  {Zhou}}, \bibinfo {author} {\bibfnamefont {K.}~\bibnamefont {Kanoda}},\ and\
  \bibinfo {author} {\bibfnamefont {T.-K.}\ \bibnamefont {Ng}},\ }\bibfield
  {title} {\bibinfo {title} {{Quantum spin liquid states}},\ }\href
  {https://doi.org/10.1103/RevModPhys.89.025003} {\bibfield  {journal}
  {\bibinfo  {journal} {Rev. Mod. Phys.}\ }\textbf {\bibinfo {volume} {89}},\
  \bibinfo {pages} {025003} (\bibinfo {year} {2017})}\BibitemShut {NoStop}%
\bibitem [{\citenamefont {Knolle}\ and\ \citenamefont
  {Moessner}(2019)}]{Knolle2019}%
  \BibitemOpen
  \bibfield  {author} {\bibinfo {author} {\bibfnamefont {J.}~\bibnamefont
  {Knolle}}\ and\ \bibinfo {author} {\bibfnamefont {R.}~\bibnamefont
  {Moessner}},\ }\bibfield  {title} {\bibinfo {title} {{A Field Guide to Spin
  Liquids}},\ }\href {https://doi.org/10.1146/annurev-conmatphys-031218-013401}
  {\bibfield  {journal} {\bibinfo  {journal} {Annu. Rev. Condens. Matter
  Phys.}\ }\textbf {\bibinfo {volume} {10}},\ \bibinfo {pages} {451} (\bibinfo
  {year} {2019})}\BibitemShut {NoStop}%
\bibitem [{\citenamefont {Broholm}\ \emph {et~al.}(2020)\citenamefont
  {Broholm}, \citenamefont {Cava}, \citenamefont {Kivelson}, \citenamefont
  {Nocera}, \citenamefont {Norman},\ and\ \citenamefont
  {Senthil}}]{Broholm2020}%
  \BibitemOpen
  \bibfield  {author} {\bibinfo {author} {\bibfnamefont {C.}~\bibnamefont
  {Broholm}}, \bibinfo {author} {\bibfnamefont {R.~J.}\ \bibnamefont {Cava}},
  \bibinfo {author} {\bibfnamefont {S.~A.}\ \bibnamefont {Kivelson}}, \bibinfo
  {author} {\bibfnamefont {D.~G.}\ \bibnamefont {Nocera}}, \bibinfo {author}
  {\bibfnamefont {M.~R.}\ \bibnamefont {Norman}},\ and\ \bibinfo {author}
  {\bibfnamefont {T.}~\bibnamefont {Senthil}},\ }\bibfield  {title} {\bibinfo
  {title} {{Quantum spin liquids}},\ }\href
  {https://www.science.org/doi/full/10.1126/science.aay0668} {\bibfield
  {journal} {\bibinfo  {journal} {Science}\ }\textbf {\bibinfo {volume}
  {367}},\ \bibinfo {pages} {eaay0668} (\bibinfo {year} {2020})}\BibitemShut
  {NoStop}%
\bibitem [{\citenamefont {Motome}\ and\ \citenamefont
  {Nasu}(2020)}]{Motome2020}%
  \BibitemOpen
  \bibfield  {author} {\bibinfo {author} {\bibfnamefont {Y.}~\bibnamefont
  {Motome}}\ and\ \bibinfo {author} {\bibfnamefont {J.}~\bibnamefont {Nasu}},\
  }\bibfield  {title} {\bibinfo {title} {{Hunting Majorana Fermions in Kitaev
  Magnets}},\ }\href {https://doi.org/10.7566/JPSJ.89.012002} {\bibfield
  {journal} {\bibinfo  {journal} {J. Phys. Soc. Jpn.}\ }\textbf {\bibinfo
  {volume} {89}},\ \bibinfo {pages} {012002} (\bibinfo {year}
  {2020})}\BibitemShut {NoStop}%
\bibitem [{\citenamefont {Trebst}\ and\ \citenamefont
  {Hickey}(2022)}]{Trebst2022}%
  \BibitemOpen
  \bibfield  {author} {\bibinfo {author} {\bibfnamefont {S.}~\bibnamefont
  {Trebst}}\ and\ \bibinfo {author} {\bibfnamefont {C.}~\bibnamefont
  {Hickey}},\ }\bibfield  {title} {\bibinfo {title} {Kitaev materials},\ }\href
  {https://doi.org/https://doi.org/10.1016/j.physrep.2021.11.003} {\bibfield
  {journal} {\bibinfo  {journal} {Phys. Rep.}\ }\textbf {\bibinfo {volume}
  {950}},\ \bibinfo {pages} {1} (\bibinfo {year} {2022})}\BibitemShut {NoStop}%
\bibitem [{\citenamefont {Takagi}\ \emph {et~al.}(2019)\citenamefont {Takagi},
  \citenamefont {Takayama}, \citenamefont {Jackeli}, \citenamefont
  {Khaliullin},\ and\ \citenamefont {Nagler}}]{Takagi2019}%
  \BibitemOpen
  \bibfield  {author} {\bibinfo {author} {\bibfnamefont {H.}~\bibnamefont
  {Takagi}}, \bibinfo {author} {\bibfnamefont {T.}~\bibnamefont {Takayama}},
  \bibinfo {author} {\bibfnamefont {G.}~\bibnamefont {Jackeli}}, \bibinfo
  {author} {\bibfnamefont {G.}~\bibnamefont {Khaliullin}},\ and\ \bibinfo
  {author} {\bibfnamefont {S.~E.}\ \bibnamefont {Nagler}},\ }\bibfield  {title}
  {\bibinfo {title} {{Concept and realization of Kitaev quantum spin
  liquids}},\ }\href {https://doi.org/10.1038/s42254-019-0038-2} {\bibfield
  {journal} {\bibinfo  {journal} {Nat. Rev. Phys.}\ }\textbf {\bibinfo {volume}
  {1}},\ \bibinfo {pages} {264} (\bibinfo {year} {2019})}\BibitemShut {NoStop}%
\bibitem [{\citenamefont {Kasahara}\ \emph {et~al.}(2018)\citenamefont
  {Kasahara}, \citenamefont {Ohnishi}, \citenamefont {Mizukami}, \citenamefont
  {Tanaka}, \citenamefont {Ma}, \citenamefont {Sugii}, \citenamefont {Kurita},
  \citenamefont {Tanaka}, \citenamefont {Nasu}, \citenamefont {Motome} \emph
  {et~al.}}]{Kasahara2018}%
  \BibitemOpen
  \bibfield  {author} {\bibinfo {author} {\bibfnamefont {Y.}~\bibnamefont
  {Kasahara}}, \bibinfo {author} {\bibfnamefont {T.}~\bibnamefont {Ohnishi}},
  \bibinfo {author} {\bibfnamefont {Y.}~\bibnamefont {Mizukami}}, \bibinfo
  {author} {\bibfnamefont {O.}~\bibnamefont {Tanaka}}, \bibinfo {author}
  {\bibfnamefont {S.}~\bibnamefont {Ma}}, \bibinfo {author} {\bibfnamefont
  {K.}~\bibnamefont {Sugii}}, \bibinfo {author} {\bibfnamefont
  {N.}~\bibnamefont {Kurita}}, \bibinfo {author} {\bibfnamefont
  {H.}~\bibnamefont {Tanaka}}, \bibinfo {author} {\bibfnamefont
  {J.}~\bibnamefont {Nasu}}, \bibinfo {author} {\bibfnamefont {Y.}~\bibnamefont
  {Motome}}, \emph {et~al.},\ }\bibfield  {title} {\bibinfo {title} {{Majorana
  quantization and half-integer thermal quantum Hall effect in a Kitaev spin
  liquid}},\ }\href {https://doi.org/10.1038/s41586-018-0274-0} {\bibfield
  {journal} {\bibinfo  {journal} {Nature}\ }\textbf {\bibinfo {volume} {559}},\
  \bibinfo {pages} {227} (\bibinfo {year} {2018})}\BibitemShut {NoStop}%
\bibitem [{\citenamefont {Satzinger}\ \emph {et~al.}(2021)\citenamefont
  {Satzinger}, \citenamefont {Liu}, \citenamefont {Smith}, \citenamefont
  {Knapp}, \citenamefont {Newman}, \citenamefont {Jones}, \citenamefont {Chen},
  \citenamefont {Quintana}, \citenamefont {Mi}, \citenamefont {Dunsworth} \emph
  {et~al.}}]{Satzinger2021}%
  \BibitemOpen
  \bibfield  {author} {\bibinfo {author} {\bibfnamefont {K.~J.}\ \bibnamefont
  {Satzinger}}, \bibinfo {author} {\bibfnamefont {Y.-J.}\ \bibnamefont {Liu}},
  \bibinfo {author} {\bibfnamefont {A.}~\bibnamefont {Smith}}, \bibinfo
  {author} {\bibfnamefont {C.}~\bibnamefont {Knapp}}, \bibinfo {author}
  {\bibfnamefont {M.}~\bibnamefont {Newman}}, \bibinfo {author} {\bibfnamefont
  {C.}~\bibnamefont {Jones}}, \bibinfo {author} {\bibfnamefont
  {Z.}~\bibnamefont {Chen}}, \bibinfo {author} {\bibfnamefont {C.}~\bibnamefont
  {Quintana}}, \bibinfo {author} {\bibfnamefont {X.}~\bibnamefont {Mi}},
  \bibinfo {author} {\bibfnamefont {A.}~\bibnamefont {Dunsworth}}, \emph
  {et~al.},\ }\bibfield  {title} {\bibinfo {title} {{Realizing topologically
  ordered states on a quantum processor}},\ }\href
  {https://www.science.org/doi/10.1126/science.abi8378} {\bibfield  {journal}
  {\bibinfo  {journal} {Science}\ }\textbf {\bibinfo {volume} {374}},\ \bibinfo
  {pages} {1237} (\bibinfo {year} {2021})}\BibitemShut {NoStop}%
\bibitem [{\citenamefont {Semeghini}\ \emph {et~al.}(2021)\citenamefont
  {Semeghini}, \citenamefont {Levine}, \citenamefont {Keesling}, \citenamefont
  {Ebadi}, \citenamefont {Wang}, \citenamefont {Bluvstein}, \citenamefont
  {Verresen}, \citenamefont {Pichler}, \citenamefont {Kalinowski},
  \citenamefont {Samajdar} \emph {et~al.}}]{Semeghini2021}%
  \BibitemOpen
  \bibfield  {author} {\bibinfo {author} {\bibfnamefont {G.}~\bibnamefont
  {Semeghini}}, \bibinfo {author} {\bibfnamefont {H.}~\bibnamefont {Levine}},
  \bibinfo {author} {\bibfnamefont {A.}~\bibnamefont {Keesling}}, \bibinfo
  {author} {\bibfnamefont {S.}~\bibnamefont {Ebadi}}, \bibinfo {author}
  {\bibfnamefont {T.~T.}\ \bibnamefont {Wang}}, \bibinfo {author}
  {\bibfnamefont {D.}~\bibnamefont {Bluvstein}}, \bibinfo {author}
  {\bibfnamefont {R.}~\bibnamefont {Verresen}}, \bibinfo {author}
  {\bibfnamefont {H.}~\bibnamefont {Pichler}}, \bibinfo {author} {\bibfnamefont
  {M.}~\bibnamefont {Kalinowski}}, \bibinfo {author} {\bibfnamefont
  {R.}~\bibnamefont {Samajdar}}, \emph {et~al.},\ }\bibfield  {title} {\bibinfo
  {title} {{Probing topological spin liquids on a programmable quantum
  simulator}},\ }\href
  {https://www.science.org/doi/full/10.1126/science.abi8794} {\bibfield
  {journal} {\bibinfo  {journal} {Science}\ }\textbf {\bibinfo {volume}
  {374}},\ \bibinfo {pages} {1242} (\bibinfo {year} {2021})}\BibitemShut
  {NoStop}%
\bibitem [{\citenamefont
  {{Google~Quantum~AI~and~Collaborators}}(2023)}]{Google2023}%
  \BibitemOpen
  \bibfield  {author} {\bibinfo {author} {\bibnamefont
  {{Google~Quantum~AI~and~Collaborators}}},\ }\bibfield  {title} {\bibinfo
  {title} {Non-abelian braiding of graph vertices in a superconducting
  processor},\ }\href {https://doi.org/10.1038/s41586-023-05954-4} {\bibfield
  {journal} {\bibinfo  {journal} {Nature}\ }\textbf {\bibinfo {volume} {618}},\
  \bibinfo {pages} {264} (\bibinfo {year} {2023})}\BibitemShut {NoStop}%
\bibitem [{\citenamefont {Iqbal}\ \emph {et~al.}(2024)\citenamefont {Iqbal},
  \citenamefont {Tantivasadakarn}, \citenamefont {Verresen}, \citenamefont
  {Campbell}, \citenamefont {Dreiling}, \citenamefont {Figgatt}, \citenamefont
  {Gaebler}, \citenamefont {Johansen}, \citenamefont {Mills}, \citenamefont
  {Moses}, \citenamefont {Pino}, \citenamefont {Ransford}, \citenamefont
  {Rowe}, \citenamefont {Siegfried}, \citenamefont {Stutz}, \citenamefont
  {Foss-Feig}, \citenamefont {Vishwanath},\ and\ \citenamefont
  {Dreyer}}]{Iqbal2023}%
  \BibitemOpen
  \bibfield  {author} {\bibinfo {author} {\bibfnamefont {M.}~\bibnamefont
  {Iqbal}}, \bibinfo {author} {\bibfnamefont {N.}~\bibnamefont
  {Tantivasadakarn}}, \bibinfo {author} {\bibfnamefont {R.}~\bibnamefont
  {Verresen}}, \bibinfo {author} {\bibfnamefont {S.~L.}\ \bibnamefont
  {Campbell}}, \bibinfo {author} {\bibfnamefont {J.~M.}\ \bibnamefont
  {Dreiling}}, \bibinfo {author} {\bibfnamefont {C.}~\bibnamefont {Figgatt}},
  \bibinfo {author} {\bibfnamefont {J.~P.}\ \bibnamefont {Gaebler}}, \bibinfo
  {author} {\bibfnamefont {J.}~\bibnamefont {Johansen}}, \bibinfo {author}
  {\bibfnamefont {M.}~\bibnamefont {Mills}}, \bibinfo {author} {\bibfnamefont
  {S.~A.}\ \bibnamefont {Moses}}, \bibinfo {author} {\bibfnamefont {J.~M.}\
  \bibnamefont {Pino}}, \bibinfo {author} {\bibfnamefont {A.}~\bibnamefont
  {Ransford}}, \bibinfo {author} {\bibfnamefont {M.}~\bibnamefont {Rowe}},
  \bibinfo {author} {\bibfnamefont {P.}~\bibnamefont {Siegfried}}, \bibinfo
  {author} {\bibfnamefont {R.~P.}\ \bibnamefont {Stutz}}, \bibinfo {author}
  {\bibfnamefont {M.}~\bibnamefont {Foss-Feig}}, \bibinfo {author}
  {\bibfnamefont {A.}~\bibnamefont {Vishwanath}},\ and\ \bibinfo {author}
  {\bibfnamefont {H.}~\bibnamefont {Dreyer}},\ }\bibfield  {title} {\bibinfo
  {title} {{Non-Abelian topological order and anyons on a trapped-ion
  processor}},\ }\href {https://doi.org/10.1038/s41586-023-06934-4} {\bibfield
  {journal} {\bibinfo  {journal} {Nature}\ }\textbf {\bibinfo {volume} {626}},\
  \bibinfo {pages} {505} (\bibinfo {year} {2024})}\BibitemShut {NoStop}%
\bibitem [{\citenamefont {Xu}\ \emph {et~al.}(2023)\citenamefont {Xu},
  \citenamefont {Sun}, \citenamefont {Wang}, \citenamefont {Xiang},
  \citenamefont {Bao}, \citenamefont {Zhu}, \citenamefont {Shen}, \citenamefont
  {Song}, \citenamefont {Zhang}, \citenamefont {Ren}, \citenamefont {Zhang},
  \citenamefont {Dong}, \citenamefont {Deng}, \citenamefont {Chen},
  \citenamefont {Wu}, \citenamefont {Tan}, \citenamefont {Gao}, \citenamefont
  {Jin}, \citenamefont {Zhu}, \citenamefont {Zhang}, \citenamefont {Wang},
  \citenamefont {Zou}, \citenamefont {Zhong}, \citenamefont {Zhang},
  \citenamefont {Li}, \citenamefont {Jiang}, \citenamefont {Yu}, \citenamefont
  {Yao}, \citenamefont {Wang}, \citenamefont {Li}, \citenamefont {Guo},
  \citenamefont {Song}, \citenamefont {Wang},\ and\ \citenamefont
  {Deng}}]{Deng2023}%
  \BibitemOpen
  \bibfield  {author} {\bibinfo {author} {\bibfnamefont {S.}~\bibnamefont
  {Xu}}, \bibinfo {author} {\bibfnamefont {Z.-Z.}\ \bibnamefont {Sun}},
  \bibinfo {author} {\bibfnamefont {K.}~\bibnamefont {Wang}}, \bibinfo {author}
  {\bibfnamefont {L.}~\bibnamefont {Xiang}}, \bibinfo {author} {\bibfnamefont
  {Z.}~\bibnamefont {Bao}}, \bibinfo {author} {\bibfnamefont {Z.}~\bibnamefont
  {Zhu}}, \bibinfo {author} {\bibfnamefont {F.}~\bibnamefont {Shen}}, \bibinfo
  {author} {\bibfnamefont {Z.}~\bibnamefont {Song}}, \bibinfo {author}
  {\bibfnamefont {P.}~\bibnamefont {Zhang}}, \bibinfo {author} {\bibfnamefont
  {W.}~\bibnamefont {Ren}}, \bibinfo {author} {\bibfnamefont {X.}~\bibnamefont
  {Zhang}}, \bibinfo {author} {\bibfnamefont {H.}~\bibnamefont {Dong}},
  \bibinfo {author} {\bibfnamefont {J.}~\bibnamefont {Deng}}, \bibinfo {author}
  {\bibfnamefont {J.}~\bibnamefont {Chen}}, \bibinfo {author} {\bibfnamefont
  {Y.}~\bibnamefont {Wu}}, \bibinfo {author} {\bibfnamefont {Z.}~\bibnamefont
  {Tan}}, \bibinfo {author} {\bibfnamefont {Y.}~\bibnamefont {Gao}}, \bibinfo
  {author} {\bibfnamefont {F.}~\bibnamefont {Jin}}, \bibinfo {author}
  {\bibfnamefont {X.}~\bibnamefont {Zhu}}, \bibinfo {author} {\bibfnamefont
  {C.}~\bibnamefont {Zhang}}, \bibinfo {author} {\bibfnamefont
  {N.}~\bibnamefont {Wang}}, \bibinfo {author} {\bibfnamefont {Y.}~\bibnamefont
  {Zou}}, \bibinfo {author} {\bibfnamefont {J.}~\bibnamefont {Zhong}}, \bibinfo
  {author} {\bibfnamefont {A.}~\bibnamefont {Zhang}}, \bibinfo {author}
  {\bibfnamefont {W.}~\bibnamefont {Li}}, \bibinfo {author} {\bibfnamefont
  {W.}~\bibnamefont {Jiang}}, \bibinfo {author} {\bibfnamefont {L.-W.}\
  \bibnamefont {Yu}}, \bibinfo {author} {\bibfnamefont {Y.}~\bibnamefont
  {Yao}}, \bibinfo {author} {\bibfnamefont {Z.}~\bibnamefont {Wang}}, \bibinfo
  {author} {\bibfnamefont {H.}~\bibnamefont {Li}}, \bibinfo {author}
  {\bibfnamefont {Q.}~\bibnamefont {Guo}}, \bibinfo {author} {\bibfnamefont
  {C.}~\bibnamefont {Song}}, \bibinfo {author} {\bibfnamefont {H.}~\bibnamefont
  {Wang}},\ and\ \bibinfo {author} {\bibfnamefont {D.-L.}\ \bibnamefont
  {Deng}},\ }\bibfield  {title} {\bibinfo {title} {{Digital Simulation of
  Projective Non-Abelian Anyons with 68 Superconducting Qubits}},\ }\href
  {https://doi.org/10.1088/0256-307X/40/6/060301} {\bibfield  {journal}
  {\bibinfo  {journal} {Chinese Phys. Lett.}\ }\textbf {\bibinfo {volume}
  {40}},\ \bibinfo {pages} {060301} (\bibinfo {year} {2023})}\BibitemShut
  {NoStop}%
\bibitem [{\citenamefont {{Google~Quantum~AI}}(2023)}]{GoogleQEC2023}%
  \BibitemOpen
  \bibfield  {author} {\bibinfo {author} {\bibnamefont {{Google~Quantum~AI}}},\
  }\bibfield  {title} {\bibinfo {title} {Suppressing quantum errors by scaling
  a surface code logical qubit},\ }\href
  {https://doi.org/10.1038/s41586-022-05434-1} {\bibfield  {journal} {\bibinfo
  {journal} {Nature}\ }\textbf {\bibinfo {volume} {614}},\ \bibinfo {pages}
  {676} (\bibinfo {year} {2023})}\BibitemShut {NoStop}%
\bibitem [{\citenamefont {Bluvstein}\ \emph {et~al.}(2024)\citenamefont
  {Bluvstein}, \citenamefont {Evered}, \citenamefont {Geim}, \citenamefont
  {Li}, \citenamefont {Zhou}, \citenamefont {Manovitz}, \citenamefont {Ebadi},
  \citenamefont {Cain}, \citenamefont {Kalinowski}, \citenamefont {Hangleiter},
  \citenamefont {Bonilla~Ataides}, \citenamefont {Maskara}, \citenamefont
  {Cong}, \citenamefont {Gao}, \citenamefont {Sales~Rodriguez}, \citenamefont
  {Karolyshyn}, \citenamefont {Semeghini}, \citenamefont {Gullans},
  \citenamefont {Greiner}, \citenamefont {Vuleti{\'c}},\ and\ \citenamefont
  {Lukin}}]{Lukin2024}%
  \BibitemOpen
  \bibfield  {author} {\bibinfo {author} {\bibfnamefont {D.}~\bibnamefont
  {Bluvstein}}, \bibinfo {author} {\bibfnamefont {S.~J.}\ \bibnamefont
  {Evered}}, \bibinfo {author} {\bibfnamefont {A.~A.}\ \bibnamefont {Geim}},
  \bibinfo {author} {\bibfnamefont {S.~H.}\ \bibnamefont {Li}}, \bibinfo
  {author} {\bibfnamefont {H.}~\bibnamefont {Zhou}}, \bibinfo {author}
  {\bibfnamefont {T.}~\bibnamefont {Manovitz}}, \bibinfo {author}
  {\bibfnamefont {S.}~\bibnamefont {Ebadi}}, \bibinfo {author} {\bibfnamefont
  {M.}~\bibnamefont {Cain}}, \bibinfo {author} {\bibfnamefont {M.}~\bibnamefont
  {Kalinowski}}, \bibinfo {author} {\bibfnamefont {D.}~\bibnamefont
  {Hangleiter}}, \bibinfo {author} {\bibfnamefont {J.~P.}\ \bibnamefont
  {Bonilla~Ataides}}, \bibinfo {author} {\bibfnamefont {N.}~\bibnamefont
  {Maskara}}, \bibinfo {author} {\bibfnamefont {I.}~\bibnamefont {Cong}},
  \bibinfo {author} {\bibfnamefont {X.}~\bibnamefont {Gao}}, \bibinfo {author}
  {\bibfnamefont {P.}~\bibnamefont {Sales~Rodriguez}}, \bibinfo {author}
  {\bibfnamefont {T.}~\bibnamefont {Karolyshyn}}, \bibinfo {author}
  {\bibfnamefont {G.}~\bibnamefont {Semeghini}}, \bibinfo {author}
  {\bibfnamefont {M.~J.}\ \bibnamefont {Gullans}}, \bibinfo {author}
  {\bibfnamefont {M.}~\bibnamefont {Greiner}}, \bibinfo {author} {\bibfnamefont
  {V.}~\bibnamefont {Vuleti{\'c}}},\ and\ \bibinfo {author} {\bibfnamefont
  {M.~D.}\ \bibnamefont {Lukin}},\ }\bibfield  {title} {\bibinfo {title}
  {Logical quantum processor based on reconfigurable atom arrays},\ }\href
  {https://doi.org/10.1038/s41586-023-06927-3} {\bibfield  {journal} {\bibinfo
  {journal} {Nature}\ }\textbf {\bibinfo {volume} {626}},\ \bibinfo {pages}
  {58} (\bibinfo {year} {2024})}\BibitemShut {NoStop}%
\bibitem [{\citenamefont {Dennis}\ \emph {et~al.}(2002)\citenamefont {Dennis},
  \citenamefont {Kitaev}, \citenamefont {Landahl},\ and\ \citenamefont
  {Preskill}}]{Preskill2002}%
  \BibitemOpen
  \bibfield  {author} {\bibinfo {author} {\bibfnamefont {E.}~\bibnamefont
  {Dennis}}, \bibinfo {author} {\bibfnamefont {A.}~\bibnamefont {Kitaev}},
  \bibinfo {author} {\bibfnamefont {A.}~\bibnamefont {Landahl}},\ and\ \bibinfo
  {author} {\bibfnamefont {J.}~\bibnamefont {Preskill}},\ }\bibfield  {title}
  {\bibinfo {title} {{Topological quantum memory}},\ }\href
  {https://doi.org/10.1063/1.1499754} {\bibfield  {journal} {\bibinfo
  {journal} {J. Math. Phys.}\ }\textbf {\bibinfo {volume} {43}},\ \bibinfo
  {pages} {4452} (\bibinfo {year} {2002})}\BibitemShut {NoStop}%
\bibitem [{\citenamefont {Freedman}\ \emph {et~al.}(2003)\citenamefont
  {Freedman}, \citenamefont {Kitaev}, \citenamefont {Larsen},\ and\
  \citenamefont {Wang}}]{Freedman2003}%
  \BibitemOpen
  \bibfield  {author} {\bibinfo {author} {\bibfnamefont {M.}~\bibnamefont
  {Freedman}}, \bibinfo {author} {\bibfnamefont {A.}~\bibnamefont {Kitaev}},
  \bibinfo {author} {\bibfnamefont {M.}~\bibnamefont {Larsen}},\ and\ \bibinfo
  {author} {\bibfnamefont {Z.}~\bibnamefont {Wang}},\ }\bibfield  {title}
  {\bibinfo {title} {Topological quantum computation},\ }\href
  {https://www.ams.org/journals/bull/2003-40-01/S0273-0979-02-00964-3/}
  {\bibfield  {journal} {\bibinfo  {journal} {Bull. Amer. Math. Soc.}\ }\textbf
  {\bibinfo {volume} {40}},\ \bibinfo {pages} {31} (\bibinfo {year}
  {2003})}\BibitemShut {NoStop}%
\bibitem [{\citenamefont {Nayak}\ \emph {et~al.}(2008)\citenamefont {Nayak},
  \citenamefont {Simon}, \citenamefont {Stern}, \citenamefont {Freedman},\ and\
  \citenamefont {Das~Sarma}}]{Nayak2008}%
  \BibitemOpen
  \bibfield  {author} {\bibinfo {author} {\bibfnamefont {C.}~\bibnamefont
  {Nayak}}, \bibinfo {author} {\bibfnamefont {S.~H.}\ \bibnamefont {Simon}},
  \bibinfo {author} {\bibfnamefont {A.}~\bibnamefont {Stern}}, \bibinfo
  {author} {\bibfnamefont {M.}~\bibnamefont {Freedman}},\ and\ \bibinfo
  {author} {\bibfnamefont {S.}~\bibnamefont {Das~Sarma}},\ }\bibfield  {title}
  {\bibinfo {title} {Non-abelian anyons and topological quantum computation},\
  }\href {https://doi.org/10.1103/RevModPhys.80.1083} {\bibfield  {journal}
  {\bibinfo  {journal} {Rev. Mod. Phys.}\ }\textbf {\bibinfo {volume} {80}},\
  \bibinfo {pages} {1083} (\bibinfo {year} {2008})}\BibitemShut {NoStop}%
\bibitem [{\citenamefont {Bonesteel}\ \emph {et~al.}(2005)\citenamefont
  {Bonesteel}, \citenamefont {Hormozi}, \citenamefont {Zikos},\ and\
  \citenamefont {Simon}}]{Bonesteel2005}%
  \BibitemOpen
  \bibfield  {author} {\bibinfo {author} {\bibfnamefont {N.~E.}\ \bibnamefont
  {Bonesteel}}, \bibinfo {author} {\bibfnamefont {L.}~\bibnamefont {Hormozi}},
  \bibinfo {author} {\bibfnamefont {G.}~\bibnamefont {Zikos}},\ and\ \bibinfo
  {author} {\bibfnamefont {S.~H.}\ \bibnamefont {Simon}},\ }\bibfield  {title}
  {\bibinfo {title} {Braid topologies for quantum computation},\ }\href
  {https://doi.org/10.1103/PhysRevLett.95.140503} {\bibfield  {journal}
  {\bibinfo  {journal} {Phys. Rev. Lett.}\ }\textbf {\bibinfo {volume} {95}},\
  \bibinfo {pages} {140503} (\bibinfo {year} {2005})}\BibitemShut {NoStop}%
\bibitem [{\citenamefont {Trebst}\ \emph {et~al.}(2008)\citenamefont {Trebst},
  \citenamefont {Troyer}, \citenamefont {Wang},\ and\ \citenamefont
  {Ludwig}}]{Trebst2008}%
  \BibitemOpen
  \bibfield  {author} {\bibinfo {author} {\bibfnamefont {S.}~\bibnamefont
  {Trebst}}, \bibinfo {author} {\bibfnamefont {M.}~\bibnamefont {Troyer}},
  \bibinfo {author} {\bibfnamefont {Z.}~\bibnamefont {Wang}},\ and\ \bibinfo
  {author} {\bibfnamefont {A.~W.~W.}\ \bibnamefont {Ludwig}},\ }\bibfield
  {title} {\bibinfo {title} {{A Short Introduction to Fibonacci Anyon
  Models}},\ }\href {https://doi.org/10.1143/PTPS.176.384} {\bibfield
  {journal} {\bibinfo  {journal} {Progress of Theoretical Physics Supplement}\
  }\textbf {\bibinfo {volume} {176}},\ \bibinfo {pages} {384} (\bibinfo {year}
  {2008})}\BibitemShut {NoStop}%
\bibitem [{\citenamefont {Bombin}(2010)}]{Bombin2010}%
  \BibitemOpen
  \bibfield  {author} {\bibinfo {author} {\bibfnamefont {H.}~\bibnamefont
  {Bombin}},\ }\bibfield  {title} {\bibinfo {title} {{Topological Order with a
  Twist: Ising Anyons from an Abelian Model}},\ }\href
  {https://doi.org/10.1103/PhysRevLett.105.030403} {\bibfield  {journal}
  {\bibinfo  {journal} {Phys. Rev. Lett.}\ }\textbf {\bibinfo {volume} {105}},\
  \bibinfo {pages} {030403} (\bibinfo {year} {2010})}\BibitemShut {NoStop}%
\bibitem [{\citenamefont {Lahtinen}\ and\ \citenamefont
  {Pachos}(2017)}]{Pachos2017}%
  \BibitemOpen
  \bibfield  {author} {\bibinfo {author} {\bibfnamefont {V.}~\bibnamefont
  {Lahtinen}}\ and\ \bibinfo {author} {\bibfnamefont {J.~K.}\ \bibnamefont
  {Pachos}},\ }\bibfield  {title} {\bibinfo {title} {{A Short Introduction to
  Topological Quantum Computation}},\ }\href
  {https://doi.org/10.21468/SciPostPhys.3.3.021} {\bibfield  {journal}
  {\bibinfo  {journal} {SciPost Phys.}\ }\textbf {\bibinfo {volume} {3}},\
  \bibinfo {pages} {021} (\bibinfo {year} {2017})}\BibitemShut {NoStop}%
\bibitem [{\citenamefont {Pachos}(2012)}]{Pachos2012}%
  \BibitemOpen
  \bibfield  {author} {\bibinfo {author} {\bibfnamefont {J.~K.}\ \bibnamefont
  {Pachos}},\ }\href@noop {} {\emph {\bibinfo {title} {Introduction to
  Topological Quantum Computation}}}\ (\bibinfo  {publisher} {Cambridge
  University Press},\ \bibinfo {year} {2012})\BibitemShut {NoStop}%
\bibitem [{\citenamefont {Simon}(2023)}]{Simon2023book}%
  \BibitemOpen
  \bibfield  {author} {\bibinfo {author} {\bibfnamefont {S.~H.}\ \bibnamefont
  {Simon}},\ }\href@noop {} {\emph {\bibinfo {title} {Topological Quantum}}}\
  (\bibinfo  {publisher} {Oxford University Press},\ \bibinfo {year}
  {2023})\BibitemShut {NoStop}%
\bibitem [{\citenamefont {Petrova}\ \emph {et~al.}(2013)\citenamefont
  {Petrova}, \citenamefont {Mellado},\ and\ \citenamefont
  {Tchernyshyov}}]{Tchernyshyov2013}%
  \BibitemOpen
  \bibfield  {author} {\bibinfo {author} {\bibfnamefont {O.}~\bibnamefont
  {Petrova}}, \bibinfo {author} {\bibfnamefont {P.}~\bibnamefont {Mellado}},\
  and\ \bibinfo {author} {\bibfnamefont {O.}~\bibnamefont {Tchernyshyov}},\
  }\bibfield  {title} {\bibinfo {title} {{Unpaired Majorana modes in the gapped
  phase of Kitaev's honeycomb model}},\ }\href
  {https://doi.org/10.1103/PhysRevB.88.140405} {\bibfield  {journal} {\bibinfo
  {journal} {Phys. Rev. B}\ }\textbf {\bibinfo {volume} {88}},\ \bibinfo
  {pages} {140405} (\bibinfo {year} {2013})}\BibitemShut {NoStop}%
\bibitem [{\citenamefont {Petrova}\ \emph {et~al.}(2014)\citenamefont
  {Petrova}, \citenamefont {Mellado},\ and\ \citenamefont
  {Tchernyshyov}}]{Tchernyshyov2014}%
  \BibitemOpen
  \bibfield  {author} {\bibinfo {author} {\bibfnamefont {O.}~\bibnamefont
  {Petrova}}, \bibinfo {author} {\bibfnamefont {P.}~\bibnamefont {Mellado}},\
  and\ \bibinfo {author} {\bibfnamefont {O.}~\bibnamefont {Tchernyshyov}},\
  }\bibfield  {title} {\bibinfo {title} {{Unpaired Majorana modes on
  dislocations and string defects in Kitaev's honeycomb model}},\ }\href
  {https://doi.org/10.1103/PhysRevB.90.134404} {\bibfield  {journal} {\bibinfo
  {journal} {Phys. Rev. B}\ }\textbf {\bibinfo {volume} {90}},\ \bibinfo
  {pages} {134404} (\bibinfo {year} {2014})}\BibitemShut {NoStop}%
\bibitem [{\citenamefont {Zheng}\ \emph {et~al.}(2015)\citenamefont {Zheng},
  \citenamefont {Dua},\ and\ \citenamefont {Jiang}}]{Jiang2015}%
  \BibitemOpen
  \bibfield  {author} {\bibinfo {author} {\bibfnamefont {H.}~\bibnamefont
  {Zheng}}, \bibinfo {author} {\bibfnamefont {A.}~\bibnamefont {Dua}},\ and\
  \bibinfo {author} {\bibfnamefont {L.}~\bibnamefont {Jiang}},\ }\bibfield
  {title} {\bibinfo {title} {{Demonstrating non-Abelian statistics of Majorana
  fermions using twist defects}},\ }\href
  {https://doi.org/10.1103/PhysRevB.92.245139} {\bibfield  {journal} {\bibinfo
  {journal} {Phys. Rev. B}\ }\textbf {\bibinfo {volume} {92}},\ \bibinfo
  {pages} {245139} (\bibinfo {year} {2015})}\BibitemShut {NoStop}%
\bibitem [{\citenamefont {Horner}\ \emph {et~al.}(2020)\citenamefont {Horner},
  \citenamefont {Farjami},\ and\ \citenamefont {Pachos}}]{Pachos2020}%
  \BibitemOpen
  \bibfield  {author} {\bibinfo {author} {\bibfnamefont {M.~D.}\ \bibnamefont
  {Horner}}, \bibinfo {author} {\bibfnamefont {A.}~\bibnamefont {Farjami}},\
  and\ \bibinfo {author} {\bibfnamefont {J.~K.}\ \bibnamefont {Pachos}},\
  }\bibfield  {title} {\bibinfo {title} {{Equivalence between vortices, twists,
  and chiral gauge fields in the Kitaev honeycomb lattice model}},\ }\href
  {https://doi.org/10.1103/PhysRevB.102.125152} {\bibfield  {journal} {\bibinfo
   {journal} {Phys. Rev. B}\ }\textbf {\bibinfo {volume} {102}},\ \bibinfo
  {pages} {125152} (\bibinfo {year} {2020})}\BibitemShut {NoStop}%
\bibitem [{\citenamefont {Tantivasadakarn}\ \emph {et~al.}(2023)\citenamefont
  {Tantivasadakarn}, \citenamefont {Verresen},\ and\ \citenamefont
  {Vishwanath}}]{Tantivasadakarn2023}%
  \BibitemOpen
  \bibfield  {author} {\bibinfo {author} {\bibfnamefont {N.}~\bibnamefont
  {Tantivasadakarn}}, \bibinfo {author} {\bibfnamefont {R.}~\bibnamefont
  {Verresen}},\ and\ \bibinfo {author} {\bibfnamefont {A.}~\bibnamefont
  {Vishwanath}},\ }\bibfield  {title} {\bibinfo {title} {{Shortest Route to
  Non-Abelian Topological Order on a Quantum Processor}},\ }\href
  {https://doi.org/10.1103/PhysRevLett.131.060405} {\bibfield  {journal}
  {\bibinfo  {journal} {Phys. Rev. Lett.}\ }\textbf {\bibinfo {volume} {131}},\
  \bibinfo {pages} {060405} (\bibinfo {year} {2023})}\BibitemShut {NoStop}%
\bibitem [{\citenamefont {Gordon}\ \emph {et~al.}(2019)\citenamefont {Gordon},
  \citenamefont {Catuneanu}, \citenamefont {S{\o}rensen},\ and\ \citenamefont
  {Kee}}]{Kee2019}%
  \BibitemOpen
  \bibfield  {author} {\bibinfo {author} {\bibfnamefont {J.~S.}\ \bibnamefont
  {Gordon}}, \bibinfo {author} {\bibfnamefont {A.}~\bibnamefont {Catuneanu}},
  \bibinfo {author} {\bibfnamefont {E.~S.}\ \bibnamefont {S{\o}rensen}},\ and\
  \bibinfo {author} {\bibfnamefont {H.-Y.}\ \bibnamefont {Kee}},\ }\bibfield
  {title} {\bibinfo {title} {{Theory of the field-revealed Kitaev spin
  liquid}},\ }\href {https://doi.org/10.1038/s41467-019-10405-8} {\bibfield
  {journal} {\bibinfo  {journal} {Nat. Commun.}\ }\textbf {\bibinfo {volume}
  {10}},\ \bibinfo {pages} {2470} (\bibinfo {year} {2019})}\BibitemShut
  {NoStop}%
\bibitem [{\citenamefont {Hickey}\ and\ \citenamefont
  {Trebst}(2019)}]{Hickey2019}%
  \BibitemOpen
  \bibfield  {author} {\bibinfo {author} {\bibfnamefont {C.}~\bibnamefont
  {Hickey}}\ and\ \bibinfo {author} {\bibfnamefont {S.}~\bibnamefont
  {Trebst}},\ }\bibfield  {title} {\bibinfo {title} {{Emergence of a
  field-driven U(1) spin liquid in the Kitaev honeycomb model}},\ }\href
  {https://doi.org/10.1038/s41467-019-08459-9} {\bibfield  {journal} {\bibinfo
  {journal} {Nat. Commun.}\ }\textbf {\bibinfo {volume} {10}},\ \bibinfo
  {pages} {530} (\bibinfo {year} {2019})}\BibitemShut {NoStop}%
\bibitem [{\citenamefont {Patel}\ and\ \citenamefont
  {Trivedi}(2019)}]{Trivedi2019}%
  \BibitemOpen
  \bibfield  {author} {\bibinfo {author} {\bibfnamefont {N.~D.}\ \bibnamefont
  {Patel}}\ and\ \bibinfo {author} {\bibfnamefont {N.}~\bibnamefont
  {Trivedi}},\ }\bibfield  {title} {\bibinfo {title} {{Magnetic field-induced
  intermediate quantum spin liquid with a spinon Fermi surface}},\ }\href
  {https://doi.org/10.1073/pnas.1821406116} {\bibfield  {journal} {\bibinfo
  {journal} {Proceedings of the National Academy of Sciences}\ }\textbf
  {\bibinfo {volume} {116}},\ \bibinfo {pages} {12199} (\bibinfo {year}
  {2019})}\BibitemShut {NoStop}%
\bibitem [{\citenamefont {Zhang}\ \emph {et~al.}(2019)\citenamefont {Zhang},
  \citenamefont {Wang}, \citenamefont {Hal\'asz},\ and\ \citenamefont
  {Batista}}]{Batista2019}%
  \BibitemOpen
  \bibfield  {author} {\bibinfo {author} {\bibfnamefont {S.-S.}\ \bibnamefont
  {Zhang}}, \bibinfo {author} {\bibfnamefont {Z.}~\bibnamefont {Wang}},
  \bibinfo {author} {\bibfnamefont {G.~B.}\ \bibnamefont {Hal\'asz}},\ and\
  \bibinfo {author} {\bibfnamefont {C.~D.}\ \bibnamefont {Batista}},\
  }\bibfield  {title} {\bibinfo {title} {{Vison Crystals in an Extended Kitaev
  Model on the Honeycomb Lattice}},\ }\href
  {https://doi.org/10.1103/PhysRevLett.123.057201} {\bibfield  {journal}
  {\bibinfo  {journal} {Phys. Rev. Lett.}\ }\textbf {\bibinfo {volume} {123}},\
  \bibinfo {pages} {057201} (\bibinfo {year} {2019})}\BibitemShut {NoStop}%
\bibitem [{\citenamefont {Rokhsar}\ and\ \citenamefont
  {Kivelson}(1988)}]{Rokhsar1988}%
  \BibitemOpen
  \bibfield  {author} {\bibinfo {author} {\bibfnamefont {D.~S.}\ \bibnamefont
  {Rokhsar}}\ and\ \bibinfo {author} {\bibfnamefont {S.~A.}\ \bibnamefont
  {Kivelson}},\ }\bibfield  {title} {\bibinfo {title} {{Superconductivity and
  the Quantum Hard-Core Dimer Gas}},\ }\href
  {https://doi.org/10.1103/PhysRevLett.61.2376} {\bibfield  {journal} {\bibinfo
   {journal} {Phys. Rev. Lett.}\ }\textbf {\bibinfo {volume} {61}},\ \bibinfo
  {pages} {2376} (\bibinfo {year} {1988})}\BibitemShut {NoStop}%
\bibitem [{\citenamefont {Moessner}\ and\ \citenamefont
  {Sondhi}(2001)}]{Moessner2001}%
  \BibitemOpen
  \bibfield  {author} {\bibinfo {author} {\bibfnamefont {R.}~\bibnamefont
  {Moessner}}\ and\ \bibinfo {author} {\bibfnamefont {S.~L.}\ \bibnamefont
  {Sondhi}},\ }\bibfield  {title} {\bibinfo {title} {{Resonating Valence Bond
  Phase in the Triangular Lattice Quantum Dimer Model}},\ }\href
  {https://doi.org/10.1103/PhysRevLett.86.1881} {\bibfield  {journal} {\bibinfo
   {journal} {Phys. Rev. Lett.}\ }\textbf {\bibinfo {volume} {86}},\ \bibinfo
  {pages} {1881} (\bibinfo {year} {2001})}\BibitemShut {NoStop}%
\bibitem [{\citenamefont {Misguich}\ \emph {et~al.}(2002)\citenamefont
  {Misguich}, \citenamefont {Serban},\ and\ \citenamefont
  {Pasquier}}]{Misguich2002}%
  \BibitemOpen
  \bibfield  {author} {\bibinfo {author} {\bibfnamefont {G.}~\bibnamefont
  {Misguich}}, \bibinfo {author} {\bibfnamefont {D.}~\bibnamefont {Serban}},\
  and\ \bibinfo {author} {\bibfnamefont {V.}~\bibnamefont {Pasquier}},\
  }\bibfield  {title} {\bibinfo {title} {{Quantum Dimer Model on the Kagome
  Lattice: Solvable Dimer-Liquid and Ising Gauge Theory}},\ }\href
  {https://doi.org/10.1103/PhysRevLett.89.137202} {\bibfield  {journal}
  {\bibinfo  {journal} {Phys. Rev. Lett.}\ }\textbf {\bibinfo {volume} {89}},\
  \bibinfo {pages} {137202} (\bibinfo {year} {2002})}\BibitemShut {NoStop}%
\bibitem [{\citenamefont {Samajdar}\ \emph {et~al.}(2021)\citenamefont
  {Samajdar}, \citenamefont {Ho}, \citenamefont {Pichler}, \citenamefont
  {Lukin},\ and\ \citenamefont {Sachdev}}]{Samajdar2020}%
  \BibitemOpen
  \bibfield  {author} {\bibinfo {author} {\bibfnamefont {R.}~\bibnamefont
  {Samajdar}}, \bibinfo {author} {\bibfnamefont {W.~W.}\ \bibnamefont {Ho}},
  \bibinfo {author} {\bibfnamefont {H.}~\bibnamefont {Pichler}}, \bibinfo
  {author} {\bibfnamefont {M.~D.}\ \bibnamefont {Lukin}},\ and\ \bibinfo
  {author} {\bibfnamefont {S.}~\bibnamefont {Sachdev}},\ }\bibfield  {title}
  {\bibinfo {title} {{Quantum phases of Rydberg atoms on a kagome lattice}},\
  }\href {https://doi.org/10.1073/pnas.2015785118} {\bibfield  {journal}
  {\bibinfo  {journal} {Proc. Natl. Acad. Sci. U.S.A.}\ }\textbf {\bibinfo
  {volume} {118}},\ \bibinfo {pages} {e2015785118} (\bibinfo {year}
  {2021})}\BibitemShut {NoStop}%
\bibitem [{\citenamefont {Verresen}\ \emph {et~al.}(2021)\citenamefont
  {Verresen}, \citenamefont {Lukin},\ and\ \citenamefont
  {Vishwanath}}]{Verresen2021}%
  \BibitemOpen
  \bibfield  {author} {\bibinfo {author} {\bibfnamefont {R.}~\bibnamefont
  {Verresen}}, \bibinfo {author} {\bibfnamefont {M.~D.}\ \bibnamefont
  {Lukin}},\ and\ \bibinfo {author} {\bibfnamefont {A.}~\bibnamefont
  {Vishwanath}},\ }\bibfield  {title} {\bibinfo {title} {{Prediction of Toric
  Code Topological Order from Rydberg Blockade}},\ }\href
  {https://doi.org/10.1103/PhysRevX.11.031005} {\bibfield  {journal} {\bibinfo
  {journal} {Phys. Rev. X}\ }\textbf {\bibinfo {volume} {11}},\ \bibinfo
  {pages} {031005} (\bibinfo {year} {2021})}\BibitemShut {NoStop}%
\bibitem [{\citenamefont {Verresen}\ and\ \citenamefont
  {Vishwanath}(2022)}]{Verresen2022}%
  \BibitemOpen
  \bibfield  {author} {\bibinfo {author} {\bibfnamefont {R.}~\bibnamefont
  {Verresen}}\ and\ \bibinfo {author} {\bibfnamefont {A.}~\bibnamefont
  {Vishwanath}},\ }\bibfield  {title} {\bibinfo {title} {{Unifying Kitaev
  Magnets, Kagom\'e Dimer Models, and Ruby Rydberg Spin Liquids}},\ }\href
  {https://doi.org/10.1103/PhysRevX.12.041029} {\bibfield  {journal} {\bibinfo
  {journal} {Phys. Rev. X}\ }\textbf {\bibinfo {volume} {12}},\ \bibinfo
  {pages} {041029} (\bibinfo {year} {2022})}\BibitemShut {NoStop}%
\bibitem [{\citenamefont {Bais}\ and\ \citenamefont
  {Slingerland}(2009)}]{Bais2009}%
  \BibitemOpen
  \bibfield  {author} {\bibinfo {author} {\bibfnamefont {F.~A.}\ \bibnamefont
  {Bais}}\ and\ \bibinfo {author} {\bibfnamefont {J.~K.}\ \bibnamefont
  {Slingerland}},\ }\bibfield  {title} {\bibinfo {title} {{Condensate-induced
  transitions between topologically ordered phases}},\ }\href
  {https://doi.org/10.1103/PhysRevB.79.045316} {\bibfield  {journal} {\bibinfo
  {journal} {Phys. Rev. B}\ }\textbf {\bibinfo {volume} {79}},\ \bibinfo
  {pages} {045316} (\bibinfo {year} {2009})}\BibitemShut {NoStop}%
\bibitem [{\citenamefont {Bombin}\ and\ \citenamefont
  {Martin-Delgado}(2008)}]{Bombin2008}%
  \BibitemOpen
  \bibfield  {author} {\bibinfo {author} {\bibfnamefont {H.}~\bibnamefont
  {Bombin}}\ and\ \bibinfo {author} {\bibfnamefont {M.~A.}\ \bibnamefont
  {Martin-Delgado}},\ }\bibfield  {title} {\bibinfo {title} {{Family of
  non-Abelian Kitaev models on a lattice: Topological condensation and
  confinement}},\ }\href {https://doi.org/10.1103/PhysRevB.78.115421}
  {\bibfield  {journal} {\bibinfo  {journal} {Phys. Rev. B}\ }\textbf {\bibinfo
  {volume} {78}},\ \bibinfo {pages} {115421} (\bibinfo {year}
  {2008})}\BibitemShut {NoStop}%
\bibitem [{\citenamefont {Barkeshli}\ and\ \citenamefont
  {Wen}(2010)}]{Barkeshli2010}%
  \BibitemOpen
  \bibfield  {author} {\bibinfo {author} {\bibfnamefont {M.}~\bibnamefont
  {Barkeshli}}\ and\ \bibinfo {author} {\bibfnamefont {X.-G.}\ \bibnamefont
  {Wen}},\ }\bibfield  {title} {\bibinfo {title} {{Anyon Condensation and
  Continuous Topological Phase Transitions in Non-Abelian Fractional Quantum
  Hall States}},\ }\href {https://doi.org/10.1103/PhysRevLett.105.216804}
  {\bibfield  {journal} {\bibinfo  {journal} {Phys. Rev. Lett.}\ }\textbf
  {\bibinfo {volume} {105}},\ \bibinfo {pages} {216804} (\bibinfo {year}
  {2010})}\BibitemShut {NoStop}%
\bibitem [{\citenamefont {Burnell}\ \emph {et~al.}(2011)\citenamefont
  {Burnell}, \citenamefont {Simon},\ and\ \citenamefont
  {Slingerland}}]{Burnell2011}%
  \BibitemOpen
  \bibfield  {author} {\bibinfo {author} {\bibfnamefont {F.~J.}\ \bibnamefont
  {Burnell}}, \bibinfo {author} {\bibfnamefont {S.~H.}\ \bibnamefont {Simon}},\
  and\ \bibinfo {author} {\bibfnamefont {J.~K.}\ \bibnamefont {Slingerland}},\
  }\bibfield  {title} {\bibinfo {title} {{Condensation of achiral simple
  currents in topological lattice models: Hamiltonian study of topological
  symmetry breaking}},\ }\href {https://doi.org/10.1103/PhysRevB.84.125434}
  {\bibfield  {journal} {\bibinfo  {journal} {Phys. Rev. B}\ }\textbf {\bibinfo
  {volume} {84}},\ \bibinfo {pages} {125434} (\bibinfo {year}
  {2011})}\BibitemShut {NoStop}%
\bibitem [{\citenamefont {Burnell}\ \emph {et~al.}(2012)\citenamefont
  {Burnell}, \citenamefont {Simon},\ and\ \citenamefont
  {Slingerland}}]{Burnell2012}%
  \BibitemOpen
  \bibfield  {author} {\bibinfo {author} {\bibfnamefont {F.~J.}\ \bibnamefont
  {Burnell}}, \bibinfo {author} {\bibfnamefont {S.~H.}\ \bibnamefont {Simon}},\
  and\ \bibinfo {author} {\bibfnamefont {J.~K.}\ \bibnamefont {Slingerland}},\
  }\bibfield  {title} {\bibinfo {title} {{Phase transitions in topological
  lattice models via topological symmetry breaking}},\ }\href
  {https://doi.org/10.1088/1367-2630/14/1/015004} {\bibfield  {journal}
  {\bibinfo  {journal} {New J. Phys.}\ }\textbf {\bibinfo {volume} {14}},\
  \bibinfo {pages} {015004} (\bibinfo {year} {2012})}\BibitemShut {NoStop}%
\bibitem [{\citenamefont {Eli\"ens}\ \emph {et~al.}(2014)\citenamefont
  {Eli\"ens}, \citenamefont {Romers},\ and\ \citenamefont {Bais}}]{Bais2014}%
  \BibitemOpen
  \bibfield  {author} {\bibinfo {author} {\bibfnamefont {I.~S.}\ \bibnamefont
  {Eli\"ens}}, \bibinfo {author} {\bibfnamefont {J.~C.}\ \bibnamefont
  {Romers}},\ and\ \bibinfo {author} {\bibfnamefont {F.~A.}\ \bibnamefont
  {Bais}},\ }\bibfield  {title} {\bibinfo {title} {{Diagrammatics for Bose
  condensation in anyon theories}},\ }\href
  {https://doi.org/10.1103/PhysRevB.90.195130} {\bibfield  {journal} {\bibinfo
  {journal} {Phys. Rev. B}\ }\textbf {\bibinfo {volume} {90}},\ \bibinfo
  {pages} {195130} (\bibinfo {year} {2014})}\BibitemShut {NoStop}%
\bibitem [{\citenamefont {Neupert}\ \emph {et~al.}(2016)\citenamefont
  {Neupert}, \citenamefont {He}, \citenamefont {von Keyserlingk}, \citenamefont
  {Sierra},\ and\ \citenamefont {Bernevig}}]{Neupert2016}%
  \BibitemOpen
  \bibfield  {author} {\bibinfo {author} {\bibfnamefont {T.}~\bibnamefont
  {Neupert}}, \bibinfo {author} {\bibfnamefont {H.}~\bibnamefont {He}},
  \bibinfo {author} {\bibfnamefont {C.}~\bibnamefont {von Keyserlingk}},
  \bibinfo {author} {\bibfnamefont {G.}~\bibnamefont {Sierra}},\ and\ \bibinfo
  {author} {\bibfnamefont {B.~A.}\ \bibnamefont {Bernevig}},\ }\bibfield
  {title} {\bibinfo {title} {{Boson condensation in topologically ordered
  quantum liquids}},\ }\href {https://doi.org/10.1103/PhysRevB.93.115103}
  {\bibfield  {journal} {\bibinfo  {journal} {Phys. Rev. B}\ }\textbf {\bibinfo
  {volume} {93}},\ \bibinfo {pages} {115103} (\bibinfo {year}
  {2016})}\BibitemShut {NoStop}%
\bibitem [{\citenamefont {Burnell}(2018)}]{Burnell2018}%
  \BibitemOpen
  \bibfield  {author} {\bibinfo {author} {\bibfnamefont {F.~J.}\ \bibnamefont
  {Burnell}},\ }\bibfield  {title} {\bibinfo {title} {{Anyon Condensation and
  Its Applications}},\ }\href
  {https://doi.org/10.1146/annurev-conmatphys-033117-054154} {\bibfield
  {journal} {\bibinfo  {journal} {Annu. Rev. Condens. Matter Phys.}\ }\textbf
  {\bibinfo {volume} {9}},\ \bibinfo {pages} {307} (\bibinfo {year}
  {2018})}\BibitemShut {NoStop}%
\bibitem [{\citenamefont {Kong}(2014)}]{Kong2014}%
  \BibitemOpen
  \bibfield  {author} {\bibinfo {author} {\bibfnamefont {L.}~\bibnamefont
  {Kong}},\ }\bibfield  {title} {\bibinfo {title} {Anyon condensation and
  tensor categories},\ }\href
  {https://doi.org/https://doi.org/10.1016/j.nuclphysb.2014.07.003} {\bibfield
  {journal} {\bibinfo  {journal} {Nuclear Physics B}\ }\textbf {\bibinfo
  {volume} {886}},\ \bibinfo {pages} {436} (\bibinfo {year}
  {2014})}\BibitemShut {NoStop}%
\bibitem [{\citenamefont {Teo}\ \emph {et~al.}(2015)\citenamefont {Teo},
  \citenamefont {Hughes},\ and\ \citenamefont {Fradkin}}]{Teo2015}%
  \BibitemOpen
  \bibfield  {author} {\bibinfo {author} {\bibfnamefont {J.~C.}\ \bibnamefont
  {Teo}}, \bibinfo {author} {\bibfnamefont {T.~L.}\ \bibnamefont {Hughes}},\
  and\ \bibinfo {author} {\bibfnamefont {E.}~\bibnamefont {Fradkin}},\
  }\bibfield  {title} {\bibinfo {title} {{Theory of twist liquids: Gauging an
  anyonic symmetry}},\ }\href
  {https://doi.org/https://doi.org/10.1016/j.aop.2015.05.012} {\bibfield
  {journal} {\bibinfo  {journal} {Ann. Phys.}\ }\textbf {\bibinfo {volume}
  {360}},\ \bibinfo {pages} {349} (\bibinfo {year} {2015})}\BibitemShut
  {NoStop}%
\bibitem [{\citenamefont {Barkeshli}\ \emph {et~al.}(2019)\citenamefont
  {Barkeshli}, \citenamefont {Bonderson}, \citenamefont {Cheng},\ and\
  \citenamefont {Wang}}]{Barkeshli2019}%
  \BibitemOpen
  \bibfield  {author} {\bibinfo {author} {\bibfnamefont {M.}~\bibnamefont
  {Barkeshli}}, \bibinfo {author} {\bibfnamefont {P.}~\bibnamefont
  {Bonderson}}, \bibinfo {author} {\bibfnamefont {M.}~\bibnamefont {Cheng}},\
  and\ \bibinfo {author} {\bibfnamefont {Z.}~\bibnamefont {Wang}},\ }\bibfield
  {title} {\bibinfo {title} {Symmetry fractionalization, defects, and gauging
  of topological phases},\ }\href {https://doi.org/10.1103/PhysRevB.100.115147}
  {\bibfield  {journal} {\bibinfo  {journal} {Phys. Rev. B}\ }\textbf {\bibinfo
  {volume} {100}},\ \bibinfo {pages} {115147} (\bibinfo {year}
  {2019})}\BibitemShut {NoStop}%
\bibitem [{\citenamefont {Wiedmann}\ \emph {et~al.}(2020)\citenamefont
  {Wiedmann}, \citenamefont {Lenke}, \citenamefont {Walther}, \citenamefont
  {M\"uhlhauser},\ and\ \citenamefont {Schmidt}}]{Schmidt2020}%
  \BibitemOpen
  \bibfield  {author} {\bibinfo {author} {\bibfnamefont {R.}~\bibnamefont
  {Wiedmann}}, \bibinfo {author} {\bibfnamefont {L.}~\bibnamefont {Lenke}},
  \bibinfo {author} {\bibfnamefont {M.~R.}\ \bibnamefont {Walther}}, \bibinfo
  {author} {\bibfnamefont {M.}~\bibnamefont {M\"uhlhauser}},\ and\ \bibinfo
  {author} {\bibfnamefont {K.~P.}\ \bibnamefont {Schmidt}},\ }\bibfield
  {title} {\bibinfo {title} {{Quantum critical phase transition between two
  topologically ordered phases in the Ising toric code bilayer}},\ }\href
  {https://doi.org/10.1103/PhysRevB.102.214422} {\bibfield  {journal} {\bibinfo
   {journal} {Phys. Rev. B}\ }\textbf {\bibinfo {volume} {102}},\ \bibinfo
  {pages} {214422} (\bibinfo {year} {2020})}\BibitemShut {NoStop}%
\bibitem [{\citenamefont {Schamri\ss{}}\ \emph {et~al.}(2022)\citenamefont
  {Schamri\ss{}}, \citenamefont {Lenke}, \citenamefont {M\"uhlhauser},\ and\
  \citenamefont {Schmidt}}]{Schmidt2022}%
  \BibitemOpen
  \bibfield  {author} {\bibinfo {author} {\bibfnamefont {L.}~\bibnamefont
  {Schamri\ss{}}}, \bibinfo {author} {\bibfnamefont {L.}~\bibnamefont {Lenke}},
  \bibinfo {author} {\bibfnamefont {M.}~\bibnamefont {M\"uhlhauser}},\ and\
  \bibinfo {author} {\bibfnamefont {K.~P.}\ \bibnamefont {Schmidt}},\
  }\bibfield  {title} {\bibinfo {title} {{Quantum phase transitions in the
  $K$-layer Ising toric code}},\ }\href
  {https://doi.org/10.1103/PhysRevB.105.184425} {\bibfield  {journal} {\bibinfo
   {journal} {Phys. Rev. B}\ }\textbf {\bibinfo {volume} {105}},\ \bibinfo
  {pages} {184425} (\bibinfo {year} {2022})}\BibitemShut {NoStop}%
\bibitem [{\citenamefont {Wang}\ \emph {et~al.}(2021)\citenamefont {Wang},
  \citenamefont {Yan}, \citenamefont {Wang}, \citenamefont {Qi},\ and\
  \citenamefont {Meng}}]{Meng2021}%
  \BibitemOpen
  \bibfield  {author} {\bibinfo {author} {\bibfnamefont {Y.-C.}\ \bibnamefont
  {Wang}}, \bibinfo {author} {\bibfnamefont {Z.}~\bibnamefont {Yan}}, \bibinfo
  {author} {\bibfnamefont {C.}~\bibnamefont {Wang}}, \bibinfo {author}
  {\bibfnamefont {Y.}~\bibnamefont {Qi}},\ and\ \bibinfo {author}
  {\bibfnamefont {Z.~Y.}\ \bibnamefont {Meng}},\ }\bibfield  {title} {\bibinfo
  {title} {{Vestigial anyon condensation in kagome quantum spin liquids}},\
  }\href {https://doi.org/10.1103/PhysRevB.103.014408} {\bibfield  {journal}
  {\bibinfo  {journal} {Phys. Rev. B}\ }\textbf {\bibinfo {volume} {103}},\
  \bibinfo {pages} {014408} (\bibinfo {year} {2021})}\BibitemShut {NoStop}%
\bibitem [{\citenamefont {Xu}\ and\ \citenamefont {Schuch}(2021)}]{Schuch2021}%
  \BibitemOpen
  \bibfield  {author} {\bibinfo {author} {\bibfnamefont {W.-T.}\ \bibnamefont
  {Xu}}\ and\ \bibinfo {author} {\bibfnamefont {N.}~\bibnamefont {Schuch}},\
  }\bibfield  {title} {\bibinfo {title} {{Characterization of topological phase
  transitions from a non-Abelian topological state and its Galois conjugate
  through condensation and confinement order parameters}},\ }\href
  {https://doi.org/10.1103/PhysRevB.104.155119} {\bibfield  {journal} {\bibinfo
   {journal} {Phys. Rev. B}\ }\textbf {\bibinfo {volume} {104}},\ \bibinfo
  {pages} {155119} (\bibinfo {year} {2021})}\BibitemShut {NoStop}%
\bibitem [{\citenamefont {Xu}\ \emph {et~al.}(2022)\citenamefont {Xu},
  \citenamefont {Garre-Rubio},\ and\ \citenamefont {Schuch}}]{Schuch2022}%
  \BibitemOpen
  \bibfield  {author} {\bibinfo {author} {\bibfnamefont {W.-T.}\ \bibnamefont
  {Xu}}, \bibinfo {author} {\bibfnamefont {J.}~\bibnamefont {Garre-Rubio}},\
  and\ \bibinfo {author} {\bibfnamefont {N.}~\bibnamefont {Schuch}},\
  }\bibfield  {title} {\bibinfo {title} {{Complete characterization of
  non-Abelian topological phase transitions and detection of anyon splitting
  with projected entangled pair states}},\ }\href
  {https://doi.org/10.1103/PhysRevB.106.205139} {\bibfield  {journal} {\bibinfo
   {journal} {Phys. Rev. B}\ }\textbf {\bibinfo {volume} {106}},\ \bibinfo
  {pages} {205139} (\bibinfo {year} {2022})}\BibitemShut {NoStop}%
\bibitem [{\citenamefont {Huxford}\ \emph {et~al.}(2023)\citenamefont
  {Huxford}, \citenamefont {Nguyen},\ and\ \citenamefont {Kim}}]{Huxford2023}%
  \BibitemOpen
  \bibfield  {author} {\bibinfo {author} {\bibfnamefont {J.}~\bibnamefont
  {Huxford}}, \bibinfo {author} {\bibfnamefont {D.~X.}\ \bibnamefont
  {Nguyen}},\ and\ \bibinfo {author} {\bibfnamefont {Y.~B.}\ \bibnamefont
  {Kim}},\ }\bibfield  {title} {\bibinfo {title} {{Gaining insights on anyon
  condensation and 1-form symmetry breaking across a topological phase
  transition in a deformed toric code model}},\ }\href
  {https://doi.org/10.21468/SciPostPhys.15.6.253} {\bibfield  {journal}
  {\bibinfo  {journal} {SciPost Phys.}\ }\textbf {\bibinfo {volume} {15}},\
  \bibinfo {pages} {253} (\bibinfo {year} {2023})}\BibitemShut {NoStop}%
\bibitem [{\citenamefont {Long}\ and\ \citenamefont
  {Doherty}(2024)}]{Doherty2024}%
  \BibitemOpen
  \bibfield  {author} {\bibinfo {author} {\bibfnamefont {D.~M.}\ \bibnamefont
  {Long}}\ and\ \bibinfo {author} {\bibfnamefont {A.~C.}\ \bibnamefont
  {Doherty}},\ }\bibfield  {title} {\bibinfo {title} {Edge theories for anyon
  condensation phase transitions},\ }\href
  {https://doi.org/10.1103/PhysRevB.109.075140} {\bibfield  {journal} {\bibinfo
   {journal} {Phys. Rev. B}\ }\textbf {\bibinfo {volume} {109}},\ \bibinfo
  {pages} {075140} (\bibinfo {year} {2024})}\BibitemShut {NoStop}%
\bibitem [{\citenamefont {Lin}\ and\ \citenamefont
  {Burnell}(2023)}]{Burnell2023}%
  \BibitemOpen
  \bibfield  {author} {\bibinfo {author} {\bibfnamefont {C.-H.}\ \bibnamefont
  {Lin}}\ and\ \bibinfo {author} {\bibfnamefont {F.~J.}\ \bibnamefont
  {Burnell}},\ }\href@noop {} {\bibinfo {title} {Anyon condensation in the
  string-net models}} (\bibinfo {year} {2023}),\ \Eprint
  {https://arxiv.org/abs/2303.07291} {arXiv:2303.07291 [cond-mat.str-el]}
  \BibitemShut {NoStop}%
\bibitem [{\citenamefont {Brown}\ \emph {et~al.}(2017)\citenamefont {Brown},
  \citenamefont {Laubscher}, \citenamefont {Kesselring},\ and\ \citenamefont
  {Wootton}}]{Brown2017}%
  \BibitemOpen
  \bibfield  {author} {\bibinfo {author} {\bibfnamefont {B.~J.}\ \bibnamefont
  {Brown}}, \bibinfo {author} {\bibfnamefont {K.}~\bibnamefont {Laubscher}},
  \bibinfo {author} {\bibfnamefont {M.~S.}\ \bibnamefont {Kesselring}},\ and\
  \bibinfo {author} {\bibfnamefont {J.~R.}\ \bibnamefont {Wootton}},\
  }\bibfield  {title} {\bibinfo {title} {{Poking Holes and Cutting Corners to
  Achieve Clifford Gates with the Surface Code}},\ }\href
  {https://doi.org/10.1103/PhysRevX.7.021029} {\bibfield  {journal} {\bibinfo
  {journal} {Phys. Rev. X}\ }\textbf {\bibinfo {volume} {7}},\ \bibinfo {pages}
  {021029} (\bibinfo {year} {2017})}\BibitemShut {NoStop}%
\bibitem [{\citenamefont {Kesselring}\ \emph {et~al.}(2018)\citenamefont
  {Kesselring}, \citenamefont {Pastawski}, \citenamefont {Eisert},\ and\
  \citenamefont {Brown}}]{Brown2018}%
  \BibitemOpen
  \bibfield  {author} {\bibinfo {author} {\bibfnamefont {M.~S.}\ \bibnamefont
  {Kesselring}}, \bibinfo {author} {\bibfnamefont {F.}~\bibnamefont
  {Pastawski}}, \bibinfo {author} {\bibfnamefont {J.}~\bibnamefont {Eisert}},\
  and\ \bibinfo {author} {\bibfnamefont {B.~J.}\ \bibnamefont {Brown}},\
  }\bibfield  {title} {\bibinfo {title} {The boundaries and twist defects of
  the color code and their applications to topological quantum computation},\
  }\href {https://doi.org/10.22331/q-2018-10-19-101} {\bibfield  {journal}
  {\bibinfo  {journal} {Quantum}\ }\textbf {\bibinfo {volume} {2}},\ \bibinfo
  {pages} {101} (\bibinfo {year} {2018})}\BibitemShut {NoStop}%
\bibitem [{\citenamefont {Benhemou}\ \emph {et~al.}(2022)\citenamefont
  {Benhemou}, \citenamefont {Pachos},\ and\ \citenamefont
  {Browne}}]{Browne2022}%
  \BibitemOpen
  \bibfield  {author} {\bibinfo {author} {\bibfnamefont {A.}~\bibnamefont
  {Benhemou}}, \bibinfo {author} {\bibfnamefont {J.~K.}\ \bibnamefont
  {Pachos}},\ and\ \bibinfo {author} {\bibfnamefont {D.~E.}\ \bibnamefont
  {Browne}},\ }\bibfield  {title} {\bibinfo {title} {{Non-Abelian statistics
  with mixed-boundary punctures on the toric code}},\ }\href
  {https://doi.org/10.1103/PhysRevA.105.042417} {\bibfield  {journal} {\bibinfo
   {journal} {Phys. Rev. A}\ }\textbf {\bibinfo {volume} {105}},\ \bibinfo
  {pages} {042417} (\bibinfo {year} {2022})}\BibitemShut {NoStop}%
\bibitem [{\citenamefont {Ellison}\ \emph {et~al.}(2022)\citenamefont
  {Ellison}, \citenamefont {Chen}, \citenamefont {Dua}, \citenamefont
  {Shirley}, \citenamefont {Tantivasadakarn},\ and\ \citenamefont
  {Williamson}}]{Ellison2022}%
  \BibitemOpen
  \bibfield  {author} {\bibinfo {author} {\bibfnamefont {T.~D.}\ \bibnamefont
  {Ellison}}, \bibinfo {author} {\bibfnamefont {Y.-A.}\ \bibnamefont {Chen}},
  \bibinfo {author} {\bibfnamefont {A.}~\bibnamefont {Dua}}, \bibinfo {author}
  {\bibfnamefont {W.}~\bibnamefont {Shirley}}, \bibinfo {author} {\bibfnamefont
  {N.}~\bibnamefont {Tantivasadakarn}},\ and\ \bibinfo {author} {\bibfnamefont
  {D.~J.}\ \bibnamefont {Williamson}},\ }\bibfield  {title} {\bibinfo {title}
  {{Pauli Stabilizer Models of Twisted Quantum Doubles}},\ }\href
  {https://doi.org/10.1103/PRXQuantum.3.010353} {\bibfield  {journal} {\bibinfo
   {journal} {PRX Quantum}\ }\textbf {\bibinfo {volume} {3}},\ \bibinfo {pages}
  {010353} (\bibinfo {year} {2022})}\BibitemShut {NoStop}%
\bibitem [{\citenamefont {Kesselring}\ \emph {et~al.}(2022)\citenamefont
  {Kesselring}, \citenamefont {de~la Fuente}, \citenamefont {Thomsen},
  \citenamefont {Eisert}, \citenamefont {Bartlett},\ and\ \citenamefont
  {Brown}}]{Brown2022}%
  \BibitemOpen
  \bibfield  {author} {\bibinfo {author} {\bibfnamefont {M.~S.}\ \bibnamefont
  {Kesselring}}, \bibinfo {author} {\bibfnamefont {J.~C.~M.}\ \bibnamefont
  {de~la Fuente}}, \bibinfo {author} {\bibfnamefont {F.}~\bibnamefont
  {Thomsen}}, \bibinfo {author} {\bibfnamefont {J.}~\bibnamefont {Eisert}},
  \bibinfo {author} {\bibfnamefont {S.~D.}\ \bibnamefont {Bartlett}},\ and\
  \bibinfo {author} {\bibfnamefont {B.~J.}\ \bibnamefont {Brown}},\ }\href@noop
  {} {\bibinfo {title} {Anyon condensation and the color code}} (\bibinfo
  {year} {2022}),\ \Eprint {https://arxiv.org/abs/2212.00042} {arXiv:2212.00042
  [quant-ph]} \BibitemShut {NoStop}%
\bibitem [{\citenamefont {Wen}(1991)}]{Wen1991}%
  \BibitemOpen
  \bibfield  {author} {\bibinfo {author} {\bibfnamefont {X.~G.}\ \bibnamefont
  {Wen}},\ }\bibfield  {title} {\bibinfo {title} {Mean-field theory of
  spin-liquid states with finite energy gap and topological orders},\ }\href
  {https://doi.org/10.1103/PhysRevB.44.2664} {\bibfield  {journal} {\bibinfo
  {journal} {Phys. Rev. B}\ }\textbf {\bibinfo {volume} {44}},\ \bibinfo
  {pages} {2664} (\bibinfo {year} {1991})}\BibitemShut {NoStop}%
\bibitem [{\citenamefont {Sachdev}(1992)}]{Sachdev1992}%
  \BibitemOpen
  \bibfield  {author} {\bibinfo {author} {\bibfnamefont {S.}~\bibnamefont
  {Sachdev}},\ }\bibfield  {title} {\bibinfo {title} {{Kagome- and
  triangular-lattice Heisenberg antiferromagnets: Ordering from quantum
  fluctuations and quantum-disordered ground states with unconfined bosonic
  spinons}},\ }\href {https://doi.org/10.1103/PhysRevB.45.12377} {\bibfield
  {journal} {\bibinfo  {journal} {Phys. Rev. B}\ }\textbf {\bibinfo {volume}
  {45}},\ \bibinfo {pages} {12377} (\bibinfo {year} {1992})}\BibitemShut
  {NoStop}%
\bibitem [{\citenamefont {Senthil}\ and\ \citenamefont
  {Fisher}(2000)}]{Senthil2000}%
  \BibitemOpen
  \bibfield  {author} {\bibinfo {author} {\bibfnamefont {T.}~\bibnamefont
  {Senthil}}\ and\ \bibinfo {author} {\bibfnamefont {M.~P.~A.}\ \bibnamefont
  {Fisher}},\ }\bibfield  {title} {\bibinfo {title} {${Z}_{2}$ gauge theory of
  electron fractionalization in strongly correlated systems},\ }\href
  {https://doi.org/10.1103/PhysRevB.62.7850} {\bibfield  {journal} {\bibinfo
  {journal} {Phys. Rev. B}\ }\textbf {\bibinfo {volume} {62}},\ \bibinfo
  {pages} {7850} (\bibinfo {year} {2000})}\BibitemShut {NoStop}%
\bibitem [{\citenamefont {Moessner}\ \emph {et~al.}(2001)\citenamefont
  {Moessner}, \citenamefont {Sondhi},\ and\ \citenamefont
  {Fradkin}}]{Fradkin2001}%
  \BibitemOpen
  \bibfield  {author} {\bibinfo {author} {\bibfnamefont {R.}~\bibnamefont
  {Moessner}}, \bibinfo {author} {\bibfnamefont {S.~L.}\ \bibnamefont
  {Sondhi}},\ and\ \bibinfo {author} {\bibfnamefont {E.}~\bibnamefont
  {Fradkin}},\ }\bibfield  {title} {\bibinfo {title} {{Short-ranged resonating
  valence bond physics, quantum dimer models, and Ising gauge theories}},\
  }\href {https://doi.org/10.1103/PhysRevB.65.024504} {\bibfield  {journal}
  {\bibinfo  {journal} {Phys. Rev. B}\ }\textbf {\bibinfo {volume} {65}},\
  \bibinfo {pages} {024504} (\bibinfo {year} {2001})}\BibitemShut {NoStop}%
\bibitem [{\citenamefont {Balents}\ \emph {et~al.}(2002)\citenamefont
  {Balents}, \citenamefont {Fisher},\ and\ \citenamefont
  {Girvin}}]{Balents2002}%
  \BibitemOpen
  \bibfield  {author} {\bibinfo {author} {\bibfnamefont {L.}~\bibnamefont
  {Balents}}, \bibinfo {author} {\bibfnamefont {M.~P.~A.}\ \bibnamefont
  {Fisher}},\ and\ \bibinfo {author} {\bibfnamefont {S.~M.}\ \bibnamefont
  {Girvin}},\ }\bibfield  {title} {\bibinfo {title} {Fractionalization in an
  easy-axis kagome antiferromagnet},\ }\href
  {https://doi.org/10.1103/PhysRevB.65.224412} {\bibfield  {journal} {\bibinfo
  {journal} {Phys. Rev. B}\ }\textbf {\bibinfo {volume} {65}},\ \bibinfo
  {pages} {224412} (\bibinfo {year} {2002})}\BibitemShut {NoStop}%
\bibitem [{\citenamefont {Wang}\ and\ \citenamefont
  {Vishwanath}(2006)}]{Wang2006}%
  \BibitemOpen
  \bibfield  {author} {\bibinfo {author} {\bibfnamefont {F.}~\bibnamefont
  {Wang}}\ and\ \bibinfo {author} {\bibfnamefont {A.}~\bibnamefont
  {Vishwanath}},\ }\bibfield  {title} {\bibinfo {title} {{Spin-liquid states on
  the triangular and Kagom\'e lattices: A projective-symmetry-group analysis of
  Schwinger boson states}},\ }\href
  {https://doi.org/10.1103/PhysRevB.74.174423} {\bibfield  {journal} {\bibinfo
  {journal} {Phys. Rev. B}\ }\textbf {\bibinfo {volume} {74}},\ \bibinfo
  {pages} {174423} (\bibinfo {year} {2006})}\BibitemShut {NoStop}%
\bibitem [{\citenamefont {Schwandt}\ \emph {et~al.}(2010)\citenamefont
  {Schwandt}, \citenamefont {Mambrini},\ and\ \citenamefont
  {Poilblanc}}]{Poilblanc2010}%
  \BibitemOpen
  \bibfield  {author} {\bibinfo {author} {\bibfnamefont {D.}~\bibnamefont
  {Schwandt}}, \bibinfo {author} {\bibfnamefont {M.}~\bibnamefont {Mambrini}},\
  and\ \bibinfo {author} {\bibfnamefont {D.}~\bibnamefont {Poilblanc}},\
  }\bibfield  {title} {\bibinfo {title} {{Generalized hard-core dimer model
  approach to low-energy Heisenberg frustrated antiferromagnets: General
  properties and application to the kagome antiferromagnet}},\ }\href
  {https://doi.org/10.1103/PhysRevB.81.214413} {\bibfield  {journal} {\bibinfo
  {journal} {Phys. Rev. B}\ }\textbf {\bibinfo {volume} {81}},\ \bibinfo
  {pages} {214413} (\bibinfo {year} {2010})}\BibitemShut {NoStop}%
\bibitem [{\citenamefont {Yan}\ \emph {et~al.}(2011)\citenamefont {Yan},
  \citenamefont {Huse},\ and\ \citenamefont {White}}]{White2011}%
  \BibitemOpen
  \bibfield  {author} {\bibinfo {author} {\bibfnamefont {S.}~\bibnamefont
  {Yan}}, \bibinfo {author} {\bibfnamefont {D.~A.}\ \bibnamefont {Huse}},\ and\
  \bibinfo {author} {\bibfnamefont {S.~R.}\ \bibnamefont {White}},\ }\bibfield
  {title} {\bibinfo {title} {{Spin-Liquid Ground State of the $S=$1/2 Kagome
  Heisenberg Antiferromagnet}},\ }\href
  {https://doi.org/10.1126/science.1201080} {\bibfield  {journal} {\bibinfo
  {journal} {Science}\ }\textbf {\bibinfo {volume} {332}},\ \bibinfo {pages}
  {1173} (\bibinfo {year} {2011})}\BibitemShut {NoStop}%
\bibitem [{\citenamefont {Depenbrock}\ \emph {et~al.}(2012)\citenamefont
  {Depenbrock}, \citenamefont {McCulloch},\ and\ \citenamefont
  {Schollw\"ock}}]{Depenbrock2012}%
  \BibitemOpen
  \bibfield  {author} {\bibinfo {author} {\bibfnamefont {S.}~\bibnamefont
  {Depenbrock}}, \bibinfo {author} {\bibfnamefont {I.~P.}\ \bibnamefont
  {McCulloch}},\ and\ \bibinfo {author} {\bibfnamefont {U.}~\bibnamefont
  {Schollw\"ock}},\ }\bibfield  {title} {\bibinfo {title} {{Nature of the
  Spin-Liquid Ground State of the $S=1/2$ Heisenberg Model on the Kagome
  Lattice}},\ }\href {https://doi.org/10.1103/PhysRevLett.109.067201}
  {\bibfield  {journal} {\bibinfo  {journal} {Phys. Rev. Lett.}\ }\textbf
  {\bibinfo {volume} {109}},\ \bibinfo {pages} {067201} (\bibinfo {year}
  {2012})}\BibitemShut {NoStop}%
\bibitem [{\citenamefont {Lu}\ \emph {et~al.}(2011)\citenamefont {Lu},
  \citenamefont {Ran},\ and\ \citenamefont {Lee}}]{Lu2011}%
  \BibitemOpen
  \bibfield  {author} {\bibinfo {author} {\bibfnamefont {Y.-M.}\ \bibnamefont
  {Lu}}, \bibinfo {author} {\bibfnamefont {Y.}~\bibnamefont {Ran}},\ and\
  \bibinfo {author} {\bibfnamefont {P.~A.}\ \bibnamefont {Lee}},\ }\bibfield
  {title} {\bibinfo {title} {{${\mathbb{Z}}_{2}$ spin liquids in the
  $S=\frac{1}{2}$ Heisenberg model on the kagome lattice: A projective
  symmetry-group study of Schwinger fermion mean-field states}},\ }\href
  {https://doi.org/10.1103/PhysRevB.83.224413} {\bibfield  {journal} {\bibinfo
  {journal} {Phys. Rev. B}\ }\textbf {\bibinfo {volume} {83}},\ \bibinfo
  {pages} {224413} (\bibinfo {year} {2011})}\BibitemShut {NoStop}%
\bibitem [{\citenamefont {Iqbal}\ \emph {et~al.}(2011)\citenamefont {Iqbal},
  \citenamefont {Becca},\ and\ \citenamefont {Poilblanc}}]{Iqbal2011}%
  \BibitemOpen
  \bibfield  {author} {\bibinfo {author} {\bibfnamefont {Y.}~\bibnamefont
  {Iqbal}}, \bibinfo {author} {\bibfnamefont {F.}~\bibnamefont {Becca}},\ and\
  \bibinfo {author} {\bibfnamefont {D.}~\bibnamefont {Poilblanc}},\ }\bibfield
  {title} {\bibinfo {title} {{Projected wave function study of
  ${\mathbb{Z}}_{2}$ spin liquids on the kagome lattice for the
  spin-$\frac{1}{2}$ quantum Heisenberg antiferromagnet}},\ }\href
  {https://doi.org/10.1103/PhysRevB.84.020407} {\bibfield  {journal} {\bibinfo
  {journal} {Phys. Rev. B}\ }\textbf {\bibinfo {volume} {84}},\ \bibinfo
  {pages} {020407} (\bibinfo {year} {2011})}\BibitemShut {NoStop}%
\bibitem [{\citenamefont {Read}\ and\ \citenamefont
  {Chakraborty}(1989)}]{Read1989}%
  \BibitemOpen
  \bibfield  {author} {\bibinfo {author} {\bibfnamefont {N.}~\bibnamefont
  {Read}}\ and\ \bibinfo {author} {\bibfnamefont {B.}~\bibnamefont
  {Chakraborty}},\ }\bibfield  {title} {\bibinfo {title} {Statistics of the
  excitations of the resonating-valence-bond state},\ }\href
  {https://doi.org/10.1103/PhysRevB.40.7133} {\bibfield  {journal} {\bibinfo
  {journal} {Phys. Rev. B}\ }\textbf {\bibinfo {volume} {40}},\ \bibinfo
  {pages} {7133} (\bibinfo {year} {1989})}\BibitemShut {NoStop}%
\bibitem [{\citenamefont {Huh}\ \emph {et~al.}(2011)\citenamefont {Huh},
  \citenamefont {Punk},\ and\ \citenamefont {Sachdev}}]{Huh2011}%
  \BibitemOpen
  \bibfield  {author} {\bibinfo {author} {\bibfnamefont {Y.}~\bibnamefont
  {Huh}}, \bibinfo {author} {\bibfnamefont {M.}~\bibnamefont {Punk}},\ and\
  \bibinfo {author} {\bibfnamefont {S.}~\bibnamefont {Sachdev}},\ }\bibfield
  {title} {\bibinfo {title} {{Vison states and confinement transitions of
  ${\mathbb{Z}}_{2}$ spin liquids on the kagome lattice}},\ }\href
  {https://doi.org/10.1103/PhysRevB.84.094419} {\bibfield  {journal} {\bibinfo
  {journal} {Phys. Rev. B}\ }\textbf {\bibinfo {volume} {84}},\ \bibinfo
  {pages} {094419} (\bibinfo {year} {2011})}\BibitemShut {NoStop}%
\bibitem [{\citenamefont {Zhang}\ \emph {et~al.}(2012)\citenamefont {Zhang},
  \citenamefont {Grover}, \citenamefont {Turner}, \citenamefont {Oshikawa},\
  and\ \citenamefont {Vishwanath}}]{Zhang2012}%
  \BibitemOpen
  \bibfield  {author} {\bibinfo {author} {\bibfnamefont {Y.}~\bibnamefont
  {Zhang}}, \bibinfo {author} {\bibfnamefont {T.}~\bibnamefont {Grover}},
  \bibinfo {author} {\bibfnamefont {A.}~\bibnamefont {Turner}}, \bibinfo
  {author} {\bibfnamefont {M.}~\bibnamefont {Oshikawa}},\ and\ \bibinfo
  {author} {\bibfnamefont {A.}~\bibnamefont {Vishwanath}},\ }\bibfield  {title}
  {\bibinfo {title} {{Quasiparticle statistics and braiding from ground-state
  entanglement}},\ }\href {https://doi.org/10.1103/PhysRevB.85.235151}
  {\bibfield  {journal} {\bibinfo  {journal} {Phys. Rev. B}\ }\textbf {\bibinfo
  {volume} {85}},\ \bibinfo {pages} {235151} (\bibinfo {year}
  {2012})}\BibitemShut {NoStop}%
\bibitem [{\citenamefont {Jiang}\ \emph {et~al.}(2012)\citenamefont {Jiang},
  \citenamefont {Wang},\ and\ \citenamefont {Balents}}]{Jiang2012}%
  \BibitemOpen
  \bibfield  {author} {\bibinfo {author} {\bibfnamefont {H.-C.}\ \bibnamefont
  {Jiang}}, \bibinfo {author} {\bibfnamefont {Z.}~\bibnamefont {Wang}},\ and\
  \bibinfo {author} {\bibfnamefont {L.}~\bibnamefont {Balents}},\ }\bibfield
  {title} {\bibinfo {title} {{Identifying topological order by entanglement
  entropy}},\ }\href {https://doi.org/10.1038/nphys2465} {\bibfield  {journal}
  {\bibinfo  {journal} {Nat. Phys.}\ }\textbf {\bibinfo {volume} {8}},\
  \bibinfo {pages} {902} (\bibinfo {year} {2012})}\BibitemShut {NoStop}%
\bibitem [{\citenamefont {Messio}\ \emph {et~al.}(2012)\citenamefont {Messio},
  \citenamefont {Bernu},\ and\ \citenamefont {Lhuillier}}]{Messio2012}%
  \BibitemOpen
  \bibfield  {author} {\bibinfo {author} {\bibfnamefont {L.}~\bibnamefont
  {Messio}}, \bibinfo {author} {\bibfnamefont {B.}~\bibnamefont {Bernu}},\ and\
  \bibinfo {author} {\bibfnamefont {C.}~\bibnamefont {Lhuillier}},\ }\bibfield
  {title} {\bibinfo {title} {{Kagome Antiferromagnet: A Chiral Topological Spin
  Liquid?}},\ }\href {https://doi.org/10.1103/PhysRevLett.108.207204}
  {\bibfield  {journal} {\bibinfo  {journal} {Phys. Rev. Lett.}\ }\textbf
  {\bibinfo {volume} {108}},\ \bibinfo {pages} {207204} (\bibinfo {year}
  {2012})}\BibitemShut {NoStop}%
\bibitem [{\citenamefont {Essin}\ and\ \citenamefont
  {Hermele}(2013)}]{Hermele2013}%
  \BibitemOpen
  \bibfield  {author} {\bibinfo {author} {\bibfnamefont {A.~M.}\ \bibnamefont
  {Essin}}\ and\ \bibinfo {author} {\bibfnamefont {M.}~\bibnamefont
  {Hermele}},\ }\bibfield  {title} {\bibinfo {title} {{Classifying
  fractionalization: Symmetry classification of gapped ${\mathbb{Z}}_{2}$ spin
  liquids in two dimensions}},\ }\href
  {https://doi.org/10.1103/PhysRevB.87.104406} {\bibfield  {journal} {\bibinfo
  {journal} {Phys. Rev. B}\ }\textbf {\bibinfo {volume} {87}},\ \bibinfo
  {pages} {104406} (\bibinfo {year} {2013})}\BibitemShut {NoStop}%
\bibitem [{\citenamefont {Wan}\ and\ \citenamefont
  {Tchernyshyov}(2013)}]{Wan2013}%
  \BibitemOpen
  \bibfield  {author} {\bibinfo {author} {\bibfnamefont {Y.}~\bibnamefont
  {Wan}}\ and\ \bibinfo {author} {\bibfnamefont {O.}~\bibnamefont
  {Tchernyshyov}},\ }\bibfield  {title} {\bibinfo {title} {{Phenomenological
  ${Z}_{2}$ lattice gauge theory of the spin-liquid state of the kagome
  Heisenberg antiferromagnet}},\ }\href
  {https://doi.org/10.1103/PhysRevB.87.104408} {\bibfield  {journal} {\bibinfo
  {journal} {Phys. Rev. B}\ }\textbf {\bibinfo {volume} {87}},\ \bibinfo
  {pages} {104408} (\bibinfo {year} {2013})}\BibitemShut {NoStop}%
\bibitem [{\citenamefont {Hwang}\ \emph {et~al.}(2015)\citenamefont {Hwang},
  \citenamefont {Huh},\ and\ \citenamefont {Kim}}]{Hwang2015}%
  \BibitemOpen
  \bibfield  {author} {\bibinfo {author} {\bibfnamefont {K.}~\bibnamefont
  {Hwang}}, \bibinfo {author} {\bibfnamefont {Y.}~\bibnamefont {Huh}},\ and\
  \bibinfo {author} {\bibfnamefont {Y.~B.}\ \bibnamefont {Kim}},\ }\bibfield
  {title} {\bibinfo {title} {{$\mathbb{Z}_2$ gauge theory for valence bond
  solids on the kagome lattice}},\ }\href
  {https://doi.org/10.1103/PhysRevB.92.205131} {\bibfield  {journal} {\bibinfo
  {journal} {Phys. Rev. B}\ }\textbf {\bibinfo {volume} {92}},\ \bibinfo
  {pages} {205131} (\bibinfo {year} {2015})}\BibitemShut {NoStop}%
\bibitem [{\citenamefont {Zaletel}\ and\ \citenamefont
  {Vishwanath}(2015)}]{Zaletel2015}%
  \BibitemOpen
  \bibfield  {author} {\bibinfo {author} {\bibfnamefont {M.~P.}\ \bibnamefont
  {Zaletel}}\ and\ \bibinfo {author} {\bibfnamefont {A.}~\bibnamefont
  {Vishwanath}},\ }\bibfield  {title} {\bibinfo {title} {{Constraints on
  Topological Order in Mott Insulators}},\ }\href
  {https://doi.org/10.1103/PhysRevLett.114.077201} {\bibfield  {journal}
  {\bibinfo  {journal} {Phys. Rev. Lett.}\ }\textbf {\bibinfo {volume} {114}},\
  \bibinfo {pages} {077201} (\bibinfo {year} {2015})}\BibitemShut {NoStop}%
\bibitem [{\citenamefont {Lu}\ \emph {et~al.}(2017)\citenamefont {Lu},
  \citenamefont {Cho},\ and\ \citenamefont {Vishwanath}}]{Lu2017}%
  \BibitemOpen
  \bibfield  {author} {\bibinfo {author} {\bibfnamefont {Y.-M.}\ \bibnamefont
  {Lu}}, \bibinfo {author} {\bibfnamefont {G.~Y.}\ \bibnamefont {Cho}},\ and\
  \bibinfo {author} {\bibfnamefont {A.}~\bibnamefont {Vishwanath}},\ }\bibfield
   {title} {\bibinfo {title} {{Unification of bosonic and fermionic theories of
  spin liquids on the kagome lattice}},\ }\href
  {https://doi.org/10.1103/PhysRevB.96.205150} {\bibfield  {journal} {\bibinfo
  {journal} {Phys. Rev. B}\ }\textbf {\bibinfo {volume} {96}},\ \bibinfo
  {pages} {205150} (\bibinfo {year} {2017})}\BibitemShut {NoStop}%
\bibitem [{\citenamefont {Sun}\ \emph {et~al.}(2018)\citenamefont {Sun},
  \citenamefont {Wang}, \citenamefont {Fang}, \citenamefont {Qi}, \citenamefont
  {Cheng},\ and\ \citenamefont {Meng}}]{Meng2018BFG}%
  \BibitemOpen
  \bibfield  {author} {\bibinfo {author} {\bibfnamefont {G.-Y.}\ \bibnamefont
  {Sun}}, \bibinfo {author} {\bibfnamefont {Y.-C.}\ \bibnamefont {Wang}},
  \bibinfo {author} {\bibfnamefont {C.}~\bibnamefont {Fang}}, \bibinfo {author}
  {\bibfnamefont {Y.}~\bibnamefont {Qi}}, \bibinfo {author} {\bibfnamefont
  {M.}~\bibnamefont {Cheng}},\ and\ \bibinfo {author} {\bibfnamefont {Z.~Y.}\
  \bibnamefont {Meng}},\ }\bibfield  {title} {\bibinfo {title} {{Dynamical
  Signature of Symmetry Fractionalization in Frustrated Magnets}},\ }\href
  {https://doi.org/10.1103/PhysRevLett.121.077201} {\bibfield  {journal}
  {\bibinfo  {journal} {Phys. Rev. Lett.}\ }\textbf {\bibinfo {volume} {121}},\
  \bibinfo {pages} {077201} (\bibinfo {year} {2018})}\BibitemShut {NoStop}%
\bibitem [{\citenamefont {Becker}\ and\ \citenamefont
  {Wessel}(2018)}]{Wessel2018BFG}%
  \BibitemOpen
  \bibfield  {author} {\bibinfo {author} {\bibfnamefont {J.}~\bibnamefont
  {Becker}}\ and\ \bibinfo {author} {\bibfnamefont {S.}~\bibnamefont
  {Wessel}},\ }\bibfield  {title} {\bibinfo {title} {{Diagnosing
  Fractionalization from the Spin Dynamics of ${Z}_{2}$ Spin Liquids on the
  Kagome Lattice by Quantum Monte Carlo Simulations}},\ }\href
  {https://doi.org/10.1103/PhysRevLett.121.077202} {\bibfield  {journal}
  {\bibinfo  {journal} {Phys. Rev. Lett.}\ }\textbf {\bibinfo {volume} {121}},\
  \bibinfo {pages} {077202} (\bibinfo {year} {2018})}\BibitemShut {NoStop}%
\bibitem [{\citenamefont {Zhu}\ and\ \citenamefont {White}(2015)}]{White2015}%
  \BibitemOpen
  \bibfield  {author} {\bibinfo {author} {\bibfnamefont {Z.}~\bibnamefont
  {Zhu}}\ and\ \bibinfo {author} {\bibfnamefont {S.~R.}\ \bibnamefont
  {White}},\ }\bibfield  {title} {\bibinfo {title} {{Spin liquid phase of the
  $S=\frac{1}{2}\phantom{\rule{4.pt}{0ex}}{J}_{1}\ensuremath{-}{J}_{2}$
  Heisenberg model on the triangular lattice}},\ }\href
  {https://doi.org/10.1103/PhysRevB.92.041105} {\bibfield  {journal} {\bibinfo
  {journal} {Phys. Rev. B}\ }\textbf {\bibinfo {volume} {92}},\ \bibinfo
  {pages} {041105} (\bibinfo {year} {2015})}\BibitemShut {NoStop}%
\bibitem [{\citenamefont {Hu}\ \emph {et~al.}(2015)\citenamefont {Hu},
  \citenamefont {Gong}, \citenamefont {Zhu},\ and\ \citenamefont
  {Sheng}}]{Sheng2015}%
  \BibitemOpen
  \bibfield  {author} {\bibinfo {author} {\bibfnamefont {W.-J.}\ \bibnamefont
  {Hu}}, \bibinfo {author} {\bibfnamefont {S.-S.}\ \bibnamefont {Gong}},
  \bibinfo {author} {\bibfnamefont {W.}~\bibnamefont {Zhu}},\ and\ \bibinfo
  {author} {\bibfnamefont {D.~N.}\ \bibnamefont {Sheng}},\ }\bibfield  {title}
  {\bibinfo {title} {{Competing spin-liquid states in the spin-$\frac{1}{2}$
  Heisenberg model on the triangular lattice}},\ }\href
  {https://doi.org/10.1103/PhysRevB.92.140403} {\bibfield  {journal} {\bibinfo
  {journal} {Phys. Rev. B}\ }\textbf {\bibinfo {volume} {92}},\ \bibinfo
  {pages} {140403} (\bibinfo {year} {2015})}\BibitemShut {NoStop}%
\bibitem [{\citenamefont {Ralko}\ \emph {et~al.}(2007)\citenamefont {Ralko},
  \citenamefont {Ferrero}, \citenamefont {Becca}, \citenamefont {Ivanov},\ and\
  \citenamefont {Mila}}]{Mila2007}%
  \BibitemOpen
  \bibfield  {author} {\bibinfo {author} {\bibfnamefont {A.}~\bibnamefont
  {Ralko}}, \bibinfo {author} {\bibfnamefont {M.}~\bibnamefont {Ferrero}},
  \bibinfo {author} {\bibfnamefont {F.}~\bibnamefont {Becca}}, \bibinfo
  {author} {\bibfnamefont {D.}~\bibnamefont {Ivanov}},\ and\ \bibinfo {author}
  {\bibfnamefont {F.}~\bibnamefont {Mila}},\ }\bibfield  {title} {\bibinfo
  {title} {Crystallization of the resonating valence bond liquid as vortex
  condensation},\ }\href {https://doi.org/10.1103/PhysRevB.76.140404}
  {\bibfield  {journal} {\bibinfo  {journal} {Phys. Rev. B}\ }\textbf {\bibinfo
  {volume} {76}},\ \bibinfo {pages} {140404} (\bibinfo {year}
  {2007})}\BibitemShut {NoStop}%
\bibitem [{\citenamefont {Misguich}\ and\ \citenamefont
  {Mila}(2008)}]{Misguich2008}%
  \BibitemOpen
  \bibfield  {author} {\bibinfo {author} {\bibfnamefont {G.}~\bibnamefont
  {Misguich}}\ and\ \bibinfo {author} {\bibfnamefont {F.}~\bibnamefont
  {Mila}},\ }\bibfield  {title} {\bibinfo {title} {{Quantum dimer model on the
  triangular lattice: Semiclassical and variational approaches to vison
  dispersion and condensation}},\ }\href
  {https://doi.org/10.1103/PhysRevB.77.134421} {\bibfield  {journal} {\bibinfo
  {journal} {Phys. Rev. B}\ }\textbf {\bibinfo {volume} {77}},\ \bibinfo
  {pages} {134421} (\bibinfo {year} {2008})}\BibitemShut {NoStop}%
\bibitem [{\citenamefont {Slagle}\ and\ \citenamefont {Xu}(2014)}]{Slagle2014}%
  \BibitemOpen
  \bibfield  {author} {\bibinfo {author} {\bibfnamefont {K.}~\bibnamefont
  {Slagle}}\ and\ \bibinfo {author} {\bibfnamefont {C.}~\bibnamefont {Xu}},\
  }\bibfield  {title} {\bibinfo {title} {Quantum phase transition between the
  ${Z}_{2}$ spin liquid and valence bond crystals on a triangular lattice},\
  }\href {https://doi.org/10.1103/PhysRevB.89.104418} {\bibfield  {journal}
  {\bibinfo  {journal} {Phys. Rev. B}\ }\textbf {\bibinfo {volume} {89}},\
  \bibinfo {pages} {104418} (\bibinfo {year} {2014})}\BibitemShut {NoStop}%
\bibitem [{\citenamefont {Yan}\ \emph {et~al.}(2021)\citenamefont {Yan},
  \citenamefont {Wang}, \citenamefont {Ma}, \citenamefont {Qi},\ and\
  \citenamefont {Meng}}]{Meng2021QDM}%
  \BibitemOpen
  \bibfield  {author} {\bibinfo {author} {\bibfnamefont {Z.}~\bibnamefont
  {Yan}}, \bibinfo {author} {\bibfnamefont {Y.-C.}\ \bibnamefont {Wang}},
  \bibinfo {author} {\bibfnamefont {N.}~\bibnamefont {Ma}}, \bibinfo {author}
  {\bibfnamefont {Y.}~\bibnamefont {Qi}},\ and\ \bibinfo {author}
  {\bibfnamefont {Z.~Y.}\ \bibnamefont {Meng}},\ }\bibfield  {title} {\bibinfo
  {title} {{Topological phase transition and single/multi anyon dynamics of Z2
  spin liquid}},\ }\href {https://doi.org/10.1038/s41535-021-00338-1}
  {\bibfield  {journal} {\bibinfo  {journal} {npj Quantum Mater.}\ }\textbf
  {\bibinfo {volume} {6}},\ \bibinfo {pages} {39} (\bibinfo {year}
  {2021})}\BibitemShut {NoStop}%
\bibitem [{\citenamefont {Yan}\ \emph {et~al.}(2022)\citenamefont {Yan},
  \citenamefont {Samajdar}, \citenamefont {Wang}, \citenamefont {Sachdev},\
  and\ \citenamefont {Meng}}]{Meng2022QDM}%
  \BibitemOpen
  \bibfield  {author} {\bibinfo {author} {\bibfnamefont {Z.}~\bibnamefont
  {Yan}}, \bibinfo {author} {\bibfnamefont {R.}~\bibnamefont {Samajdar}},
  \bibinfo {author} {\bibfnamefont {Y.-C.}\ \bibnamefont {Wang}}, \bibinfo
  {author} {\bibfnamefont {S.}~\bibnamefont {Sachdev}},\ and\ \bibinfo {author}
  {\bibfnamefont {Z.~Y.}\ \bibnamefont {Meng}},\ }\bibfield  {title} {\bibinfo
  {title} {Triangular lattice quantum dimer model with variable dimer
  density},\ }\href {https://doi.org/10.1038/s41467-022-33431-5} {\bibfield
  {journal} {\bibinfo  {journal} {Nat. Commun.}\ }\textbf {\bibinfo {volume}
  {13}},\ \bibinfo {pages} {5799} (\bibinfo {year} {2022})}\BibitemShut
  {NoStop}%
\bibitem [{\citenamefont {Hwang}\ \emph {et~al.}(2022)\citenamefont {Hwang},
  \citenamefont {Go}, \citenamefont {Seong}, \citenamefont {Shibauchi},\ and\
  \citenamefont {Moon}}]{Hwang2022}%
  \BibitemOpen
  \bibfield  {author} {\bibinfo {author} {\bibfnamefont {K.}~\bibnamefont
  {Hwang}}, \bibinfo {author} {\bibfnamefont {A.}~\bibnamefont {Go}}, \bibinfo
  {author} {\bibfnamefont {J.~H.}\ \bibnamefont {Seong}}, \bibinfo {author}
  {\bibfnamefont {T.}~\bibnamefont {Shibauchi}},\ and\ \bibinfo {author}
  {\bibfnamefont {E.-G.}\ \bibnamefont {Moon}},\ }\bibfield  {title} {\bibinfo
  {title} {{Identification of a Kitaev quantum spin liquid by magnetic field
  angle dependence}},\ }\href {https://doi.org/10.1038/s41467-021-27943-9}
  {\bibfield  {journal} {\bibinfo  {journal} {Nat. Commun.}\ }\textbf {\bibinfo
  {volume} {13}},\ \bibinfo {pages} {323} (\bibinfo {year} {2022})}\BibitemShut
  {NoStop}%
\bibitem [{\citenamefont {Wen}(1990)}]{Wen1990}%
  \BibitemOpen
  \bibfield  {author} {\bibinfo {author} {\bibfnamefont {X.~G.}\ \bibnamefont
  {Wen}},\ }\bibfield  {title} {\bibinfo {title} {Topological orders in rigid
  states},\ }\href {https://doi.org/10.1142/S0217979290000139} {\bibfield
  {journal} {\bibinfo  {journal} {International Journal of Modern Physics B}\
  }\textbf {\bibinfo {volume} {04}},\ \bibinfo {pages} {239} (\bibinfo {year}
  {1990})}\BibitemShut {NoStop}%
\bibitem [{\citenamefont {Wen}\ and\ \citenamefont {Niu}(1990)}]{WenNiu1990}%
  \BibitemOpen
  \bibfield  {author} {\bibinfo {author} {\bibfnamefont {X.~G.}\ \bibnamefont
  {Wen}}\ and\ \bibinfo {author} {\bibfnamefont {Q.}~\bibnamefont {Niu}},\
  }\bibfield  {title} {\bibinfo {title} {{Ground-state degeneracy of the
  fractional quantum Hall states in the presence of a random potential and on
  high-genus Riemann surfaces}},\ }\href
  {https://doi.org/10.1103/PhysRevB.41.9377} {\bibfield  {journal} {\bibinfo
  {journal} {Phys. Rev. B}\ }\textbf {\bibinfo {volume} {41}},\ \bibinfo
  {pages} {9377} (\bibinfo {year} {1990})}\BibitemShut {NoStop}%
\bibitem [{\citenamefont {Zhang}\ \emph {et~al.}(2020)\citenamefont {Zhang},
  \citenamefont {Batista},\ and\ \citenamefont {Hal\'asz}}]{Zhang2020}%
  \BibitemOpen
  \bibfield  {author} {\bibinfo {author} {\bibfnamefont {S.-S.}\ \bibnamefont
  {Zhang}}, \bibinfo {author} {\bibfnamefont {C.~D.}\ \bibnamefont {Batista}},\
  and\ \bibinfo {author} {\bibfnamefont {G.~B.}\ \bibnamefont {Hal\'asz}},\
  }\bibfield  {title} {\bibinfo {title} {{Toward Kitaev's sixteenfold way in a
  honeycomb lattice model}},\ }\href
  {https://doi.org/10.1103/PhysRevResearch.2.023334} {\bibfield  {journal}
  {\bibinfo  {journal} {Phys. Rev. Res.}\ }\textbf {\bibinfo {volume} {2}},\
  \bibinfo {pages} {023334} (\bibinfo {year} {2020})}\BibitemShut {NoStop}%
\bibitem [{\citenamefont {Fuchs}\ \emph {et~al.}(2020)\citenamefont {Fuchs},
  \citenamefont {Patil},\ and\ \citenamefont {Vidal}}]{Vidal2020}%
  \BibitemOpen
  \bibfield  {author} {\bibinfo {author} {\bibfnamefont {J.-N.}\ \bibnamefont
  {Fuchs}}, \bibinfo {author} {\bibfnamefont {S.}~\bibnamefont {Patil}},\ and\
  \bibinfo {author} {\bibfnamefont {J.}~\bibnamefont {Vidal}},\ }\bibfield
  {title} {\bibinfo {title} {{Parity of Chern numbers in the Kitaev honeycomb
  model and the sixteenfold way}},\ }\href
  {https://doi.org/10.1103/PhysRevB.102.115130} {\bibfield  {journal} {\bibinfo
   {journal} {Phys. Rev. B}\ }\textbf {\bibinfo {volume} {102}},\ \bibinfo
  {pages} {115130} (\bibinfo {year} {2020})}\BibitemShut {NoStop}%
\bibitem [{\citenamefont {Chulliparambil}\ \emph {et~al.}(2020)\citenamefont
  {Chulliparambil}, \citenamefont {Seifert}, \citenamefont {Vojta},
  \citenamefont {Janssen},\ and\ \citenamefont {Tu}}]{Tu2020}%
  \BibitemOpen
  \bibfield  {author} {\bibinfo {author} {\bibfnamefont {S.}~\bibnamefont
  {Chulliparambil}}, \bibinfo {author} {\bibfnamefont {U.~F.~P.}\ \bibnamefont
  {Seifert}}, \bibinfo {author} {\bibfnamefont {M.}~\bibnamefont {Vojta}},
  \bibinfo {author} {\bibfnamefont {L.}~\bibnamefont {Janssen}},\ and\ \bibinfo
  {author} {\bibfnamefont {H.-H.}\ \bibnamefont {Tu}},\ }\bibfield  {title}
  {\bibinfo {title} {{Microscopic models for Kitaev's sixteenfold way of anyon
  theories}},\ }\href {https://doi.org/10.1103/PhysRevB.102.201111} {\bibfield
  {journal} {\bibinfo  {journal} {Phys. Rev. B}\ }\textbf {\bibinfo {volume}
  {102}},\ \bibinfo {pages} {201111} (\bibinfo {year} {2020})}\BibitemShut
  {NoStop}%
\bibitem [{\citenamefont {Jin}\ \emph {et~al.}(2023)\citenamefont {Jin},
  \citenamefont {Miao},\ and\ \citenamefont {Zhou}}]{Zhou2023}%
  \BibitemOpen
  \bibfield  {author} {\bibinfo {author} {\bibfnamefont {J.-T.}\ \bibnamefont
  {Jin}}, \bibinfo {author} {\bibfnamefont {J.-J.}\ \bibnamefont {Miao}},\ and\
  \bibinfo {author} {\bibfnamefont {Y.}~\bibnamefont {Zhou}},\ }\bibfield
  {title} {\bibinfo {title} {{Lacing topological orders in two dimensions:
  Exactly solvable models for Kitaev's sixteen-fold way}},\ }\href
  {https://doi.org/10.21468/SciPostPhys.14.5.087} {\bibfield  {journal}
  {\bibinfo  {journal} {SciPost Phys.}\ }\textbf {\bibinfo {volume} {14}},\
  \bibinfo {pages} {087} (\bibinfo {year} {2023})}\BibitemShut {NoStop}%
\bibitem [{\citenamefont {Herbut}(2006)}]{Herbut2006}%
  \BibitemOpen
  \bibfield  {author} {\bibinfo {author} {\bibfnamefont {I.~F.}\ \bibnamefont
  {Herbut}},\ }\bibfield  {title} {\bibinfo {title} {{Interactions and Phase
  Transitions on Graphene's Honeycomb Lattice}},\ }\href
  {https://doi.org/10.1103/PhysRevLett.97.146401} {\bibfield  {journal}
  {\bibinfo  {journal} {Phys. Rev. Lett.}\ }\textbf {\bibinfo {volume} {97}},\
  \bibinfo {pages} {146401} (\bibinfo {year} {2006})}\BibitemShut {NoStop}%
\bibitem [{\citenamefont {Yang}\ \emph {et~al.}(2007)\citenamefont {Yang},
  \citenamefont {Zhou},\ and\ \citenamefont {Sun}}]{Yang2007}%
  \BibitemOpen
  \bibfield  {author} {\bibinfo {author} {\bibfnamefont {S.}~\bibnamefont
  {Yang}}, \bibinfo {author} {\bibfnamefont {D.~L.}\ \bibnamefont {Zhou}},\
  and\ \bibinfo {author} {\bibfnamefont {C.~P.}\ \bibnamefont {Sun}},\
  }\bibfield  {title} {\bibinfo {title} {{Mosaic spin models with topological
  order}},\ }\href {https://doi.org/10.1103/PhysRevB.76.180404} {\bibfield
  {journal} {\bibinfo  {journal} {Phys. Rev. B}\ }\textbf {\bibinfo {volume}
  {76}},\ \bibinfo {pages} {180404} (\bibinfo {year} {2007})}\BibitemShut
  {NoStop}%
\bibitem [{\citenamefont {Yao}\ and\ \citenamefont {Kivelson}(2007)}]{Yao2007}%
  \BibitemOpen
  \bibfield  {author} {\bibinfo {author} {\bibfnamefont {H.}~\bibnamefont
  {Yao}}\ and\ \bibinfo {author} {\bibfnamefont {S.~A.}\ \bibnamefont
  {Kivelson}},\ }\bibfield  {title} {\bibinfo {title} {{Exact Chiral Spin
  Liquid with Non-Abelian Anyons}},\ }\href
  {https://doi.org/10.1103/PhysRevLett.99.247203} {\bibfield  {journal}
  {\bibinfo  {journal} {Phys. Rev. Lett.}\ }\textbf {\bibinfo {volume} {99}},\
  \bibinfo {pages} {247203} (\bibinfo {year} {2007})}\BibitemShut {NoStop}%
\bibitem [{\citenamefont {Shibata}\ and\ \citenamefont
  {Katsura}(2019)}]{Katsura2019}%
  \BibitemOpen
  \bibfield  {author} {\bibinfo {author} {\bibfnamefont {N.}~\bibnamefont
  {Shibata}}\ and\ \bibinfo {author} {\bibfnamefont {H.}~\bibnamefont
  {Katsura}},\ }\bibfield  {title} {\bibinfo {title} {{Dissipative spin chain
  as a non-Hermitian Kitaev ladder}},\ }\href
  {https://doi.org/10.1103/PhysRevB.99.174303} {\bibfield  {journal} {\bibinfo
  {journal} {Phys. Rev. B}\ }\textbf {\bibinfo {volume} {99}},\ \bibinfo
  {pages} {174303} (\bibinfo {year} {2019})}\BibitemShut {NoStop}%
\bibitem [{\citenamefont {Yang}\ \emph {et~al.}(2021)\citenamefont {Yang},
  \citenamefont {Morampudi},\ and\ \citenamefont {Bergholtz}}]{Bergholtz2021}%
  \BibitemOpen
  \bibfield  {author} {\bibinfo {author} {\bibfnamefont {K.}~\bibnamefont
  {Yang}}, \bibinfo {author} {\bibfnamefont {S.~C.}\ \bibnamefont
  {Morampudi}},\ and\ \bibinfo {author} {\bibfnamefont {E.~J.}\ \bibnamefont
  {Bergholtz}},\ }\bibfield  {title} {\bibinfo {title} {{Exceptional Spin
  Liquids from Couplings to the Environment}},\ }\href
  {https://doi.org/10.1103/PhysRevLett.126.077201} {\bibfield  {journal}
  {\bibinfo  {journal} {Phys. Rev. Lett.}\ }\textbf {\bibinfo {volume} {126}},\
  \bibinfo {pages} {077201} (\bibinfo {year} {2021})}\BibitemShut {NoStop}%
\bibitem [{\citenamefont {Garratt}\ \emph {et~al.}(2023)\citenamefont
  {Garratt}, \citenamefont {Weinstein},\ and\ \citenamefont
  {Altman}}]{Altman2023}%
  \BibitemOpen
  \bibfield  {author} {\bibinfo {author} {\bibfnamefont {S.~J.}\ \bibnamefont
  {Garratt}}, \bibinfo {author} {\bibfnamefont {Z.}~\bibnamefont {Weinstein}},\
  and\ \bibinfo {author} {\bibfnamefont {E.}~\bibnamefont {Altman}},\
  }\bibfield  {title} {\bibinfo {title} {{Measurements Conspire Nonlocally to
  Restructure Critical Quantum States}},\ }\href
  {https://doi.org/10.1103/PhysRevX.13.021026} {\bibfield  {journal} {\bibinfo
  {journal} {Phys. Rev. X}\ }\textbf {\bibinfo {volume} {13}},\ \bibinfo
  {pages} {021026} (\bibinfo {year} {2023})}\BibitemShut {NoStop}%
\bibitem [{\citenamefont {Lee}\ \emph {et~al.}(2023)\citenamefont {Lee},
  \citenamefont {Jian},\ and\ \citenamefont {Xu}}]{Lee2023}%
  \BibitemOpen
  \bibfield  {author} {\bibinfo {author} {\bibfnamefont {J.~Y.}\ \bibnamefont
  {Lee}}, \bibinfo {author} {\bibfnamefont {C.-M.}\ \bibnamefont {Jian}},\ and\
  \bibinfo {author} {\bibfnamefont {C.}~\bibnamefont {Xu}},\ }\bibfield
  {title} {\bibinfo {title} {{Quantum Criticality Under Decoherence or Weak
  Measurement}},\ }\href {https://doi.org/10.1103/PRXQuantum.4.030317}
  {\bibfield  {journal} {\bibinfo  {journal} {PRX Quantum}\ }\textbf {\bibinfo
  {volume} {4}},\ \bibinfo {pages} {030317} (\bibinfo {year}
  {2023})}\BibitemShut {NoStop}%
\bibitem [{\citenamefont {Bao}\ \emph {et~al.}(2023)\citenamefont {Bao},
  \citenamefont {Fan}, \citenamefont {Vishwanath},\ and\ \citenamefont
  {Altman}}]{Bao2023mixedstate}%
  \BibitemOpen
  \bibfield  {author} {\bibinfo {author} {\bibfnamefont {Y.}~\bibnamefont
  {Bao}}, \bibinfo {author} {\bibfnamefont {R.}~\bibnamefont {Fan}}, \bibinfo
  {author} {\bibfnamefont {A.}~\bibnamefont {Vishwanath}},\ and\ \bibinfo
  {author} {\bibfnamefont {E.}~\bibnamefont {Altman}},\ }\href@noop {}
  {\bibinfo {title} {{Mixed-state topological order and the errorfield double
  formulation of decoherence-induced transitions}}} (\bibinfo {year} {2023}),\
  \Eprint {https://arxiv.org/abs/2301.05687} {arXiv:2301.05687 [quant-ph]}
  \BibitemShut {NoStop}%
\bibitem [{\citenamefont {Hwang}(2023)}]{Hwang2023mixedstate}%
  \BibitemOpen
  \bibfield  {author} {\bibinfo {author} {\bibfnamefont {K.}~\bibnamefont
  {Hwang}},\ }\href@noop {} {\bibinfo {title} {{Mixed-State Quantum Spin
  Liquids and Dynamical Anyon Condensations in Kitaev Lindbladians}}} (\bibinfo
  {year} {2023}),\ \Eprint {https://arxiv.org/abs/2305.09197} {arXiv:2305.09197
  [cond-mat.str-el]} \BibitemShut {NoStop}%
\bibitem [{\citenamefont {Duan}\ \emph {et~al.}(2003)\citenamefont {Duan},
  \citenamefont {Demler},\ and\ \citenamefont {Lukin}}]{Duan2003}%
  \BibitemOpen
  \bibfield  {author} {\bibinfo {author} {\bibfnamefont {L.-M.}\ \bibnamefont
  {Duan}}, \bibinfo {author} {\bibfnamefont {E.}~\bibnamefont {Demler}},\ and\
  \bibinfo {author} {\bibfnamefont {M.~D.}\ \bibnamefont {Lukin}},\ }\bibfield
  {title} {\bibinfo {title} {{Controlling Spin Exchange Interactions of
  Ultracold Atoms in Optical Lattices}},\ }\href
  {https://doi.org/10.1103/PhysRevLett.91.090402} {\bibfield  {journal}
  {\bibinfo  {journal} {Phys. Rev. Lett.}\ }\textbf {\bibinfo {volume} {91}},\
  \bibinfo {pages} {090402} (\bibinfo {year} {2003})}\BibitemShut {NoStop}%
\bibitem [{\citenamefont {Dai}\ \emph {et~al.}(2017)\citenamefont {Dai},
  \citenamefont {Yang}, \citenamefont {Reingruber}, \citenamefont {Sun},
  \citenamefont {Xu}, \citenamefont {Chen}, \citenamefont {Yuan},\ and\
  \citenamefont {Pan}}]{Pan2017}%
  \BibitemOpen
  \bibfield  {author} {\bibinfo {author} {\bibfnamefont {H.-N.}\ \bibnamefont
  {Dai}}, \bibinfo {author} {\bibfnamefont {B.}~\bibnamefont {Yang}}, \bibinfo
  {author} {\bibfnamefont {A.}~\bibnamefont {Reingruber}}, \bibinfo {author}
  {\bibfnamefont {H.}~\bibnamefont {Sun}}, \bibinfo {author} {\bibfnamefont
  {X.-F.}\ \bibnamefont {Xu}}, \bibinfo {author} {\bibfnamefont {Y.-A.}\
  \bibnamefont {Chen}}, \bibinfo {author} {\bibfnamefont {Z.-S.}\ \bibnamefont
  {Yuan}},\ and\ \bibinfo {author} {\bibfnamefont {J.-W.}\ \bibnamefont
  {Pan}},\ }\bibfield  {title} {\bibinfo {title} {{Four-body ring-exchange
  interactions and anyonic statistics within a minimal toric-code
  Hamiltonian}},\ }\href {https://doi.org/10.1038/nphys4243} {\bibfield
  {journal} {\bibinfo  {journal} {Nat. Phys.}\ }\textbf {\bibinfo {volume}
  {13}},\ \bibinfo {pages} {1195} (\bibinfo {year} {2017})}\BibitemShut
  {NoStop}%
\bibitem [{\citenamefont {Sun}\ \emph {et~al.}(2023)\citenamefont {Sun},
  \citenamefont {Goldman}, \citenamefont {Aidelsburger},\ and\ \citenamefont
  {Bukov}}]{Bukov2023}%
  \BibitemOpen
  \bibfield  {author} {\bibinfo {author} {\bibfnamefont {B.-Y.}\ \bibnamefont
  {Sun}}, \bibinfo {author} {\bibfnamefont {N.}~\bibnamefont {Goldman}},
  \bibinfo {author} {\bibfnamefont {M.}~\bibnamefont {Aidelsburger}},\ and\
  \bibinfo {author} {\bibfnamefont {M.}~\bibnamefont {Bukov}},\ }\bibfield
  {title} {\bibinfo {title} {{Engineering and Probing Non-Abelian Chiral Spin
  Liquids Using Periodically Driven Ultracold Atoms}},\ }\href
  {https://doi.org/10.1103/PRXQuantum.4.020329} {\bibfield  {journal} {\bibinfo
   {journal} {PRX Quantum}\ }\textbf {\bibinfo {volume} {4}},\ \bibinfo {pages}
  {020329} (\bibinfo {year} {2023})}\BibitemShut {NoStop}%
\bibitem [{\citenamefont {Kalinowski}\ \emph {et~al.}(2023)\citenamefont
  {Kalinowski}, \citenamefont {Maskara},\ and\ \citenamefont
  {Lukin}}]{Lukin2023}%
  \BibitemOpen
  \bibfield  {author} {\bibinfo {author} {\bibfnamefont {M.}~\bibnamefont
  {Kalinowski}}, \bibinfo {author} {\bibfnamefont {N.}~\bibnamefont
  {Maskara}},\ and\ \bibinfo {author} {\bibfnamefont {M.~D.}\ \bibnamefont
  {Lukin}},\ }\bibfield  {title} {\bibinfo {title} {{Non-Abelian Floquet Spin
  Liquids in a Digital Rydberg Simulator}},\ }\href
  {https://doi.org/10.1103/PhysRevX.13.031008} {\bibfield  {journal} {\bibinfo
  {journal} {Phys. Rev. X}\ }\textbf {\bibinfo {volume} {13}},\ \bibinfo
  {pages} {031008} (\bibinfo {year} {2023})}\BibitemShut {NoStop}%
\bibitem [{\citenamefont {Chen}\ \emph {et~al.}(2023)\citenamefont {Chen},
  \citenamefont {Wang}, \citenamefont {Poon}, \citenamefont {Zhou},
  \citenamefont {Liu},\ and\ \citenamefont {Liu}}]{Liu2023}%
  \BibitemOpen
  \bibfield  {author} {\bibinfo {author} {\bibfnamefont {Y.-H.}\ \bibnamefont
  {Chen}}, \bibinfo {author} {\bibfnamefont {B.-Z.}\ \bibnamefont {Wang}},
  \bibinfo {author} {\bibfnamefont {T.-F.~J.}\ \bibnamefont {Poon}}, \bibinfo
  {author} {\bibfnamefont {X.-C.}\ \bibnamefont {Zhou}}, \bibinfo {author}
  {\bibfnamefont {Z.-X.}\ \bibnamefont {Liu}},\ and\ \bibinfo {author}
  {\bibfnamefont {X.-J.}\ \bibnamefont {Liu}},\ }\href@noop {} {\bibinfo
  {title} {{Realization and detection of Kitaev quantum spin liquid with
  Rydberg atoms}}} (\bibinfo {year} {2023}),\ \Eprint
  {https://arxiv.org/abs/2310.12905} {arXiv:2310.12905 [cond-mat.str-el]}
  \BibitemShut {NoStop}%
\bibitem [{\citenamefont {Winkler}\ \emph {et~al.}(2003)\citenamefont
  {Winkler}, \citenamefont {Papadakis}, \citenamefont {De~Poortere},\ and\
  \citenamefont {Shayegan}}]{Winkler2003}%
  \BibitemOpen
  \bibfield  {author} {\bibinfo {author} {\bibfnamefont {R.}~\bibnamefont
  {Winkler}}, \bibinfo {author} {\bibfnamefont {S.}~\bibnamefont {Papadakis}},
  \bibinfo {author} {\bibfnamefont {E.}~\bibnamefont {De~Poortere}},\ and\
  \bibinfo {author} {\bibfnamefont {M.}~\bibnamefont {Shayegan}},\ }\href@noop
  {} {\emph {\bibinfo {title} {Spin-Orbit Coupling in Two-Dimensional Electron
  and Hole Systems}}},\ Vol.~\bibinfo {volume} {41}\ (\bibinfo  {publisher}
  {Springer},\ \bibinfo {year} {2003})\BibitemShut {NoStop}%
\end{thebibliography}

%

\end{document}